\documentclass[useAMS,usenatbib,twocolumn]{mn2e}

\usepackage{textcomp,graphicx,amsmath,pdflscape,amssymb,setspace,subfig,rotating,multirow,psfrag,color}
\usepackage[table]{xcolor}

\def\G{\textit{g$^\prime$}}
\def\R{\textit{r$^\prime$}}
\def\I{\textit{i$^\prime$}}

\def\vespa{{\scriptsize VESPA}} 
\def\snid{{\scriptsize SNID}}
\def\salt{{\scriptsize SALT2}}

\def\volrate{0.247^{+0.029}_{-0.026}~{\rm(statistical)}~^{+0.016}_{-0.031}~{\rm(systematic)}~\times 10^{-4}~{\rm SNe}~{\rm yr}^{-1}~{\rm Mpc}^{-3}}
\def\snumrate{0.10 \pm 0.01~(\rm statistical)~\pm 0.01~(\rm systematic)~{\rm SNuM}}
\def\dtdrate{$4.5 \pm 0.6~{\rm(statistical)}~^{+0.3}_{-0.5}~{\rm(systematic)}~\times 10^{-14}~{\rm SNe}~{\rm M_\odot}^{-1}~{\rm yr}^{-1}$}
\def\NMdtd{0.35 \pm 0.14~{\rm(statistical)}~^{+0.02}_{-0.04}~{\rm(systematic)}~\times 10^{-3}~{\rm SNe}~{\rm M_\odot}^{-1}}
\def\NMkis{0.43 ^{+0.04}_{-0.10}\times 10^{-3}~{\rm SNe}~{\rm M_\odot}^{-1}}

\newcommand\aj{{AJ}}%
\newcommand\apj{{ApJ}}%
\newcommand\apjl{{ApJ}}%
\newcommand\apjs{{ApJS}}%
\newcommand\aap{{A\&A}}%
\newcommand\mnras{{MNRAS}}%
\newcommand\pasp{{PASP}}%
\newcommand\pasj{{PASJ}}%
\newcommand\nat{{Nature}}%
\newcommand\nar{{NewAR}}%
\newcommand\pasa{{PASA}}%
\newcommand\scchg{ScChG}

\title[Supernovae in SDSS DR7 Spectra]{Discovery of $90$ Type Ia supernovae among $700{,}000$ Sloan spectra: the Type-Ia supernova rate versus galaxy mass and star-formation rate at redshift $\sim 0.1$}

\author[Graur \& Maoz]
{Or~Graur$^{1,2}$\thanks{E-mail: orgraur@astro.tau.ac.il}
and Dan~Maoz$^1$
\\
$^1$School of Physics and Astronomy, Tel-Aviv University, Tel-Aviv 69978, Israel \\
$^2$Department of Astrophysics, American Museum of Natural History, Central Park West and 79th Street, New York, NY 10024-5192, USA \\
}

\begin{document}

\maketitle


\setstretch{1}

\begin{abstract}
\noindent
Using a method to discover and classify supernovae (SNe) in galaxy spectra, we find 90 Type Ia SNe (SNe~Ia) and 10 Type II SNe among the $\sim 700{,}000$ galaxy spectra in the Sloan Digital Sky Survey Data Release 7 that have \vespa-derived star-formation histories (SFHs).
We use the SN~Ia sample to measure SN~Ia rates per unit stellar mass.
We confirm, at the median redshift of the sample, $z=0.1$, the inverse dependence on galaxy mass of the SN~Ia rate per unit mass, previously reported by \citet{li2011rates} for a local sample.
We further confirm, following \citet{2011arXiv1106.3115K}, that this relation can be explained by the combination of galaxy `downsizing' and a power-law delay-time distribution (DTD; the distribution of times that elapse between a hypothetical burst of star formation and the subsequent SN~Ia explosions) with an index of $-1$, inherent to the double-degenerate progenitor scenario.
We use the method of \citet{maoz2010loss} to recover the DTD by comparing the number of SNe~Ia hosted by each galaxy in our sample with the \vespa-derived SFH of the stellar population within the spectral aperture. 
In this galaxy sample, which is dominated by old and massive galaxies, we recover a `delayed' component to the DTD of \dtdrate\ for delays in the range $>2.4$~Gyr.
The mass-normalised SN~Ia rate, averaged over all masses and redshifts in our galaxy sample, is $R_{{\rm Ia,M}}(z=0.1) = \snumrate$, and the volumetric rate is $R_{{\rm Ia,V}}(z=0.1) = \volrate$.
This rate is consistent with the rates and rate evolution from other recent SN~Ia surveys, which together also indicate a $\sim t^{-1}$ DTD.

\end{abstract}



\section{INTRODUCTION}
\label{sec:intro}

The nature of the stellar system that ends up exploding as a type Ia supernova (SN Ia) is still uncertain (see \citealt{2011NatCo...2E.350H} and \citealt{Maoz2012review} for recent reviews).
While the progenitor is most probably a carbon-oxygen white dwarf (CO WD; \citealt{2011Natur.480..344N, 2012ApJ...744L..17B}), in order for it to explode it must be ignited.
The current consensus is that the WD is in a binary system, through which it accretes mass until the pressure or the temperature in the WD core become high enough to ignite carbon and initiate a thermonuclear runaway.
The two leading scenarios for the nature of the progenitor system are the single degenerate scenario (SD; \citealt{1973ApJ...186.1007W}) in which the WD accretes mass from a main-sequence, helium, or giant star, and the double degenerate scenario (DD; \citealt{1984ApJS...54..335I, 1984ApJ...277..355W}) in which the WD merges with a second CO WD through loss of energy and angular momentum to gravitational waves.

Theoretical and observational evidence exists both for, and against, each of the progenitor scenarios.
The SD scenario, long held to be the more likely progenitor option, has been called into question by some recent observations.
Pre-explosion images and early multi-wavelength data for SN2011fe have ruled out a red giant as the binary companion in that explosion \citep{2011Natur.480..348L,2011Natur.480..344N,2012ApJ...750..164C,2012ApJ...746...21H,2012ApJ...753...22B}.
All possible SD companions, including main sequence stars, have been ruled out in the case of the SN~Ia remnant SNR 0509-67.5 in the Large Magellanic Cloud (\citealt{2012Natur.481..164S}, but see \citealt{2012arXiv1205.3168D} and \citealt*{2012arXiv1205.5028S}).
\citet{2011Sci...333..856S} have argued, based on the statistics of sodium absorption lines in SN~Ia spectra, that at least 20--25 per cent of SNe~Ia in spiral galaxies interact with circumstellar material that could originate from a wind from a SD companion (see also \citealt{2007Sci...317..924P,2009ApJ...702.1157S}).
\citet{2012arXiv1207.1306D} have found evidence of circumstellar material around the SN~Ia PTF11kx, which they explain as the result of a symbiotic nova progenitor, similar to RS Ophiuchi.
However, such events probably constitute only 0.1--1 per cent of all SNe~Ia.
Following upon ideas by \citet{2003ApJ...598.1229P}, \citet{2011ApJ...730L..34J}, \citet*{2011ApJ...738L...1D}, and \citet{2012MNRAS.419.1695I} have recently sketched scenarios wherein the lack of signs of interaction or hydrogen in SN~Ia spectra are due to spinup of the accreting WD, resulting in a delay in the explosion by enough time, e.g., for a SD companion to exhaust its hydrogen envelope.

As for the DD scenario, \citet{2012ApJ...749L..11B} have recently measured the Galactic merger rate of WD--WD binary systems in the Sloan Digital Sky Survey (SDSS; \citealt{2000AJ....120.1579Y}) and found it to be similar to the Galactic SN~Ia rate (although some fraction of those mergers, a fraction not constrained by the data, involves mergers below the Chandrasekhar mass).
The DD scenario has long been disfavored by theoretical studies that predict that the merger of WDs of unequal mass would lead to an accretion-induced collapse to a neutron star, rather than to a SN~Ia (\citealt{1985ApJ...297..531N,2012ApJ...748...35S}; see, however, \citealt{2010Natur.463...61P,2011A&A...528A.117P,2012ApJ...747L..10P}; \citealt*{2010ApJ...722L.157V}, for some recent reassessments of this prediction).

Some evidence favoring the DD scenario has emerged from reconstructions of the delay-time distribution (DTD; the distribution of times that elapse between a hypothetical $\delta$-function-like burst of star formation and the subsequent SN~Ia explosions).
Different progenitor models predict different forms for the DTD (see \citealt{2012NewAR..56..122W} for a recent review).
A power-law DTD with an index of $\sim -1$ is generic to the DD scenario, if the distribution of initial separations between the merging WDs is roughly a power law of that index (see, e.g., \citealt{Maoz2012review}).
DTDs from SD models generally cut off after a few Gyr, although extended DTDs in the SD context have been predicted as well (e.g., \citealt{2011ApJ...738L...1D}; \citealt*{2012ApJ...756L...4H}).

Recent studies, using various SN samples, environments, and redshift ranges, have been consistently measuring a power-law DTD of the form $t^{-1}$ (see \citealt{Maoz2012review} for a review).
Specifically, a measurement of the SN~Ia rates in elliptical galaxies as a function of luminosity-weighted galaxy age by \citet{2008PASJ...60.1327T} was best-fitted by a power-law DTD of this form.
Galaxy clusters, where most of the star formation is thought to have occurred in one burst at $z \approx 3$, have also yielded SN~Ia rates consistent with such a power-law DTD (\citealt{2010ApJ...718..876S}; \citealt*{maoz2010clusters}; \citealt{2012ApJ...745...32B,2012ApJ...746..163S}).
Measurements of the volumetric rate of SNe~Ia in field galaxies as a function of the cosmic star formation history (SFH) also point to a $t^{-1}$ DTD [see \citealt{Graur2011} (G11) for a compilation of such surveys and their results up to 2011, and \citealt{2012AJ....144...59P} (P12) for new survey results out to $z \approx 1$].

The dependence of SN~Ia rates on host-galaxy mass and star-formation rate provides a further probe of the DTD.
\citet{2005A&A...433..807M} and \citet*{2006MNRAS.370..773M} showed that a broad distribution of delay times is required in order to explain the increase in the SN~Ia rate per unit stellar mass in bluer galaxies, a colour dependence that \citet{Maoz2012review} showed can be reproduced with a $t^{-1}$ DTD.
\citet[L11]{li2011rates} measured SN~Ia rates per unit stellar mass in the local Universe and found that they followed a `rate-size' relation, whereby less-massive galaxies have higher mass-normalised SN~Ia rates than more massive galaxies.
\citet{2011arXiv1106.3115K} showed that the rate-size relation could be explained by the combination of galaxy `downsizing' (i.e., older galaxies tend to be more massive than younger galaxies) and a power-law DTD of index $-1$.
Using the SDSS-II SN~Ia sample \citep{dilday2010a,2011ApJ...738..162S}, \citet{2012ApJ...755...61S} confirmed that the SN~Ia rate per unit stellar mass decreased with increasing host-galaxy stellar mass in passive galaxies.

\citet{2010AJ....140..804B} and \citet[M11]{maoz2010loss} introduced a method for recovering the DTD while accounting for the detailed SFHs of the individual monitored galaxies.
M11 used this method on a subsample of the Lick Observatory SN Search sample (LOSS; \citealt{2011MNRAS.412.1419L,li2011LF}; L11), from which they recovered a DTD with both `prompt' ($t<420$~Myr) and `delayed' ($t>2.4$~Gyr) components.
In \citet*[M12]{2012arXiv1206.0465M}, this method was applied to the SDSS-II sample of SNe, from which a similar DTD with prompt, delayed, and `intermediate' ($0.42<t<2.4$~Gyr) components was recovered, consistent with the $t^{-1}$ form of previous studies.
Despite this agreement in the form of the DTD between observations and models, in general the DD DTDs produced by binary population synthesis models are an order of magnitude lower than measurements (\citealt*{2012arXiv1212.0313M,2012A&A...546A..70T}; see \citealt{Ruiter2012} for an exception).

Most SN surveys have been based on imaging, whether by monitoring specific galaxy samples or volumes of space and the galaxies that they include.
However, it is also possible to discover SNe serendipituously among large samples of galaxy spectra, in which the galaxy region covered by the spectral aperture happens to host a SN during the time of exposure.
Advantages of such samples are that the spectra, and hence classification, of transients is immediately available.
Furthermore, the availability of the spectra of the stellar populations in the specific galactic regions that were effectively `monitored' permits deriving accurate rates and DTDs.

In this paper, we discover a new SN sample by searching for, and classifying, transients (including, but not limited to, SNe~Ia) among $\sim 700{,}000$ galaxy spectra from the 7th data release of the SDSS (SDSS DR7; \citealt{SDSS-DR7}).
\citet[M03]{Madgwick2003} were the first to discover SNe~Ia in SDSS galaxy spectra (in SDSS DR1).
They found 19 SNe~Ia among $\sim 100{,}000$ galaxy spectra, from which they estimated the SN~Ia rate at $z \approx 0.1$.
\citet[K11]{Krughoff2011} made a similar measurement, using the larger galaxy sample in SDSS DR5, finding 52 SNe~Ia among $\sim 350{,}000$ galaxies.
Working with a subsample of $\sim 300{,}000$ galaxies from SDSS DR7, \citet[T10]{Tu2010} found 36 SNe.
As detailed below, in this paper we analyse a larger sample of SDSS spectra, using a number of improvements in transient detection, classification, and rate analysis.
We provide the details of our galaxy sample in Section~\ref{sec:galaxies}.
In Section~\ref{sec:method}, we describe our discovery and classification routine and assess its efficiency and purity. 
In Section~\ref{sec:sample}, we present our SN sample.
Using our SN~Ia sample, in Section~\ref{sec:results} we measure mass-normalised SN~Ia rates, convert them to a volumetric rate, and compare them to previous results.
In this section we also recover the delayed component of the DTD using the direct recovery method of M11.
We summarize and discuss our results in Section~\ref{sec:discuss}.

Throughout this paper, we assume a cosmological model with parameters $\Omega_{\Lambda} = 0.7$, $\Omega_m = 0.3$, and $H_0 = 70$~km~s$^{-1}$~Mpc$^{-1}$.
Magnitudes are on the AB system \citep{1983ApJ...266..713O}, unless noted otherwise.


\section{GALAXY SAMPLE}
\label{sec:galaxies}
For the purpose of this work, we used the $776{,}447$ galaxy spectra in SDSS DR7 that have SFHs obtained by \citet[T09]{Tojeiro2009} using the VErsatile SPectral Analysis code (\vespa\footnote{http://www-wfau.roe.ac.uk/vespa/}; \citealt{Tojeiro2007}).
As in M11 and M12, we use the total formed stellar masses fitted by \vespa\ to the SDSS spectra in three time bins: 0--0.42~Gyr, 0.42--2.4~Gyr, and $>2.4$~Gyr (corresponding to bins $24+25$, 26, and 27 in T09).
We specifically use the \vespa~reconstruction that assumes a single dust component and utilises the \citet{2005MNRAS.362..799M} spectral synthesis models.
A \citet{Kroupa2007} stellar initial-mass function (IMF) is assumed in T09, and this assumption therefore propagates into the mass normalisation of the SN~Ia rates we report.
T09 measured the total mass in the SDSS fiber aperture and then applied an aperture correction to obtain the total mass in the entire galaxy (see their equation~22).
Since, for rate purposes, we are interested in the stellar mass inside the fiber aperture (which was the stellar mass that was effectively monitored for SNe), we reverse this aperture correction, and propagate the uncertainties accordingly.
We reduce the masses in each time bin by a further factor of 0.55, due to the different flux calibrations between SDSS DR5 and DR7 (R. Tojeiro, private communication; see M12 for a detailed discussion).
Of the $776{,}447$ galaxies in our sample, $52{,}016$ galaxies, constituting $6.7$ per cent of the sample, have \vespa-derived masses only in a coarse bin in the time range $>0.42$~Gyr (i.e., in bin 29).
This precludes using these galaxies, as we cannot know how to distribute these masses between the 0.42--2.4~Gyr and $>2.4$~Gyr bins.
Furthermore, we exclude those galaxies for which the \vespa~SFH has a reduced $\chi^2$ value ($\chi^2_r$) in the range $\chi^2_r>10$, are at redshift $z>1$, or that have less than 500 good pixels in their spectra, according to the MASK array included by SDSS for each spectrum [we use only those pixels that have a `0' (good) or `40000000' (emission line) bit flag].
These cuts reduce our sample to $707{,}792$ galaxies.

Of the galaxies in the sample, $29{,}066$ galaxies have more than one spectrum, observed on different dates.
Of these galaxies, after applying the above cuts, we consider in our rates analysis only the spectra that are spaced at least 60 days apart, in order to assure independent search epochs.
Five of the host galaxies of the SNe~Ia that we present below have multiple observations (namely, 0394-51812-554, 0418-51884-144, 1298-52964-304, 1782-53383-517, and 2019-53430-010).
Throughout this work, we refer to SNe discovered in SDSS spectra according to the plate, modified Julian date (MJD), and fiber in which they are discovered (e.g., 0814-52443-249 was observed on MJD 52443, or June 18 2002, in the galaxy targeted by the 249th fiber of plate 814, where each SDSS plate contains 640 fibers).
The above cut leaves $26{,}592$ multiply-observed galaxies: $12{,}991$ with two observations; 197 with three epochs; 11 with four; three with five; and two galaxies with six epochs.

\begin{figure}
 \begin{center}
  \includegraphics[width=0.5\textwidth]{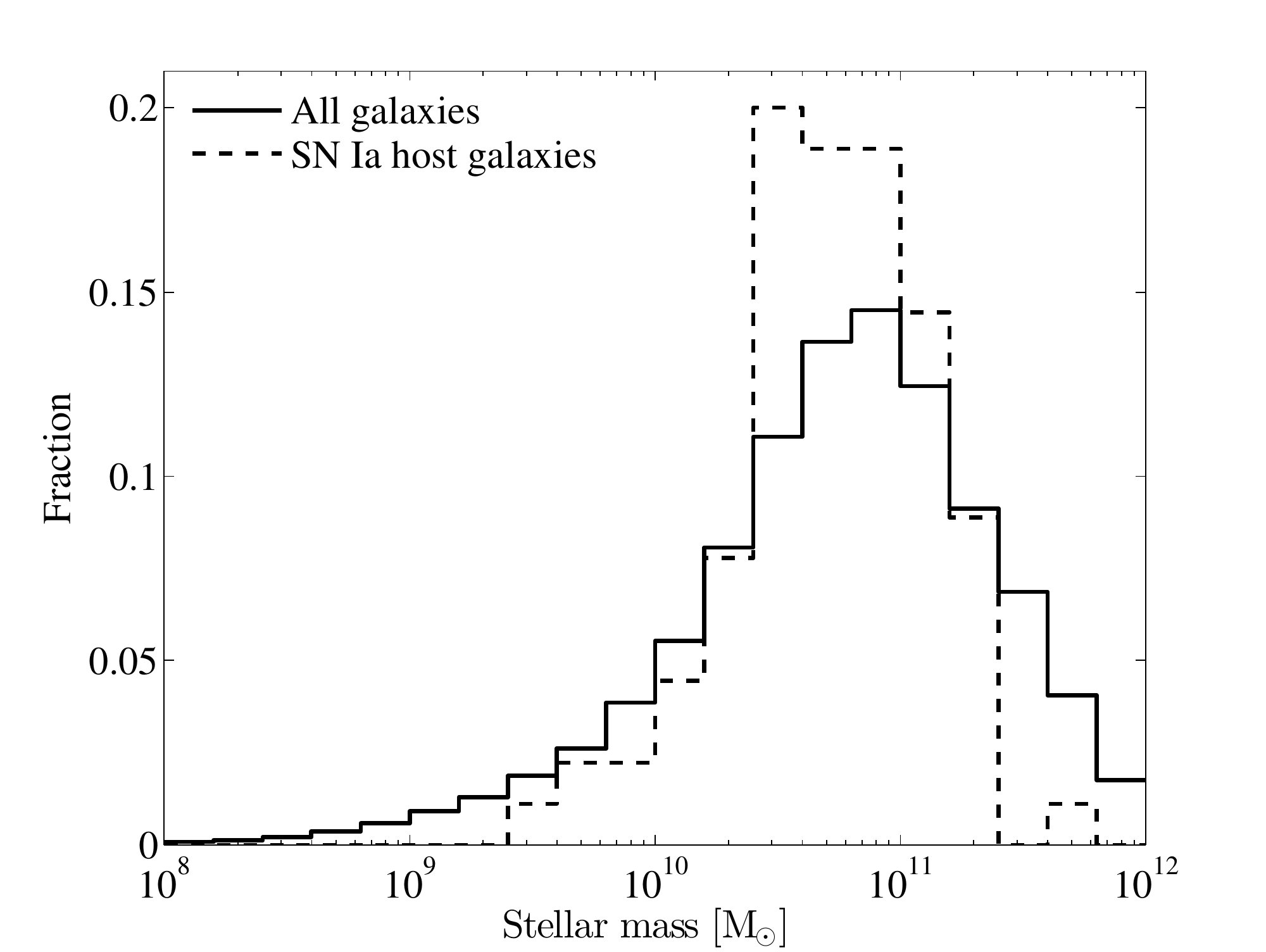} \\ 
  \includegraphics[width=0.5\textwidth]{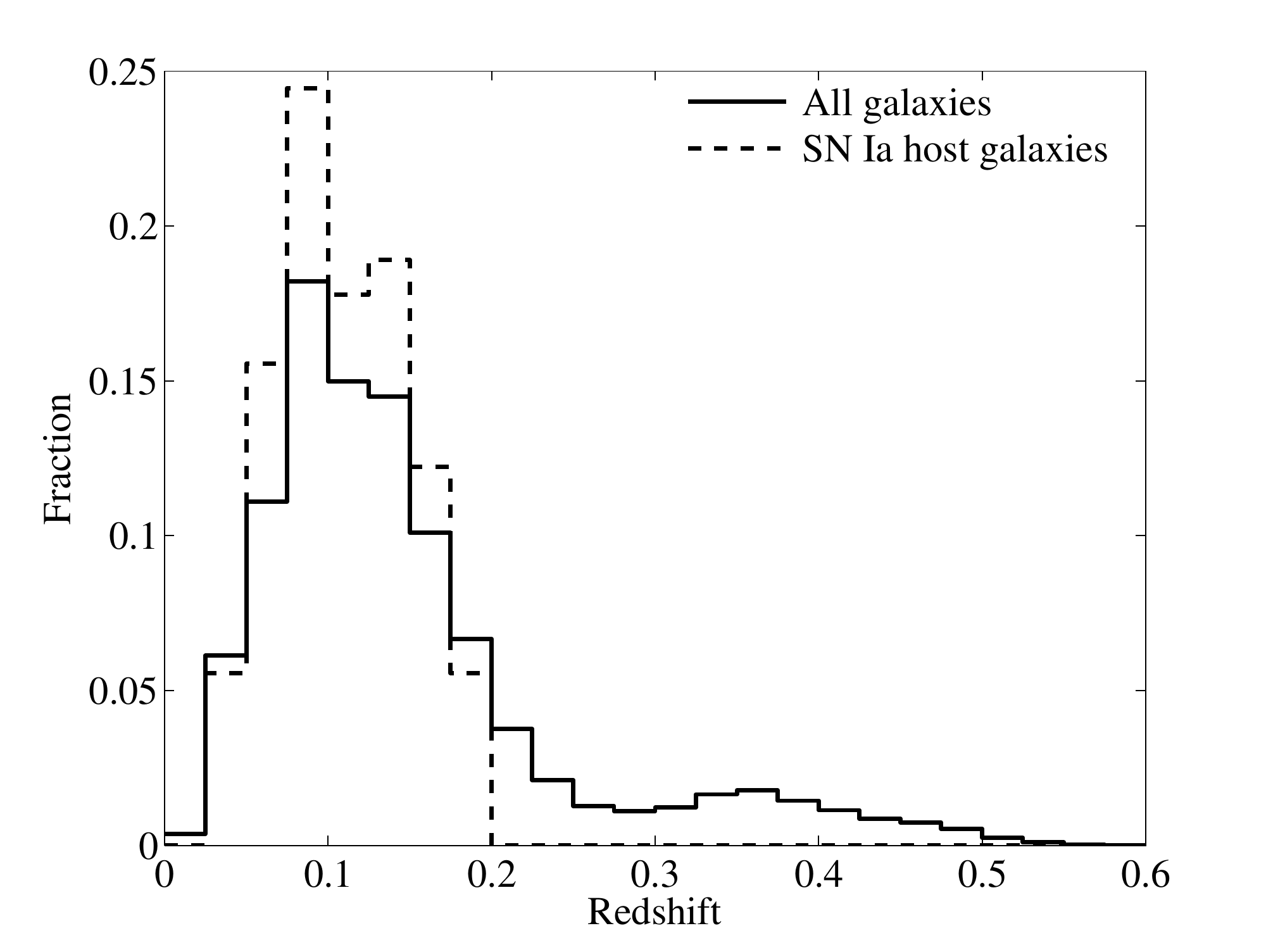} \\
  \caption{Total stellar mass (upper panel) and redshift (lower panel) distributions for all galaxies in our sample (solid curve) and SN~Ia host galaxies (dashed curve).}
  \label{fig:DR7_gals} 
 \end{center}
\end{figure}

We classify the galaxies in our sample according to their specific star-formation rate (sSFR), which is computed as the \vespa~mass formed in bins 24+25, divided by 420~Myr, and divided by the total formed mass. 
Following \citet{2006ApJ...648..868S}, we separate the galaxies into highly star-forming [$\rm{log(sSFR/{\rm yr}^{-1})}\geq-9.5$], star-forming [$-12\leq\rm{log(sSFR/{\rm yr}^{-1})}<-9.5$], and passive [$\rm{log(sSFR/{\rm yr}^{-1})}<-12$] galaxies.
As we show in the upper panel of Fig.~\ref{fig:DR7_gals}, our galaxy sample is dominated by massive, old galaxies, and consists of $54.5$, $44.7$, and $0.8$ per cent passive, star-forming, and highly star-forming galaxies, respectively.
The SN~Ia host galaxy sample is similarly divided into $47.8$, $52.2$, and $0$ per cent passive, star-forming, and highly star-forming galaxies, respectively.
Since only $0.8$ per cent of our galaxies are highly star-forming, we combine the star-forming and highly star-forming galaxies into one category, which we label `star-forming galaxies' (i.e., with $\rm{log(sSFR/{\rm yr}^{-1})}\geq-12$).
The lower panel of Fig.~\ref{fig:DR7_gals} shows the redshift distribution of our galaxy sample and SN~Ia host-galaxy sample.
The median redshifts of the galaxy and SN~Ia host galaxy samples are $z=0.11$ and $z=0.09$, respectively.
Due to the limits of our detection and classification process, we do not find SNe~Ia above $z \approx 0.2$.


\section{Supernova discovery and classification}
\label{sec:method}

\subsection{Method}
\label{subsec:how}

The galaxy spectra are first corrected for Galactic extinction according to \citet*{cardelli1989} and the $E(B-V)$ values listed in the SDSS spectrum headers.
The spectra are then masked according to the MASK array included for each spectrum.

In order to identify SNe in the SDSS DR7 galaxy spectra, we first subtract a model of the galaxy constructed from ten SDSS DR2 galaxy eigenspectra \citep{Yip2004a}, using singular value decomposition (SVD).
At this stage, we discard spectra with best-fitting galaxy models with $\chi^2_r \leq 1$, as these spectra either do not host a transient, or have a too low signal-to-noise ratio (S/N).
The residual spectrum obtained by subtracting the galaxy model is fit with a straight line in $f_{\lambda}$ vs. wavelength, which is then subtracted.
This `flattened' residual is searched for `features' by counting the number of continuous pixels above (or below) zero.
The residual spectrum must have at least ten features, each $\geq 30$ pixels wide.
These values were found to be optimal for discovering faint SNe, while reducing the number of false positives.

Every galaxy spectrum that satisfies the above criteria is then re-analysed by adding SN templates to the SVD-constructed galaxy model.
We use the spectral library from the Supernova Identification code (\snid\footnote{http://marwww.in2p3.fr/$\sim$blondin/software/snid/index.html}; \citealt{Blondin2007}), which contains SN (Ia and core collapse), active galactic nucleus (AGN), luminous blue variable (LBV), flaring M dwarf, and galaxy spectra.
For each SVD fit, we compute a figure of merit, $\chi^2_\lambda$, composed of the $\chi^2_r$ value, divided by the wavelength range covered by the specific transient template.
This is necessary, as some transient templates cover small wavelength ranges due to the instrumental setups with which they were taken.
Such templates would receive a low $\chi^2_r$ value if they fit specific features in the residual spectrum, even though a different template from the same transient class that covers a larger wavelength range may not fit the rest of the residual.
This can lead to the misclassification of transients (e.g., SNe~Ia as core-collapse~SNe).
To overcome this problem, we keep only those transient templates in the \snid~library that cover a wavelength range of at least $2{,}900$~\AA{}, reducing the \snid~spectral library from 349 objects with $3{,}056$ spectra to 304 objects with $2{,}952$ spectra.
Only transient models with a $\chi^2_r$ value smaller than the original-galaxy-fit $\chi^2_r$ are considered for further analysis.

The residual spectrum with the best minimal $\chi^2_\lambda$ is input to \snid~for a second, independent classification, where the following criteria must be met:
\begin{enumerate}
 \item\noindent The residual must fit more than one template.
 This eliminates a large number of low-S/N false positives.
 \item\noindent The best-fitting template must have rlap$\geq 5$, and lap$\geq 0.4$.
 The `lap' parameter quantifies the amount of overlap between the residual spectrum and the transient templates at the correlation redshift, while `rlap' is a quality parameter, where higher values denote better correlations (see \citealt{Blondin2007} for a full description).
 \item\noindent The best-fitting templates must be at a redshift within $\Delta z = 0.015$ of the SDSS-measured redshift.
\end{enumerate}
A large number of false positives are culled at this stage, as some residual spectra with low S/N ratios might be fit with transient templates using SVD, but will be disqualified by \snid.

Finally, the candidate is classified into eight transient categories, according to the SVD and \snid~best-fitting templates, and according to the percentage of \snid~templates belonging to a specific category, which we call \snid{\scriptsize \%}.
The transient candidates are classified as Ia, Ib/c, IIb, II, LBV, or AGN.
We also include two additional categories: Ia/Ic (Ic/Ia), which include SNe that are classified as SNe~Ia by one classification method and as SNe~Ib/c by the other.
For example, a SN can be classified as a SN~Ia in the SVD phase and then as a SN~Ic by \snid.
If less (more) than 50 per cent of the best-fitting \snid\ templates are those of SNe~Ia, we classify the SN as a Ic/Ia (Ia/Ic).
Without follow-up spectroscopy or imaging, we cannot tell whether such a SN is a SN~Ia or a SN~Ib/c, so SNe classified as Ia/Ic (Ic/Ia) would be added to the SN~Ia (Ib/c) sample, but would carry a systematic uncertainty (e.g., if two SNe Ia/Ic were discovered, they would add a $-2$ systematic error to the number of SNe Ia).
However, as detailed in Section~\ref{sec:sample}, we find no such SNe.
The classification rules are summarized in Table~\ref{table:classify}.

\begin{table}
\center
\hspace{1.5in}\parbox{5.8in}{\caption{Transient classification criteria}\label{table:classify}}
 \begin{tabular}{l c c c}
  \hline
  \hline
  {Transient} & {SVD} & {\snid} & {\snid{\scriptsize \%}} \\
  \hline
  Ia & Ia & Ia & Ia$\geq 0.5$ \\
  Ia/Ic & Ia or Ib/c & Ia or Ib/c & Ia$\geq 0.5$ \\
  Ib/c & Ib/c & Ib/c & Ib/c$\geq 0.5$ \\
  Ic/Ia & Ia or Ib/c & Ia or Ib/c & Ia$< 0.5$ \\
  IIb & Ib/c or IIb & Ib/c or IIb & Ib/c$\geq 0.5$ \\
  II & IIP, IIL, or IIn & IIP, IIL, or IIn & II$\geq 0.5$ \\
  LBV & LBV & LBV &  \\
  AGN$^a$ & AGN or other & AGN or other &  \\
  \hline 
  \multicolumn{4}{l}{Note -- In order to be classified into one of the SN subtypes} \\
  \multicolumn{4}{l}{in column 1, a candidate must meet both the SVD and} \\
  \multicolumn{4}{l}{\snid\ criteria in columns 2 and 3, and the fraction of best-} \\
  \multicolumn{4}{l}{fitting \snid\ templates in column 4. For example, if a} \\
  \multicolumn{4}{l}{candidate is fit as a SN~IIP in the SVD stage, and as a SN} \\
  \multicolumn{4}{l}{IIL by \snid, and the fraction of best-fitting \snid\ templates} \\
  \multicolumn{4}{l}{is $\geq 0.5$, that candidate would be classified as a SN~II.} \\
  \multicolumn{4}{l}{Otherwise, it would be discarded.} \\
  \multicolumn{4}{l}{$^a$AGNs were not required to be classified as such by both} \\
  \multicolumn{4}{l}{the SVD and \snid~stages.}
 \end{tabular}
\end{table}

Along with the SNe in our sample, we also find $\sim 980$ AGNs.
In order to check whether any of these AGN-hosting galaxies also host other types of transients, we re-analyse them one more time, using a combination of the galaxy eigenspectra, the \citet{2001AJ....122..549V} composite AGN template (which is also included in the \snid\ spectral library), and all other transients in the \snid\ spectral library. In other words, whereas in the previous stage the data were fit with either an AGN or a different transient template, in this stage they are fit with both simultaneously.
We find no other transients in these galaxies.

\begin{figure}
 \begin{center}
  \includegraphics[width=0.5\textwidth]{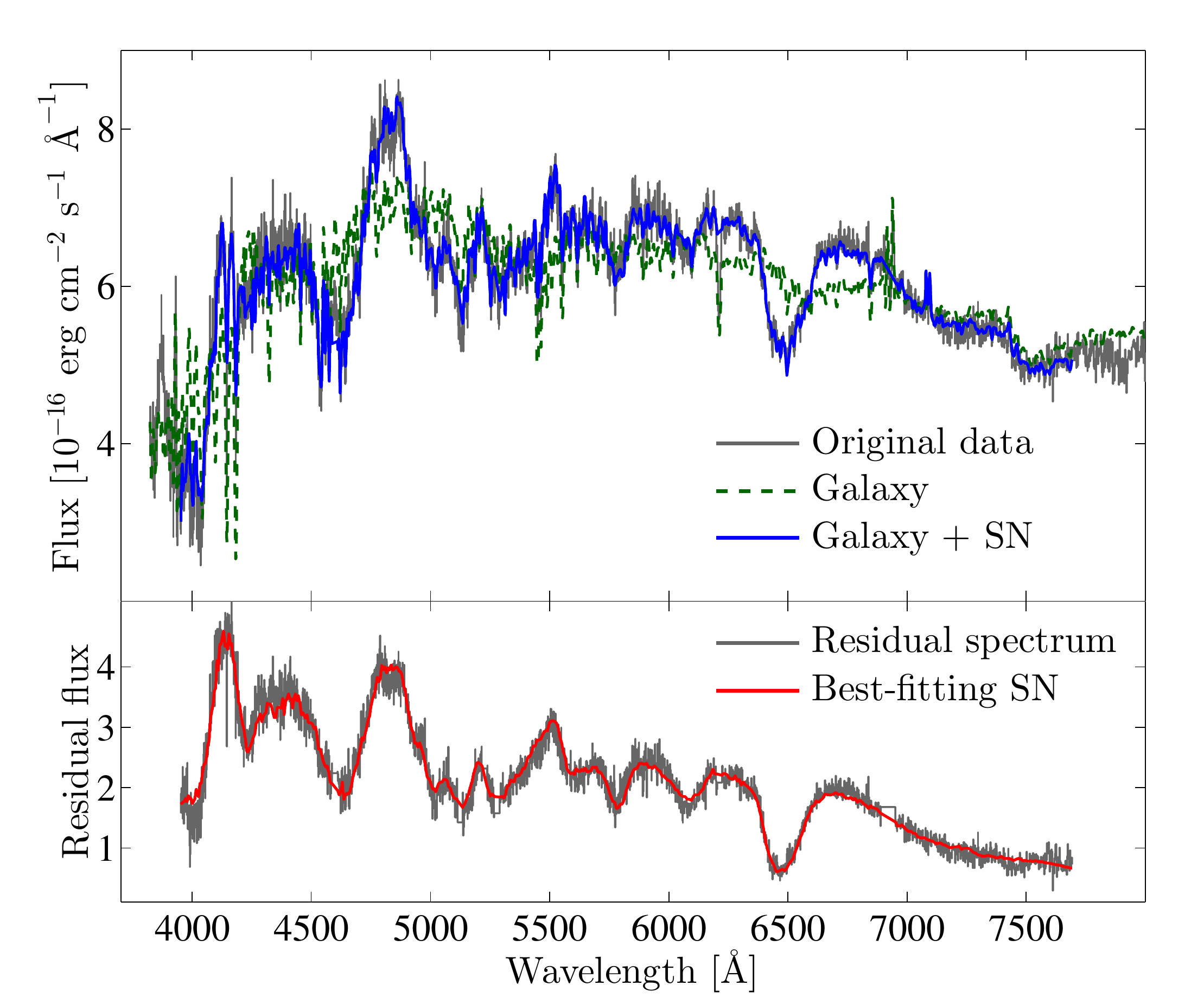}
  \caption{A SN~Ia discovered in the SDSS galaxy spectrum 0424-51893-355 (plate-MJD-fiber) at $z=0.054$. Top: the original galaxy spectrum (grey) is fit with a galaxy model (dashed green) composed of ten eigenspectra, resulting in $\chi^2_r=7.6$. When fit with both eigenspectra and transient templates (solid blue), a SN~Ia template produces $\chi^2_r=1.1$. Bottom: residual spectrum (grey) produced by subtracting the galaxy model from the original data. The best-fitting SN~Ia template is shown in red.}
  \label{fig:example}
 \end{center}
\end{figure}

We do not mask galaxy emission lines in the first stage of analysis, as we have found that the emission lines have a negligible effect on our discovery success rate.
However, since the strength of individual emission lines changes from galaxy to galaxy, they are not perfectly fit by the eigenspectra, contaminating the residual spectrum and causing \snid~to misclassify SN candidates.
In the re-analysis stage, we therefore mask the emission lines of H$\alpha$, H$\beta$, [OI] $\lambda 6300$, [OII] $\lambda 3727$, [OIII] $\lambda 4363, 4959, 5007$, [NII] $\lambda 6548, 6583$, and the Na D $\lambda 5890, 5896$ absorption feature.
In Fig.~\ref{fig:example}, we show an example of a SN~Ia discovered by our methodology.

\subsection{Comparison with previous spectral supernova searches}
\label{subsec:compare}
As in this work, M03, K11, and T10 used galaxy eigenspectra, obtained through principal component analysis (PCA), to subtract a galaxy model from the data. 
The residual spectrum, however, was analysed differently. 
M03 performed a wavelet transform on the residual to separate the SN signal from the noise. 
This resulted in a `noiseless' signal which they compared, through cross-correlation, to a small set of SN Ia spectral templates. 
Such manipulation of the data carries the risk of introducing artificial features, and indeed we believe that some of the SNe reported by M03 may have been false detections (see below). 
K11 used the same \citet{Yip2004a} eigenspectra as in this work, together with an AGN template composed of the first two QSO eigenspectra from \citet{Yip2004b}, and the \citet*{Nugent2002} SN~Ia template. 
Their comparison to SN templates was done solely through SVD.
T10 combined the \citet{Yip2004a} galaxy eigenspectra with two SN~Ia eigenspectra derived with PCA from the \citet{Nugent2002} SN~Ia template and compared the resultant residual spectrum to SN templates using cross-correlation.

Our method differs in four main ways:
\begin{enumerate}
\item \noindent We use the \snid~as well as the SVD-derived $\chi^2$ criterion. 
This allows us to cull false SN~Ia candidates (as detailed in Section~\ref{subsec:how}) and other false positives. 
\item \noindent We have found that using an AGN template together with the galaxy eigenspectra `robs' some of the continuum from a transient in the galaxy spectrum, making it harder to discover fainter SNe. 
Instead, as detailed in Section~\ref{subsec:how}, we first find $\sim 980$ AGN-hosting galaxies, and search them for additional transients.
By limiting the combination of the AGN template and the galaxy eigenspectra to these AGN-hosting galaxies alone, we assure that the AGN template does not affect the shape of the residual spectra in non-AGN-hosting galaxies.
\item \noindent By using \snid, and the \snid~templates in the SVD phase, we can discover other types of transients, from core-collapse~SNe to LBVs. 
Furthermore, \snid~and the SVD phase are modular: new spectra can be added as they become available (e.g., for new types of transients, such as the super-luminous SNe).
\item \noindent The data are not manipulated in any way (e.g., there is no binning).
\end{enumerate}

We recover all of the eight examples reported by K11 out of the 52 SNe~Ia they discovered, except 0472-51955-247 (see their figure 8) which is hosted by a galaxy that is not included in our galaxy sample.
Of the 19 SNe~Ia reported in M03, we recover ten.
Two of the 19 SNe are hosted by galaxies that are not included in our galaxy sample, as they do not have associated \vespa\ SFHs.
Of the remaining seven SNe~Ia, the residual spectra of the four galaxies hosting SN2000ga, SN2001ki, SN2001kl, and SN2001kr have no discernible SN~Ia features, and we deem them false detections.
SN2001kn and SN2001ks may be real SNe~Ia, but were not detected by our methodology due to the low S/N of their residual spectra.
Finally, SN2001kq is detected with a redshift that is higher by $\Delta z > 0.015$ than that of its host galaxy, and so is rejected by our detection and classification process.
The redshift difference appears to be due to a rather poor best fit to a noisy residual spectrum, indicating a peculiar SN~Ia.
Of the 36 SNe reported by T10, we recover 31.
As with the M03 sample, three of the T10 SNe (1304-52993-552, 1059-52618-553, and 1266-52709-024) are hosted by galaxies that are not in our galaxy sample.
The final two SNe (1274-52995-638 and 1818-54539-399) are real, but were not detected due to the low S/N of their residual spectra.
Differences in the SN~Ia rate analysis between this work and M03 and K11 are discussed in Section~\ref{subsec:rate_vol}.

\subsection{Detection and classification efficiency}
\label{subsec:eff}

\begin{table}
\center
{\caption{SN~Ia population fractions and luminosity functions}\label{table:LF}}
 \begin{tabular}{l c c}
  \hline
  \hline
  SN~Ia subtype & {fraction} & {adopted LF} \\
  {}            &            & [Vega mag]   \\
  \hline
  normal Ia     & 0.761 & $-18.67 \pm 0.51$ \\
  Ia-91bg       & 0.024 & $-17.55 \pm 0.52$ \\
  Ia-91T$^a$    & 0.197 & $-19.15 \pm 0.53$ \\
  Ia-02cx$^b$   & 0.018 &                   \\
  \hline
  IIP           & 0.394 & $-15.66 \pm 1.23$ \\
  IIL           & 0.275 & $-17.44 \pm 0.64$ \\
  IIn           & 0.230 & $-16.86 \pm 1.61$ \\
  IIb           & 0.101 & $-16.65 \pm 1.30$ \\
  \hline
  Ib            & 0.324 & $-17.01 \pm 0.41$ \\
  Ic            & 0.528 & $-16.04 \pm 1.28$ \\
  peculiar Ib/c & 0.148 & $-15.50 \pm 1.21$ \\
  \hline 
  \multicolumn{3}{l}{$^a$This category includes SN2000cx and SN2006gz.} \\
  \multicolumn{3}{l}{$^b$As SN2002cx and SN2005hk were both as faint} \\
  \multicolumn{3}{l}{as SN-91bg-like SNe~Ia, they were given its LF.} \\
 \end{tabular}
\end{table}

SNe may be missed by our discovery and classification process for numerous reasons, including poor data (such as spectra with large masked areas), low S/N, low contrast relative to the galaxy light, or failure to meet our classification criteria.
In order to quantify these and other systematic effects, we simulate the discovery and classification efficiency of our method by planting fake SN spectra in a random selection of real galaxy spectra. 

The fake SN spectra are drawn from the subsample of \snid~spectra spanning $\geq2{,}900$~\AA{}.
In order to simulate the real distribution of SNe~Ia in our sample, the host galaxies are chosen according to their sSFR.
As the rate of SNe~Ia is higher in more luminous galaxies, we favor in our simulation the selection of host galaxies based on luminosity, according to the dependence of the rate of SNe~Ia on host-galaxy \R-band luminosity in figure 12 of \citet{yasuda2010}.
Furthermore, as the SN~Ia rate in passive galaxies is $\sim1/3$ the rate in star-forming galaxies (see figure 6 in \citealt{2006ApJ...648..868S}), 25 (75) per cent of the host galaxies in the fake sample are chosen to be passive (star-forming) galaxies.
Next, the SN~Ia spectra are planted according to the luminosity functions (LFs) and subtype populations from \citet{li2011LF}.
Table~\ref{table:LF} lists the population fractions and adopted LFs we use for each SN subtype.
The \citet{li2011LF} LF is in the $R$ band, so we convert it to the $B$ band by doing synthetic photometry on the \citet{Hsiao2007} SN~Ia template spectrum at peak, from which we get a $(B-R)=-0.38$ mag correction.

For each fake spectrum we choose a random peak $B$-band magnitude from the appropriate LF and translate it into a stretch value, $s$, according to the \citet{1999ApJ...517..565P} stretch-luminosity relation
\begin{equation}\label{eq:stretch}
 m_{B,s} = m_{B,1} + \alpha(s-1),
\end{equation}
where $m_{B,s}$ is the peak $B$-band magnitude of the stretched light curve, $m_{B,1}$ is the peak $B$-band magnitude of a normal $s=1$ SN~Ia, and $\alpha=1.52$ \citep{2006A&A...447...31A}.
We use the \citet{Hsiao2007} SN~Ia spectral series to construct a template, $s=1$ light curve and normalise it so that $m_{B,1}=0$.
From this stretched light curve we choose a random epoch, and assign the fake spectrum the \R-band magnitude on that epoch.
Following \citet{2006AJ....131..960S}, we limit the stretch range to $0.6 < s < 1.4$.

\begin{table}
\center
\hspace{1.5in}\parbox{5.8in}{\caption{Classification purity}\label{table:purity}}
 \begin{tabular}{l l c c c}
  \hline
  \hline
  \multirow{7}{*}{\rotatebox{90}{\mbox{Classified as:}}} & & \multicolumn{3}{c}{Planted as:} \\
   &  & Ia & Ib/c & II \\
 \cline{3-5}
  & \multicolumn{1}{l|}{Ia}    & 98.7 &  1.7 &  0.1 \\
  & \multicolumn{1}{l|}{Ia/Ic} &  1.0 &  4.4 &  0.3 \\
  & \multicolumn{1}{l|}{Ib/c}  &  0.1 & 92.1 &  0.4 \\
  & \multicolumn{1}{l|}{Ic/Ia} &  0.1 &  1.7 &  0.2 \\
  & \multicolumn{1}{l|}{II}    &  0.1 &  0.1 & 99.0 \\
  \hline
 \multicolumn{5}{l}{Note -- All values are percentages.} \\ 
 \end{tabular}
\end{table}

We conduct three separate simulations, in each of which $\sim 10{,}000$ fake spectra of a specific SN subtype are planted in $\sim 75{,}000$ random galaxy spectra.
The SN~Ia simulation includes all normal SNe~Ia, SN-91T-like SNe [including SN2000cx (\citealt{2001PASP..113.1178L}) and SN2006gz (\citealt{2007ApJ...669L..17H})], SN-91bg-like SNe, and SN2002cx-like SNe [i.e., SN2002cx (\citealt{2003PASP..115..453L}) and SN2005hk (\citealt{2007PASP..119..360P})].
We do not include the peculiar SNe SN2002ic \citep{2003Natur.424..651H}, SN2003fg \citep{2006Natur.443..308H}, and SN2005gj \citep{2007arXiv0706.4088P}.
The SN~Ib/c simulation includes normal SNe~Ib, SNe~Ic, and SNe~IIb, along with peculiar SNe~Ib and broad-lined SNe~Ic.
The SN~II simulation includes all SN~IIP, SN~IIL, and SN~IIn spectra included in \snid.

Similarly to the SNe~Ia, the SN~Ib/c and SN~II fake spectra are planted according to the LFs and population fractions presented in tables 6 and 7 of \citet{li2011LF}, reproduced here in Table~\ref{table:LF}.
Contrary to the fake SNe~Ia, the fake SNe~Ib/c and SNe~II are only planted in star-forming galaxies, i.e., those galaxies with $\rm{log(sSFR/yr^{-1})\geq-12}$.

Table~\ref{table:purity} lists the purity of our classification method, i.e., what fraction of fake SNe were classified into the different subtypes listed in Table~\ref{table:classify}.
The table shows that our code is efficient at classifying SNe~Ia and SNe~II. 
SNe~Ib/c, however, are harder to classify, as they may appear similar to SNe~Ia (especially SNe~Ic that exhibit no hydrogen or helium in their spectra).
For every SN classified as a SN~Ia, there is a 1.7-per-cent chance that it is in fact a misclassified SN~Ib/c.
We take this into account by adding a systematic uncertainty of $-1.5$ to our sample of 90 SNe~Ia, described in Section~\ref{sec:sample}.

\begin{figure}
 \begin{center}
  \includegraphics[width=0.5\textwidth]{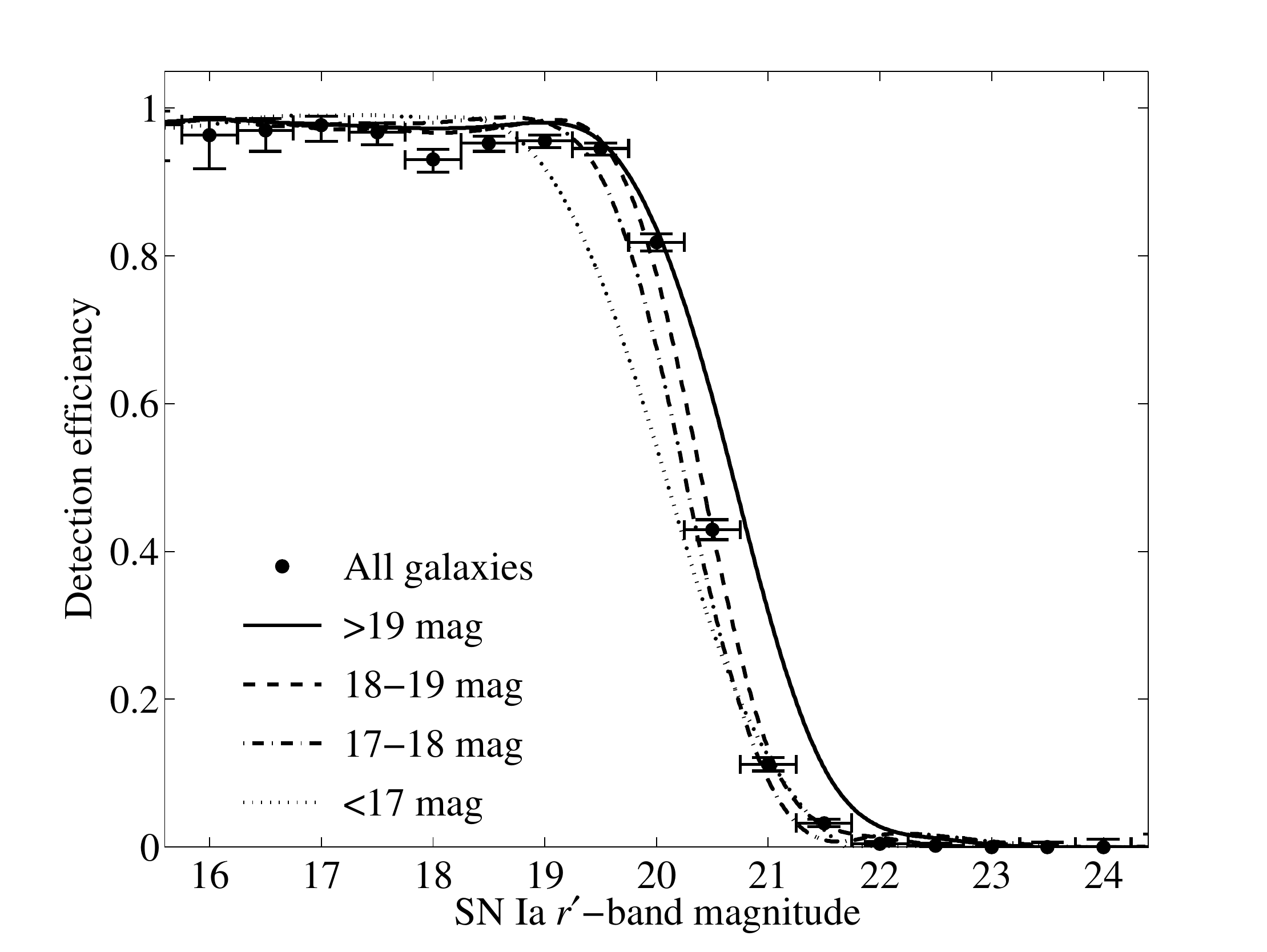}
  \caption{SN~Ia detection efficiency as a function of SN~Ia \R-band magnitude. Filled circles denote the fraction of fake SNe~Ia discovered in all galaxies in 0.5-mag-wide bins. Error bars indicate 1$\sigma$ binomial uncertainties. Curves are cubic spline fits to the detection-efficiency measurements of SNe~Ia in host galaxies with \R-band magnitudes in different ranges, as marked.}
  \label{fig:fakes_eff}
 \end{center}
\end{figure}

\begin{figure}
 \begin{center}
  \includegraphics[width=0.5\textwidth]{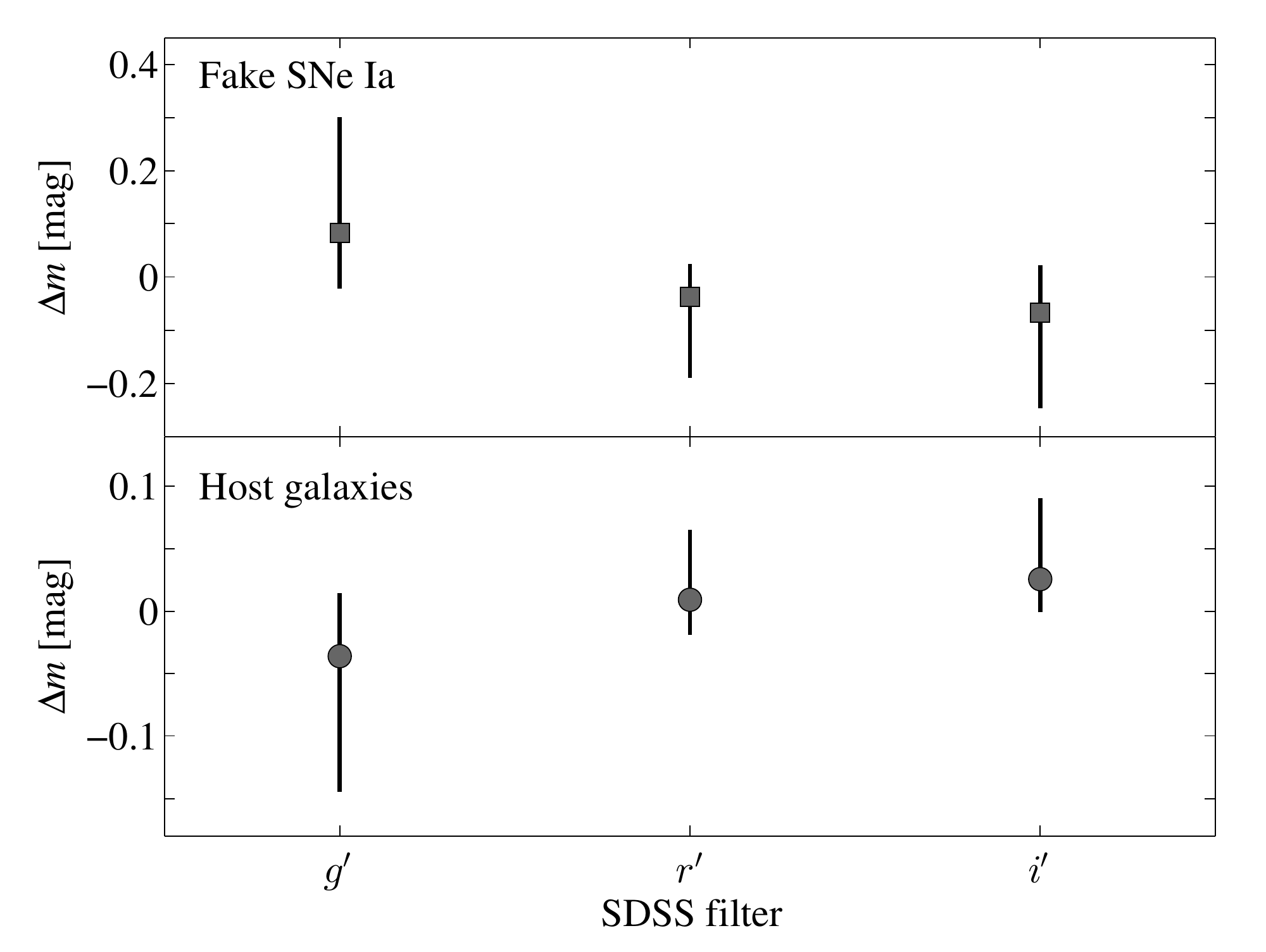}
  \caption{Distributions of differences between the planted and measured \G-, \R-, and \I-band magnitudes of the fake SNe~Ia (filled squares) and their host galaxies (filled circles), respectively. The markers denote the medians of the distributions, and the vertical error bars denote the 16th and 84th percentiles.}
  \label{fig:fakes_mags}
 \end{center}
\end{figure}

We present the results of our SN~Ia simulation in Figures~\ref{fig:fakes_eff}--\ref{fig:fakes_days}.
Fig.~\ref{fig:fakes_eff} shows our SN~Ia detection efficiency as a function of the \R-band apparent magnitude of the planted SNe.
We reach 50 per cent efficiency at $\R=20.4$~mag.
Since the brightness of the host galaxy affects the probability of detecting the SN (i.e., it is easier to discover SNe in fainter galaxies, and vice versa), we divide the SN~Ia detection efficiency measurements shown in Fig.~\ref{fig:fakes_eff} into four subsets, according to the host-galaxy \R-band magnitudes.
These subsets are: $\R_h>19, 18<\R_h\leq19, 17<\R_h\leq18$, and $\R_h\leq17$ mag.
To the efficiency measurements of each of these subsets we fit a cubic spline.

We find that our detection efficiency is strongly dependent on the S/N of the original spectrum and on the brightness contrast between the SN and its host.
Other variables, such as extinction, stretch, or the LF used to plant the fake SNe, have no discernible effect.

In Fig.~\ref{fig:fakes_mags}, we show the distributions of the differences between the planted and recovered apparent magnitudes, $\Delta m$, of the fake SNe~Ia, and their host galaxies, in the \G, \R, and \I\ bands.
Because the transient spectra are not orthogonal with the set of galaxy eigenspectra, there is a flow of flux between them, which causes the SNe to appear brighter, on average, in the \G\ band, and fainter in the \R\ and \I\ bands.
The host galaxies, conversely, appear fainter and brighter, respectively, in these bands, as shown.
Throughout this paper, we report SN and host-galaxy apparent magnitudes in the \R\ band, where the dispersion in $\Delta m$ is smallest.
Moreover, as the average offset in the measured \R-band magnitude of the SNe~Ia caused by the flux flow is negligible, and as the uncertainties in brightness are smaller than the 0.5-mag-wide bins we use for the detection efficiency measurements of Fig.~\ref{fig:fakes_eff}, we do not consider this offset any further. 

\subsection{Supernova age and stretch recovery efficiency}
\label{subsec:stretch}

\begin{figure*}
 \begin{minipage}{\textwidth}
  \vspace{0.2cm}
  \begin{tabular}{c}
   \includegraphics[width=\textwidth]{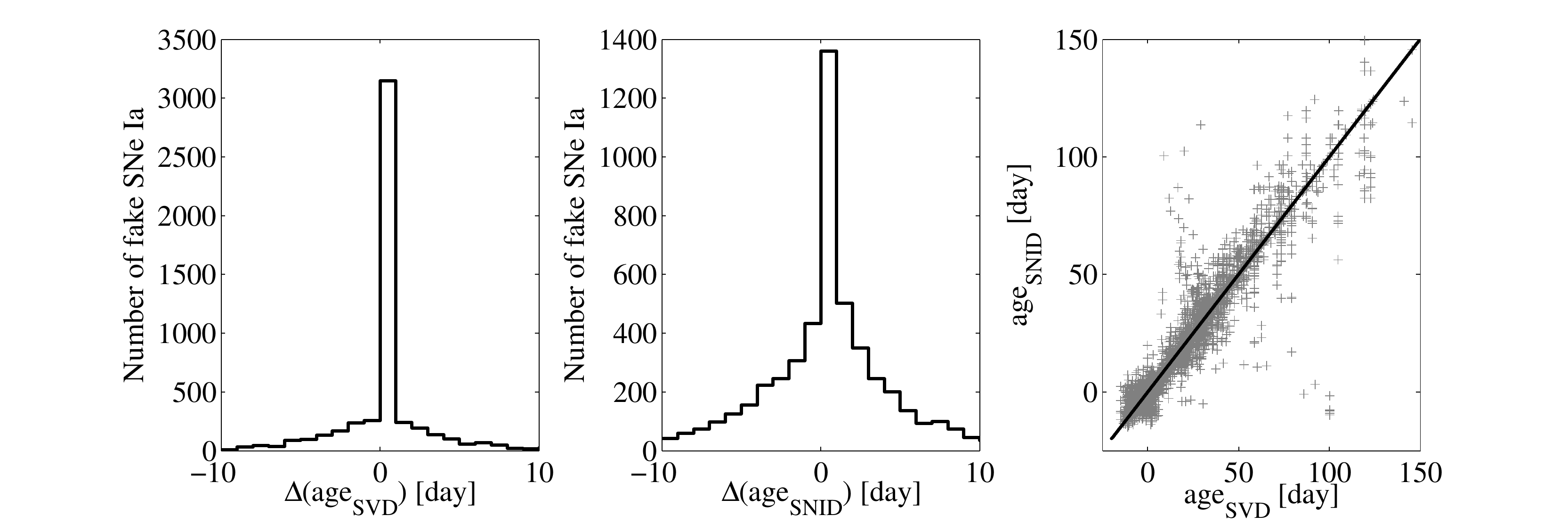} \\ 
  \end{tabular}
  \caption{Left and centre: distributions of differences between the planted and recovered SVD- and \snid-derived ages of the fake SNe~Ia, respectively, in 1-day-sized bins. Right: comparison of the SVD- and \snid-derived ages (crosses). The solid curve represents the 1:1 line.}
  \label{fig:fakes_days}
 \end{minipage}
\end{figure*}

The left and centre panels of Fig.~\ref{fig:fakes_days} show the distributions of differences between the planted and SVD- and \snid-derived ages, respectively, in 1-day-sized bins.
The right panel of Fig.~\ref{fig:fakes_days} shows a comparison of the SVD- and \snid-derived ages.
Whereas the derived ages are almost always within $\pm 1$ day of the planted values, the comparison between the SVD- and \snid-derived ages in the right panel of Fig.~\ref{fig:fakes_days} shows that there is a larger source of systematic uncertainty contributing to the discrepancy between them.
If we assume that the overall uncertainty is identical for all the SNe~Ia, then we can estimate it by fitting the SVD- to the \snid-derived ages and setting $\chi^2=1$.
After limiting the ages of the fake SNe~Ia to the range of the real SNe~Ia in our sample, $-10$ to 55 days relative to $B$-band maximum light, we estimate the overall uncertainty in the age of the SNe~Ia to be $\pm 6$~days.
In a similar manner, we estimate the uncertainty in the ages of the SNe~II to be $\pm 33$ days.

In order to measure the stretches of the SNe Ia in our sample, we substitute the \snid~templates used in the discovery phase with the \citet{guy2007} \salt\footnote{http://supernovae.in2p3.fr/$\sim$guy/salt/} templates.
Similarly to the \citet{Yip2004a} eigenspectra, the \salt~templates are composed of two base spectral sequences: $M_0(p,\lambda)$ and $M_1(p,\lambda)$, where $p$ is time since maximum light in the rest-frame $B$-band, $\lambda$ is the rest-frame wavelength, $M_0(p,\lambda)$ is the average time sequence of eigenspectra, and $M_1(p,\lambda)$ is the spectral sequence of variations induced by the stretch of the SNe.
Following equation~1 in \citet{guy2007}, we add the \salt~eigenspectra linearly:                                                                                                                                                                                                                                                                                                                                                                                                   
\begin{equation}\label{eq:salt}
F(p,\lambda) = x_0 \times [M_0(p,\lambda) + x_1 M_1(p,\lambda)],
\end{equation}
but without the exponential colour correction.
The variable $x_1$ can be converted into the stretch parameter $s$ by using the \citet{guy2007} transformation
\begin{equation}\label{eq:salt_s}
s = 0.98 + 0.091x_1 + 0.003x_1^2 -0.00075x_1^3.
\end{equation}

\begin{figure}
 \includegraphics[width=0.5\textwidth]{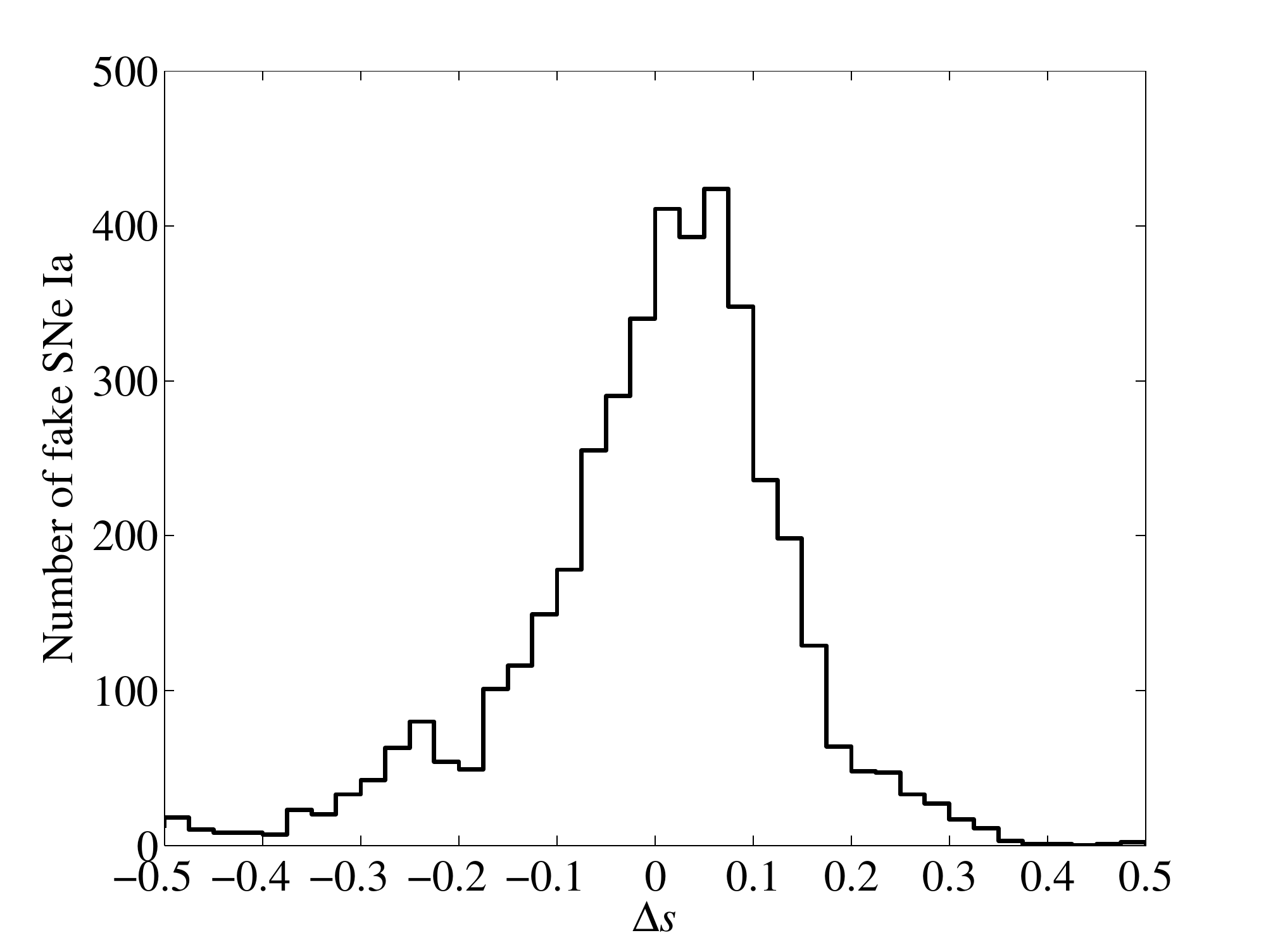}
 \caption{Distribution of differences between the planted and recovered stretch values for a sample of $7{,}000$ fake SN~Ia spectra made up of $835$ spectra of $80$ SNe~Ia with known stretch values \citep{Hsiao2007}.}
 \label{fig:salt}
\end{figure}

We first gauge the accuracy of our stretch measurements by planting $7{,}000$ fake SN~Ia spectra made up of a subsample of $835$ spectra of $80$ of the SNID SNe~Ia with known stretch values \citep{Hsiao2007}.
Fig.~\ref{fig:salt} shows the distribution of differences between the planted and recovered stretch values for a sample of $7{,}000$ fake SNe~Ia planted using this subsample, in bins of $\Delta s=0.025$.
The distribution has a median of $0.0025$, and the range of stretches enclosing 68 per cent of the area is $^{+0.10}_{-0.14}$.
We treat this range as the uncertainty of our stretch measurements.
As found before, fitting the \salt-derived ages to those derived by SVD results in the same uncertainty of $\pm 6$~days.  


\section{Supernova sample}
\label{sec:sample}

We find a total of 100 SNe, of which 90 are SNe~Ia and 10 are SNe~II.
Table~\ref{table:SNe} lists the SNe and their properties. 
The residual spectra and best-fitting SN templates are shown in Figures~\ref{fig:SNe_Ia} and \ref{fig:SNe_IIP}.
Each SN is listed according to the plate, MJD, and fiber in which it was discovered.
We measure the \R-band magnitudes of the SNe by performing synthetic photometry on the SN residual spectrum, as obtained by subtracting the galaxy model from the original data.
Similarly, the \R-band magnitudes of the SN host galaxies are measured using synthetic photometry on the original data, after subtracting the SN residual.
Finally, we measure the stretch of each SN using the \salt\ templates, as detailed in Section~\ref{subsec:stretch}.

SNe II can be spectroscopically classified into those that show signs of interaction with circumstellar material (IIn) and those that do not (IIP and IIL).
Of the non-interacting SNe II, SNe IIP are characterized by prominent P-Cygni lines, such as those visible in our SN II sample.
However, while most SNe IIL do not exhibit P-Cygni lines, the Palomar Transient Factory (PTF; \citealt{2009PASP..121.1334R}) has observed several examples of SNe IIL with such lines (I. Arcavi, private communication).
Without light curves, we cannot break the degeneracy between SNe IIP and IIL, and thus classify our 10 SNe II as non-interacting SNe II.

Apart from these SNe, we detect, as already noted, $\sim 980$ AGNs and several dozen transients which might be either SNe~IIn or AGNs.
Some SN~IIn spectra, which exhibit broad-winged hydrogen lines, are indistinguishable, without further spectroscopic or photometric follow-up, from AGNs, which have similar emission-line profiles.

Some of the SNe~Ia in our sample are sufficiently bright so as to significantly distort the shape of the combined galaxy$+$SN spectrum.
This results in a systematically biased \vespa\ fit and an incorrect SFH for the host galaxy.
In order to correct this, we replace the SFHs of such host galaxies with those of `surrogate' galaxies that did not host SNe, by fitting the continuum of the SN host galaxy with those of all galaxies within our galaxy sample that have the same redshift, within $\Delta z=5\times10^{-4}$, and \R-band magnitude within $\Delta m=0.05$~mag.
We inspect by eye the resulting five best-fitting galaxies.
In most cases, we select the galaxy with the minimal $\chi^2_r$ value as the surrogate galaxy.
In some cases, a better fit, with a $\chi^2_r$ value that is slightly larger (up to two tenths) of the minimal $\chi^2_r$, is discernible by eye, and we choose it instead.
We find that no surrogates are necessary for host galaxies where the contrast between the host and the SN~Ia is $\leq 0.1$. 
For SN-hosting galaxies with multiple, SN-free observations (i.e., 0394-51812-554, 0418-51884-144, 1298-52964-304, and 2019-53430-010), we use the formed masses derived from the \vespa\ fits to the SN-free galaxy spectra.

One member of our SN~Ia sample, 1710-53504-488, can also be fit as a SN~Ic with $5<$rlap$<7$, though the highest rlap values for the SN~Ic fits are still lower than the best SN~Ia fits (the 10 best-fitting SN Ia fits have rlap$>8$; Y. Liu, private communication).
We count 1710-53504-488 as a SN~Ia, but take the possibility that it is actually a SN~Ic into account through the systematic uncertainty of $-1.5$ SNe described in Section~\ref{subsec:eff}.

It is possible that some of the SNe in our sample exploded outside of the area covered by the spectral aperture, but some of their light leaked in due to the point-spread function.
The contamination of our SN Ia sample by such SNe may cause us to over-estimate the SN Ia rates.
To gauge any such contamination, we compare the observed \R-band magnitudes of the SNe Ia and the estimated \R-band magnitudes the SNe would have, based on their measured age and stretch values, the extinction along the line of sight to their host galaxies, and host-galaxy extinction, simulated according to the \citet{neill2006} extinction model (a one-sided positive Gaussian centred on $A_V=0$ with a standard deviation of $\sigma=0.62$).
We find that two SNe Ia are abnormally faint: 418-51884-144 and 2594-54177-348, which are $\sim 3.5$ and $\sim 2$ magnitudes too faint.
We treat these SNe as an added systematic uncertainty of $-2$ on the size of the SN Ia sample, bringing the total systematic uncertainty up to $-3.5$.

\begin{figure}
 \begin{center}
 \includegraphics[width=0.5\textwidth]{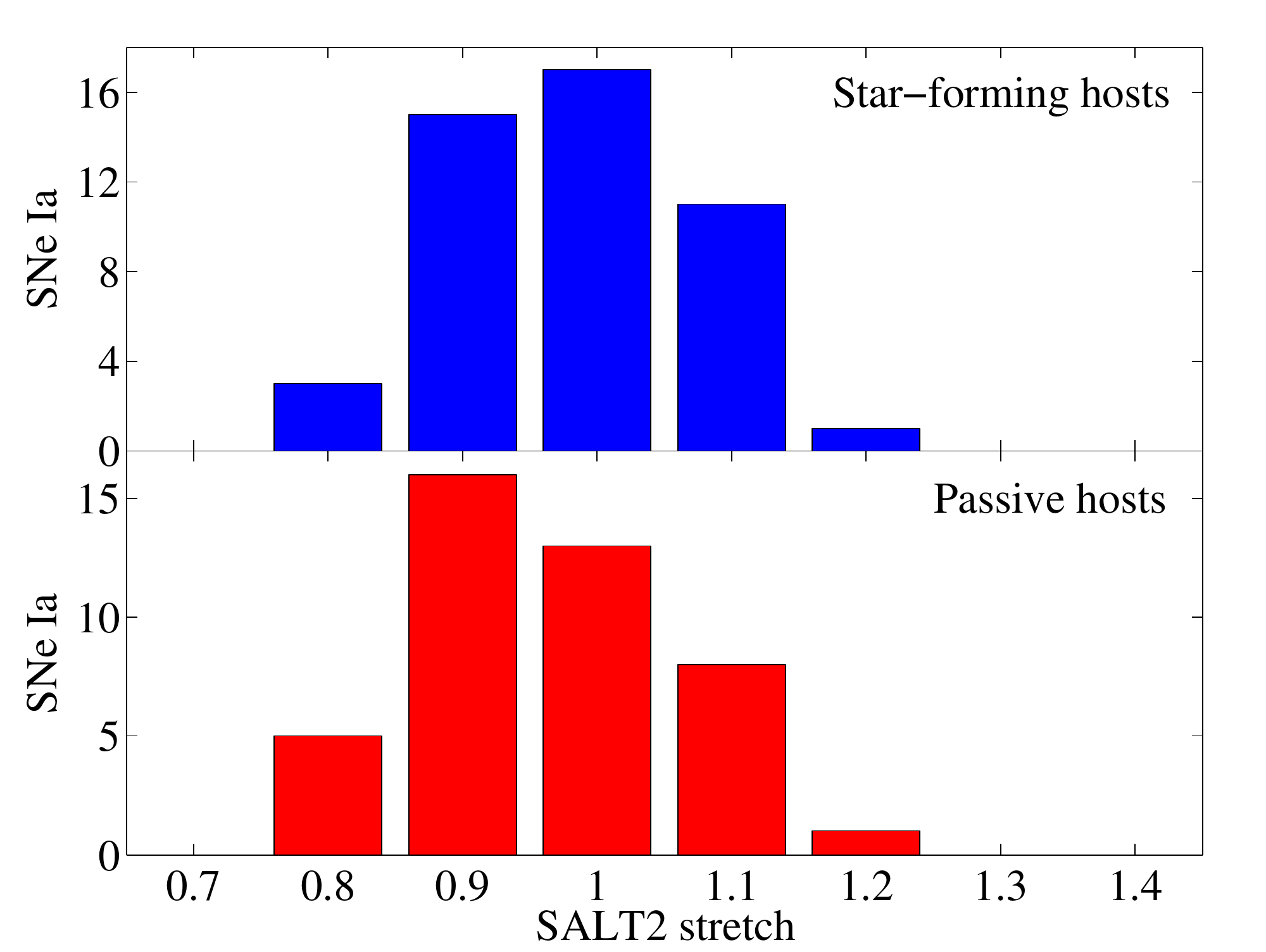}
 \caption{Stretch values for 90 SNe Ia, as measured using the \salt~templates. The stretch values in the upper (lower) panel are for the 47 (43) SNe~Ia in star-forming (passive) galaxies, where star-forming (passive) galaxies are chosen to have ${\rm log(sSFR/yr^{-1})}$ above (below) $-12$.}
 \label{fig:sne_stretch}
\end{center}
\end{figure}

\begin{figure*}
 \begin{minipage}{\textwidth}
  \vspace{0.2cm}
  \begin{tabular}{cc}
   \includegraphics[width=0.5\textwidth]{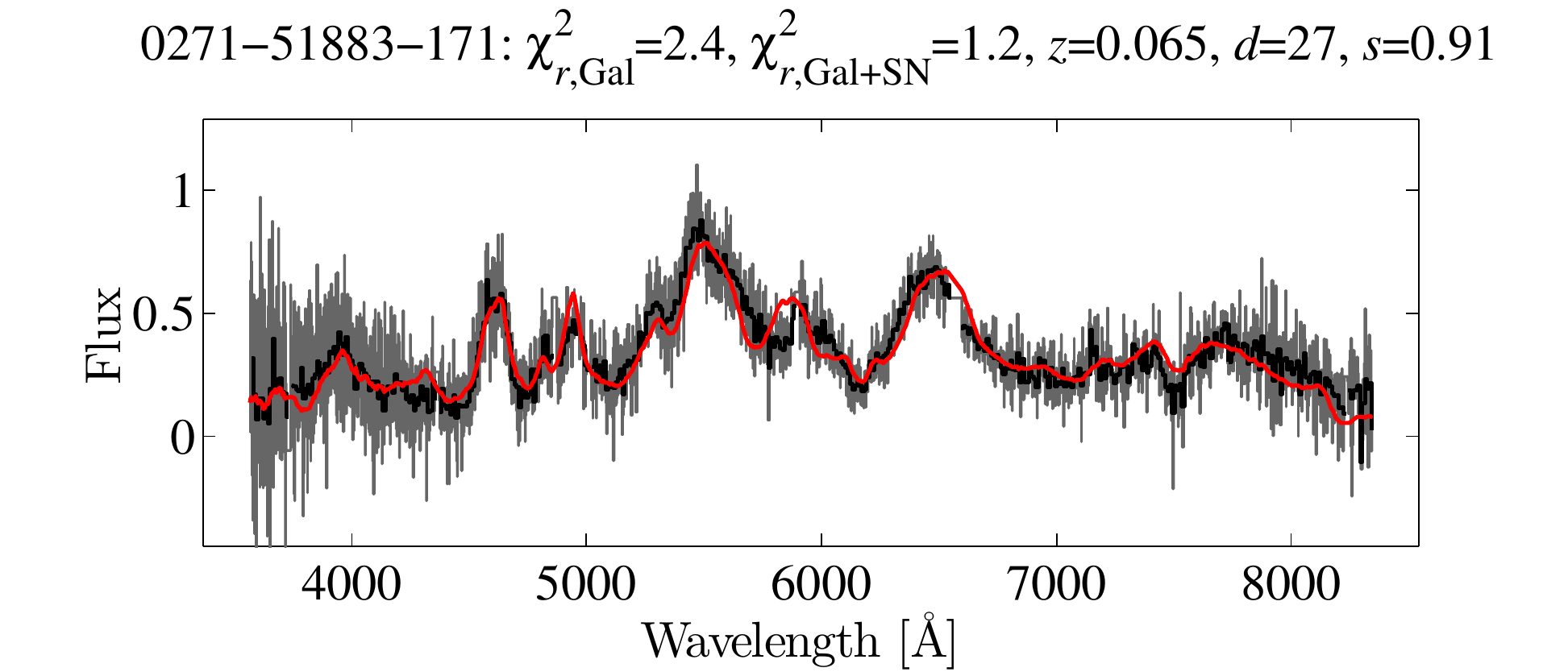} & 
   \includegraphics[width=0.5\textwidth]{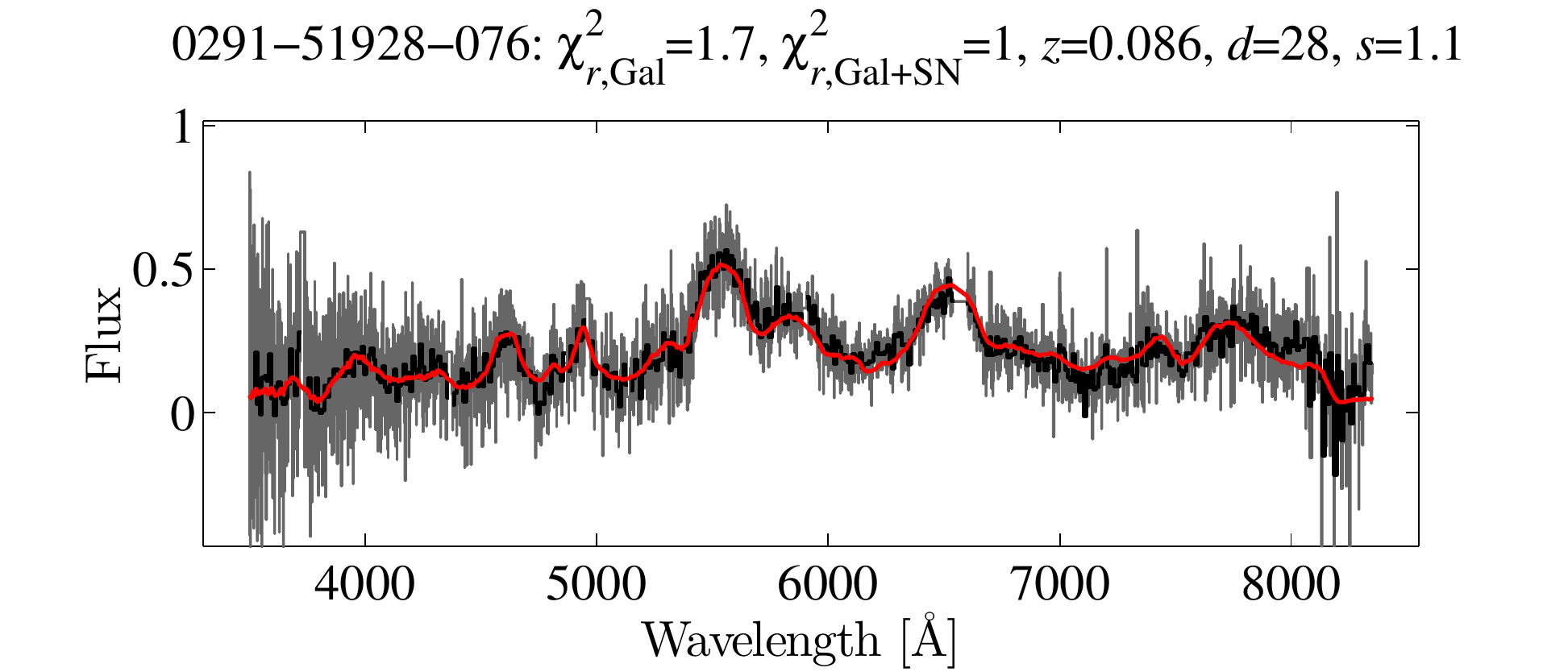} \\
   \includegraphics[width=0.5\textwidth]{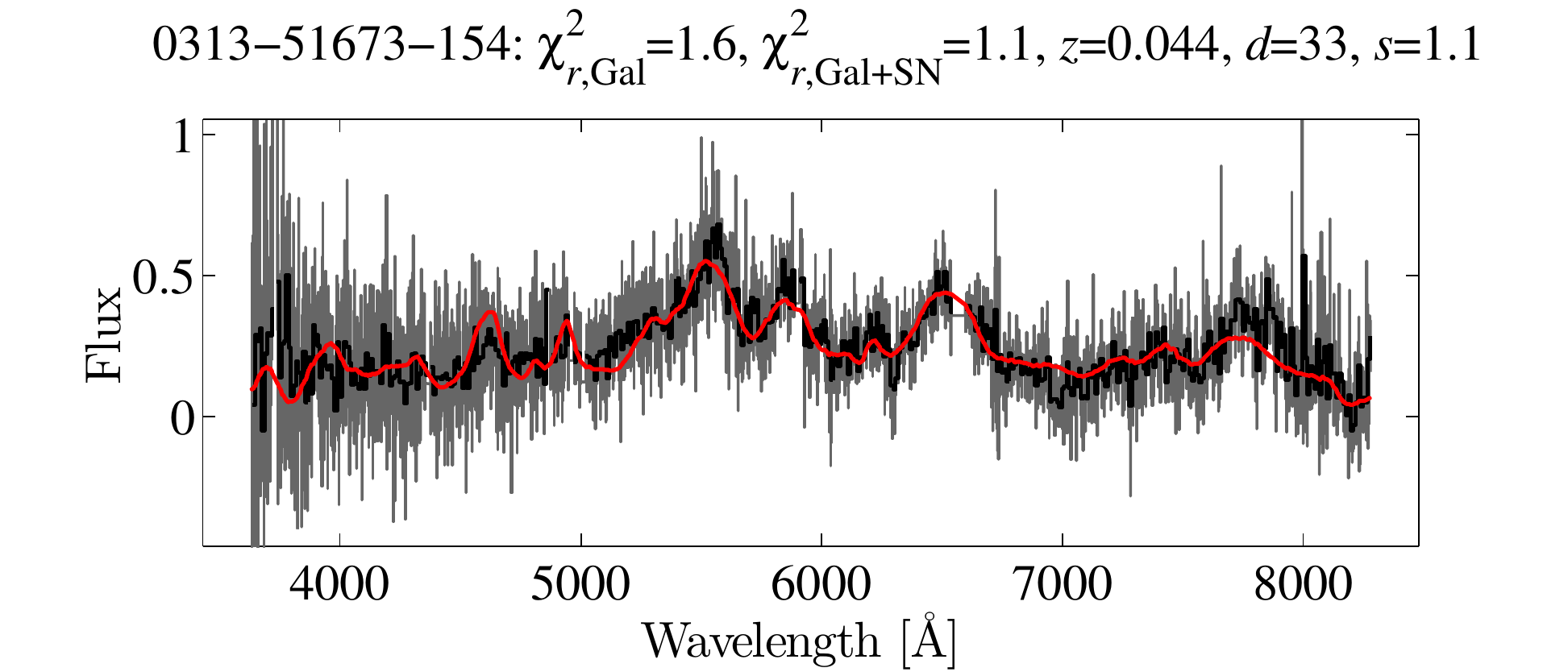} & 
   \includegraphics[width=0.5\textwidth]{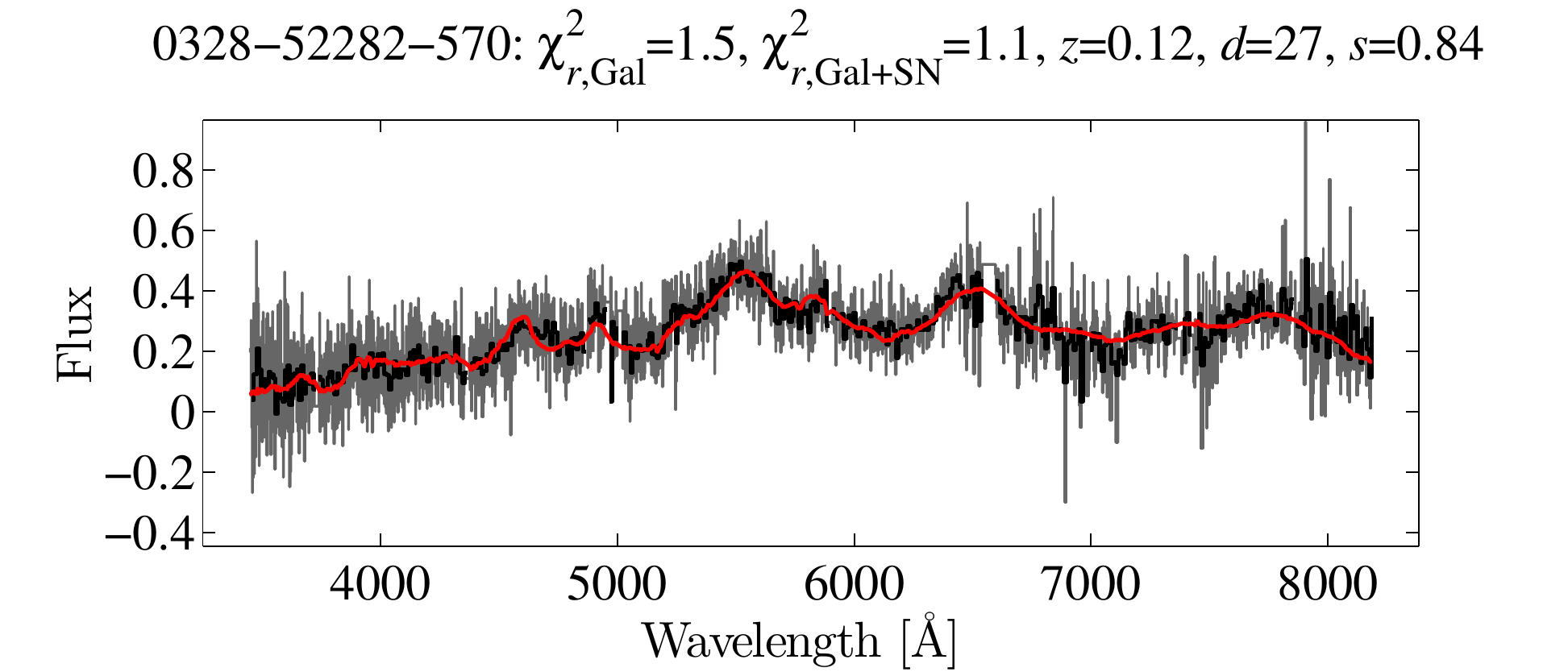} \\
   \includegraphics[width=0.5\textwidth]{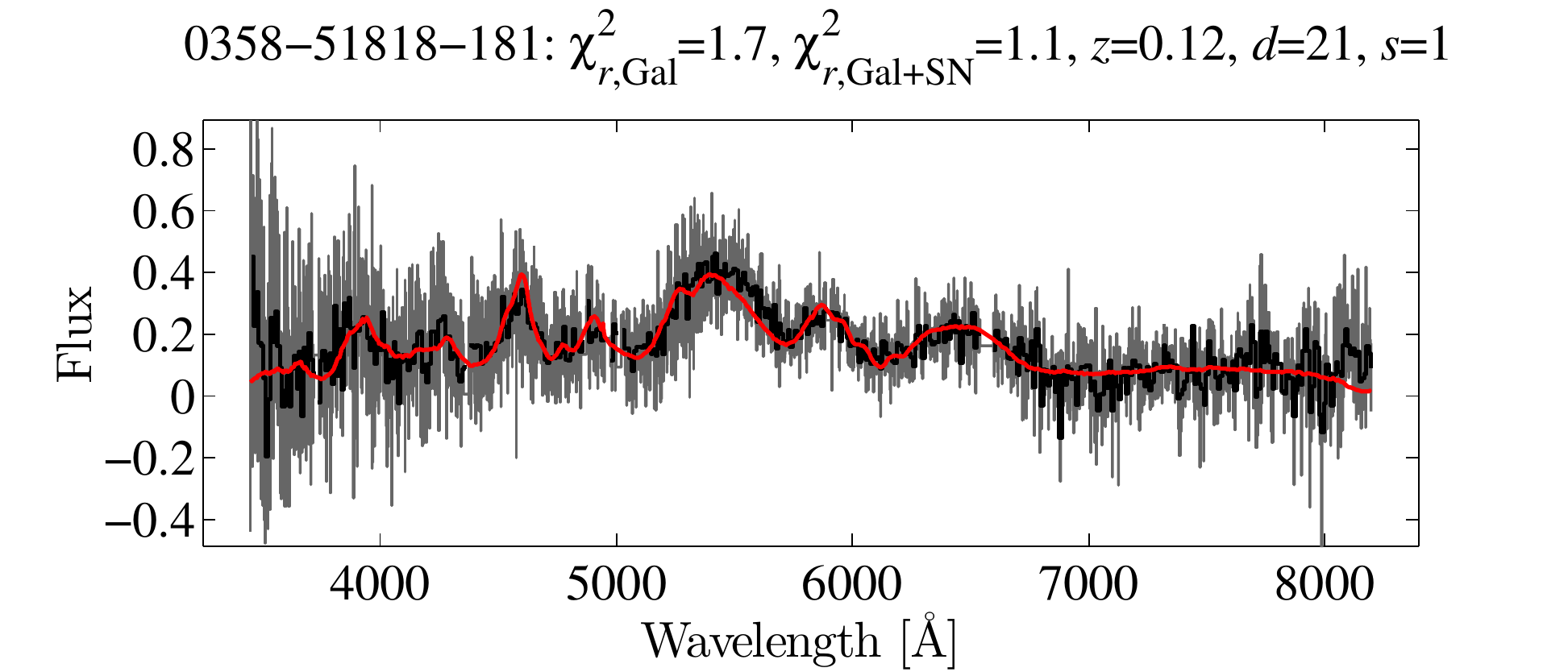} & 
   \includegraphics[width=0.5\textwidth]{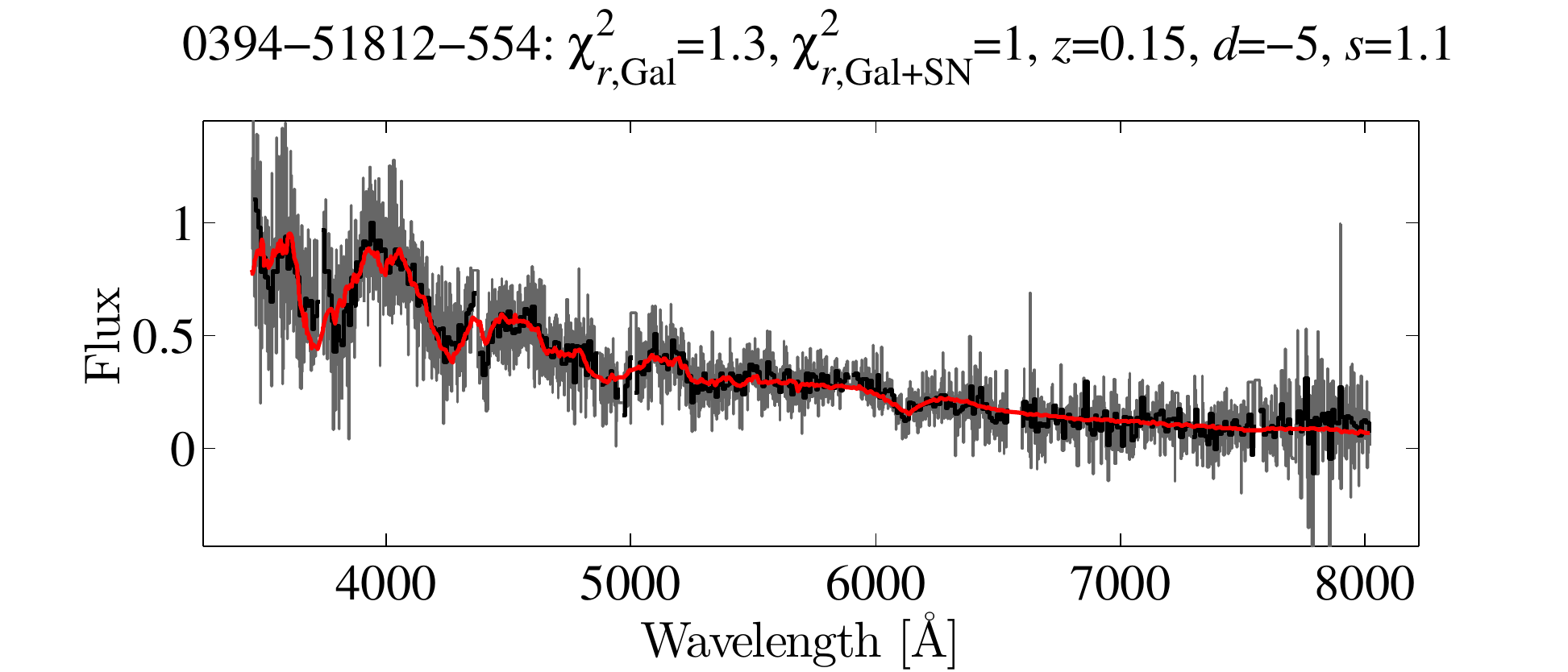} \\
   \includegraphics[width=0.5\textwidth]{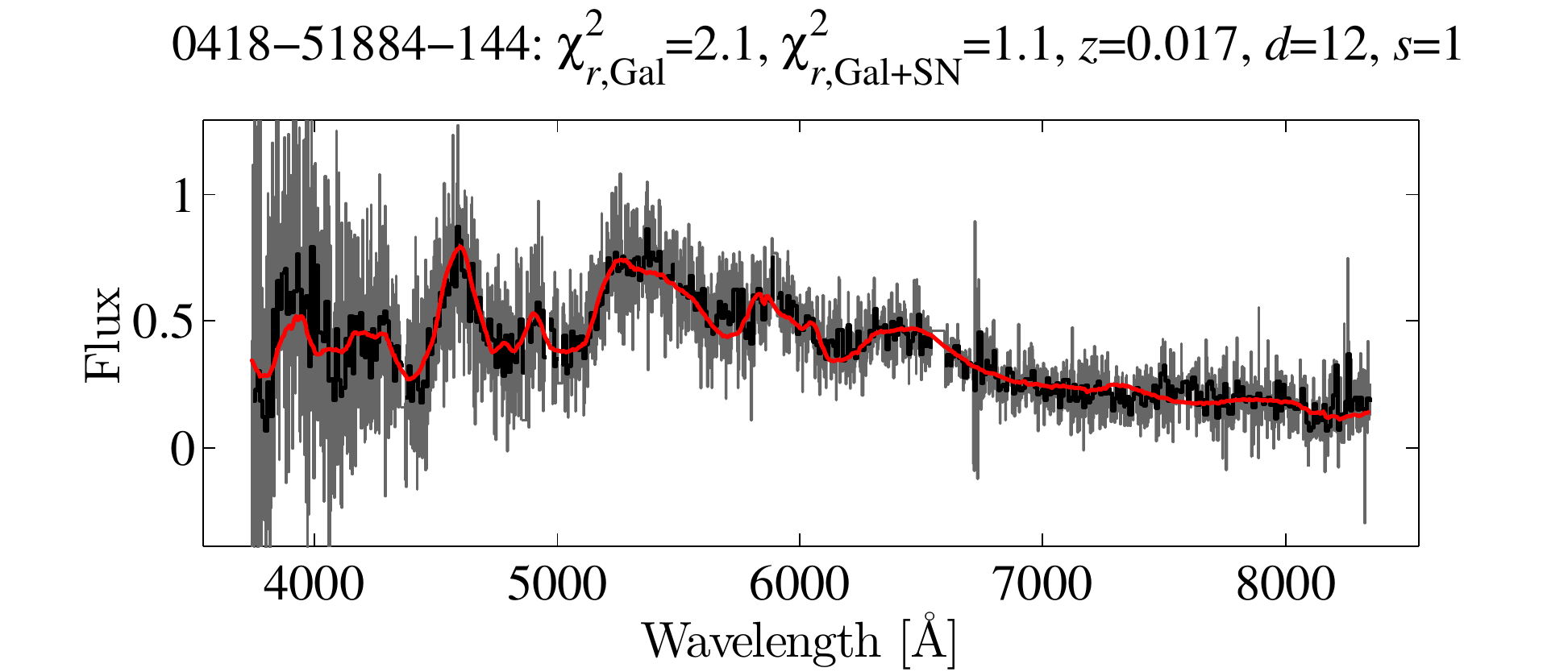} & 
   \includegraphics[width=0.5\textwidth]{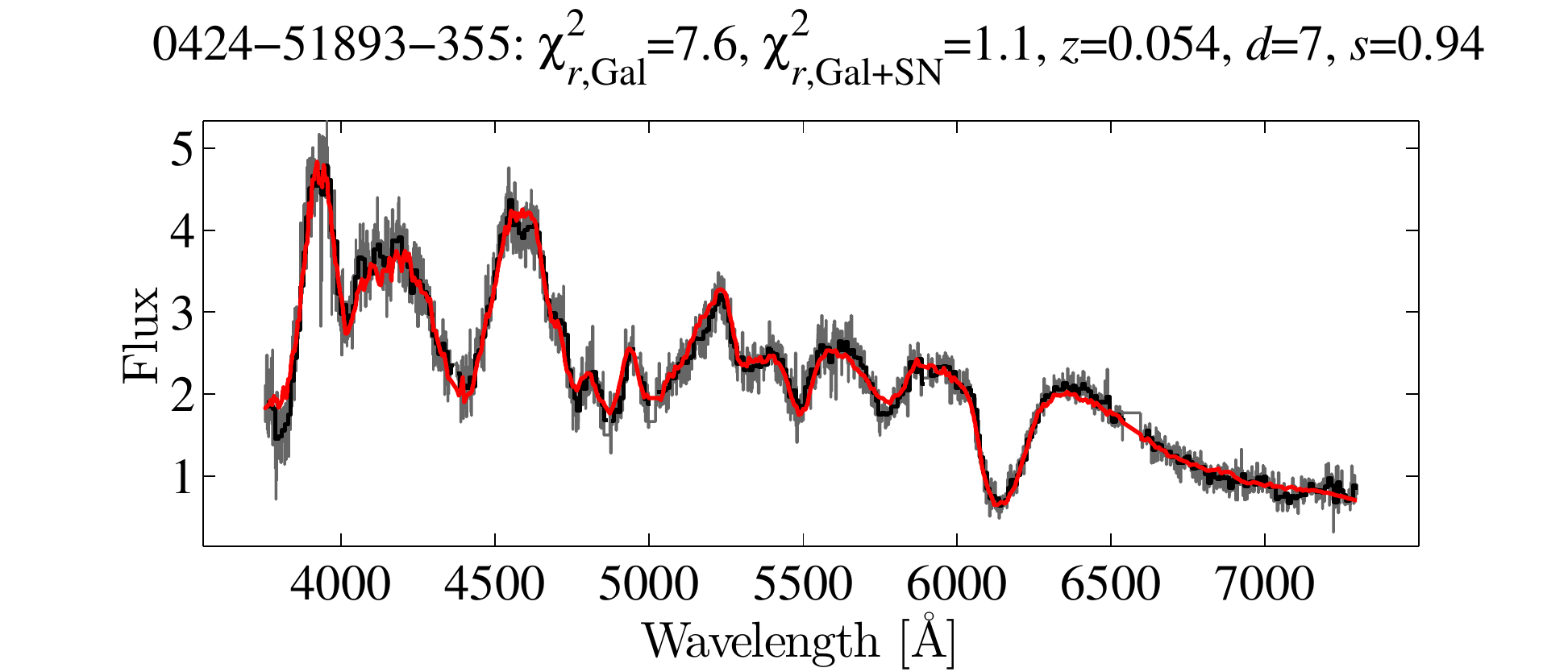} \\
   \includegraphics[width=0.5\textwidth]{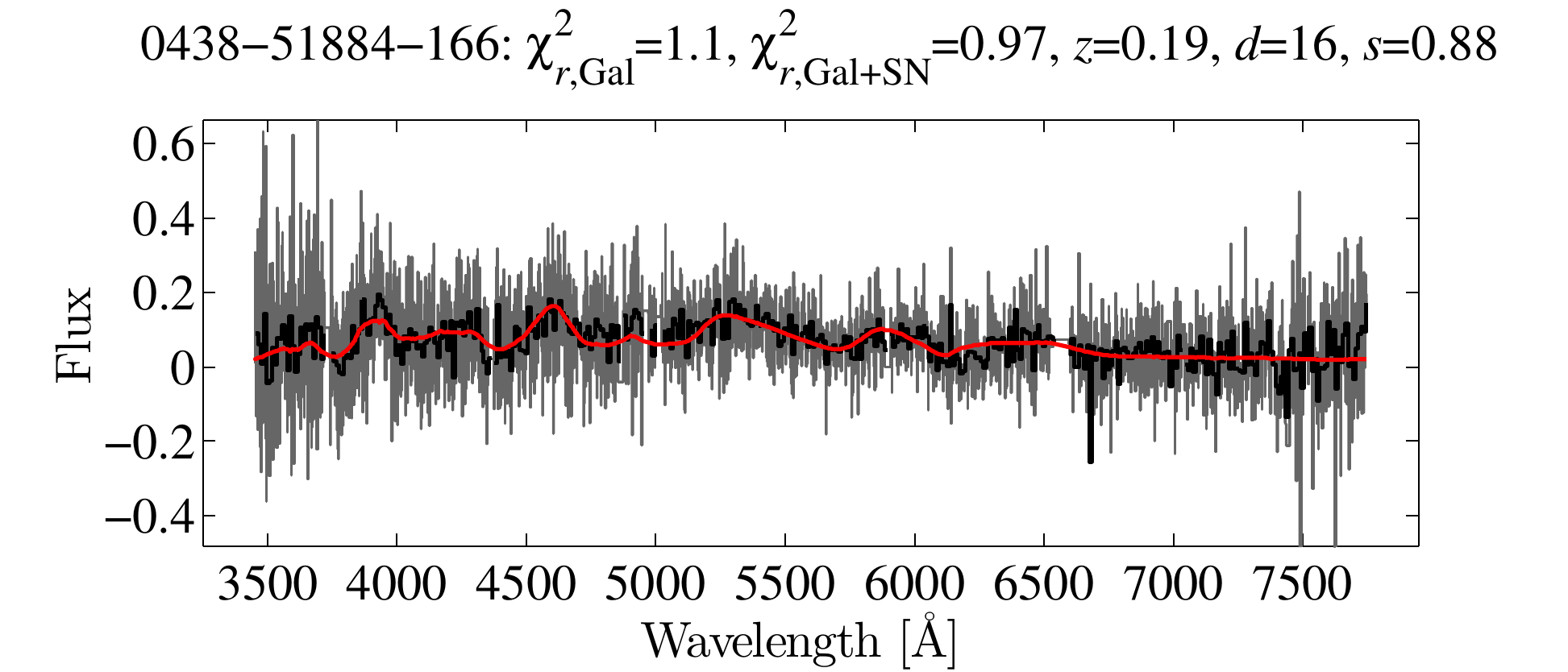} & 
   \includegraphics[width=0.5\textwidth]{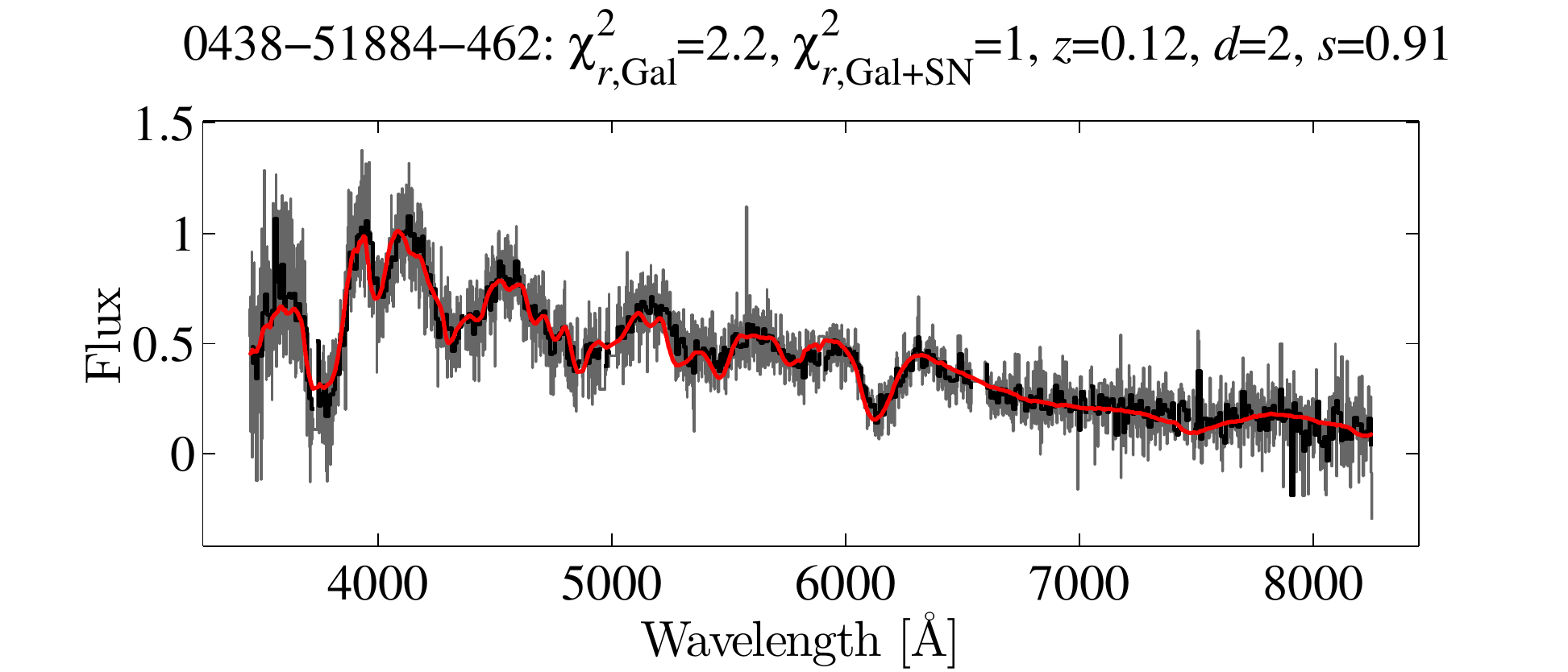} \\
  \end{tabular}
  \caption{SNe~Ia discovered in SDSS DR7 galaxy spectra. The residual spectrum, obtained by first fitting galaxy eigenspectra and transient templates to the original spectrum, and then subtracting the resulting galaxy model, is shown in grey. The same residual, rebinned into 10 \AA{} bins, is shown in black. The best-fitting SN~Ia template is overlaid in red. The flux is in units of $10^{-16}$~erg cm$^{-2}$~s$^{-1}$~\AA{}$^{-1}$, and the wavelengths are in the rest frames of the SNe. The title of each panel details the plate, MJD, and fiber in which it was discovered; the $\chi^2_r$ value obtained when fitting only galaxy eigenspectra to the spectrum, $\chi^2_r$(Gal); the $\chi^2_r$ value obtained from the best-fitting combination of galaxy eigenspectra and transient templates, $\chi^2_r$(Gal+SN); the redshift of the SN-host galaxy, $z$; the SVD-derived age, $d$; and the stretch parameter, $s$, obtained by using the \salt\ templates. Similar figures for the rest of our SN~Ia sample are available in the electronic version of the paper -- see Supporting Information.}
  \label{fig:SNe_Ia} 
 \end{minipage}
\end{figure*}

\begin{figure*}
 \begin{minipage}{\textwidth}
  \vspace{0.2cm}
  \begin{tabular}{cc}
   \includegraphics[width=0.5\textwidth]{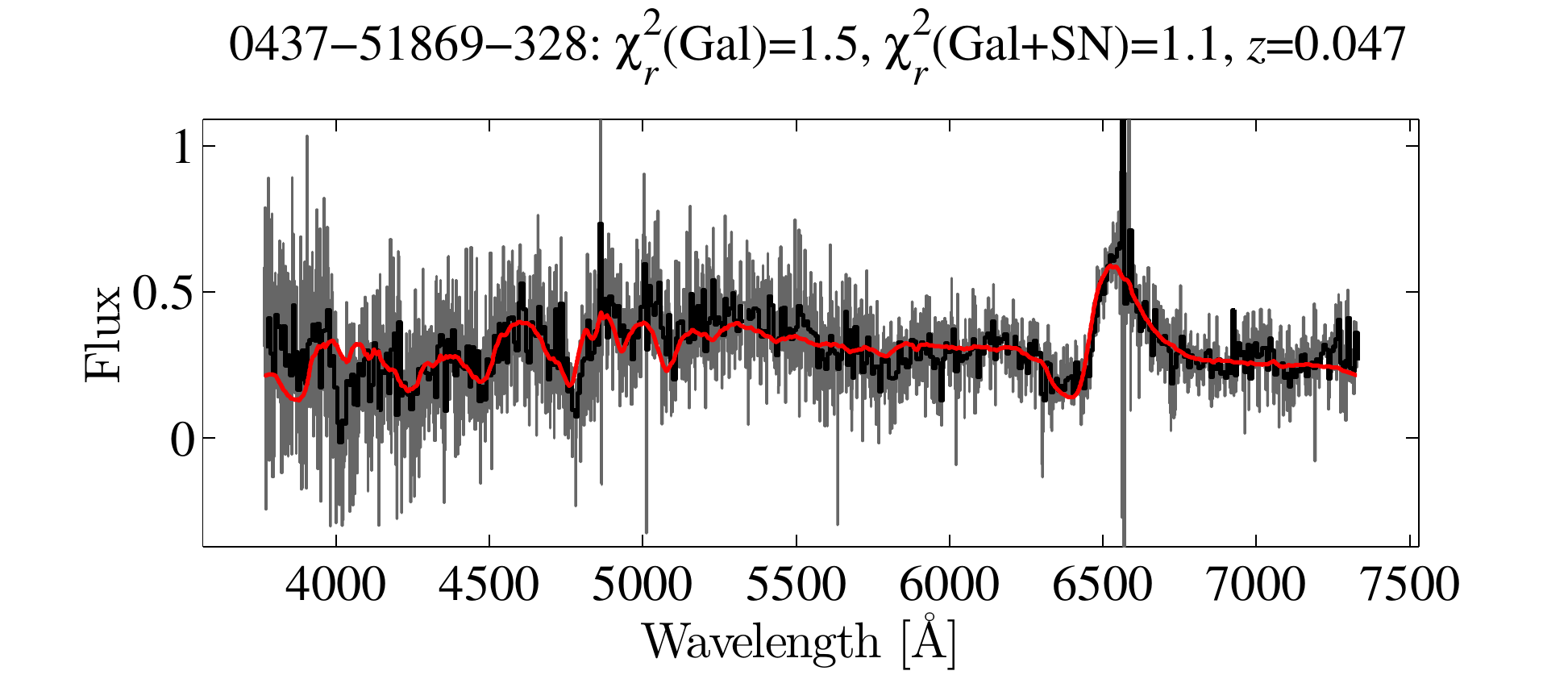} & 
   \includegraphics[width=0.5\textwidth]{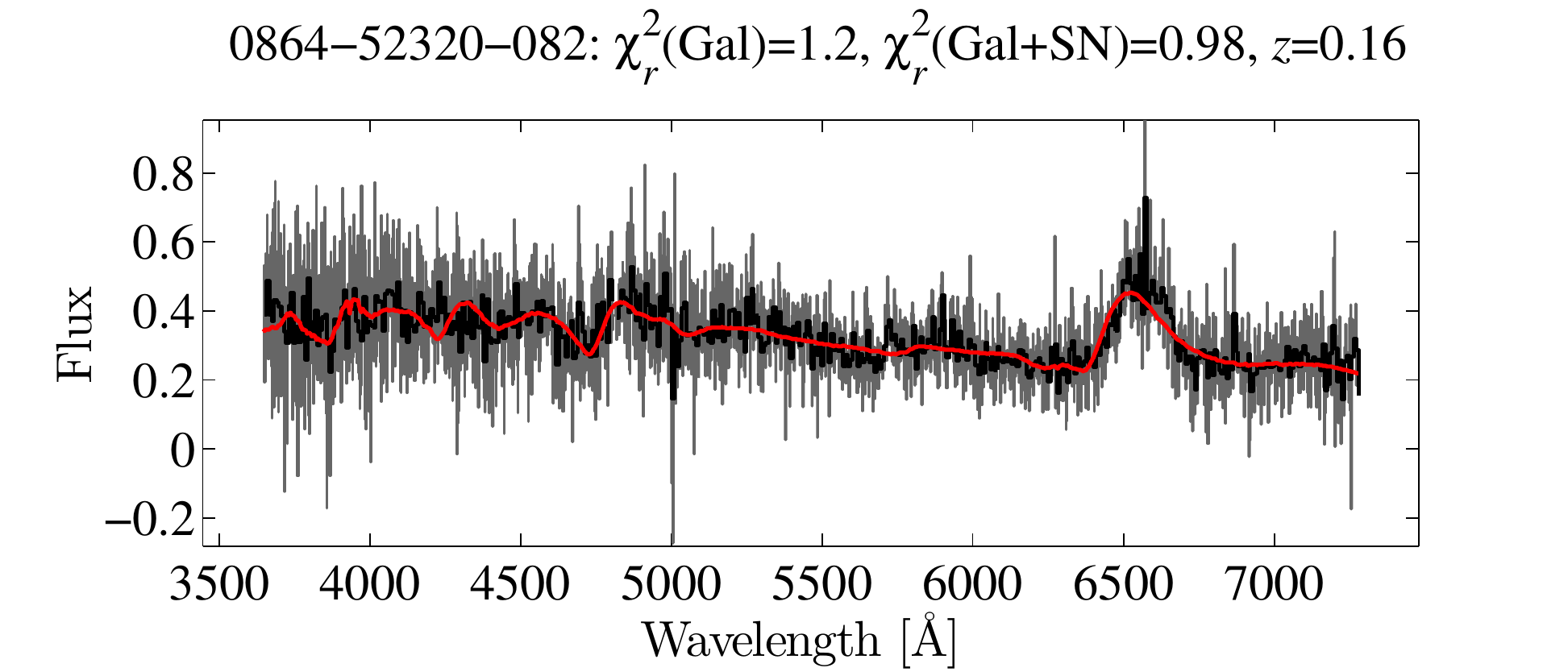} \\
   \includegraphics[width=0.5\textwidth]{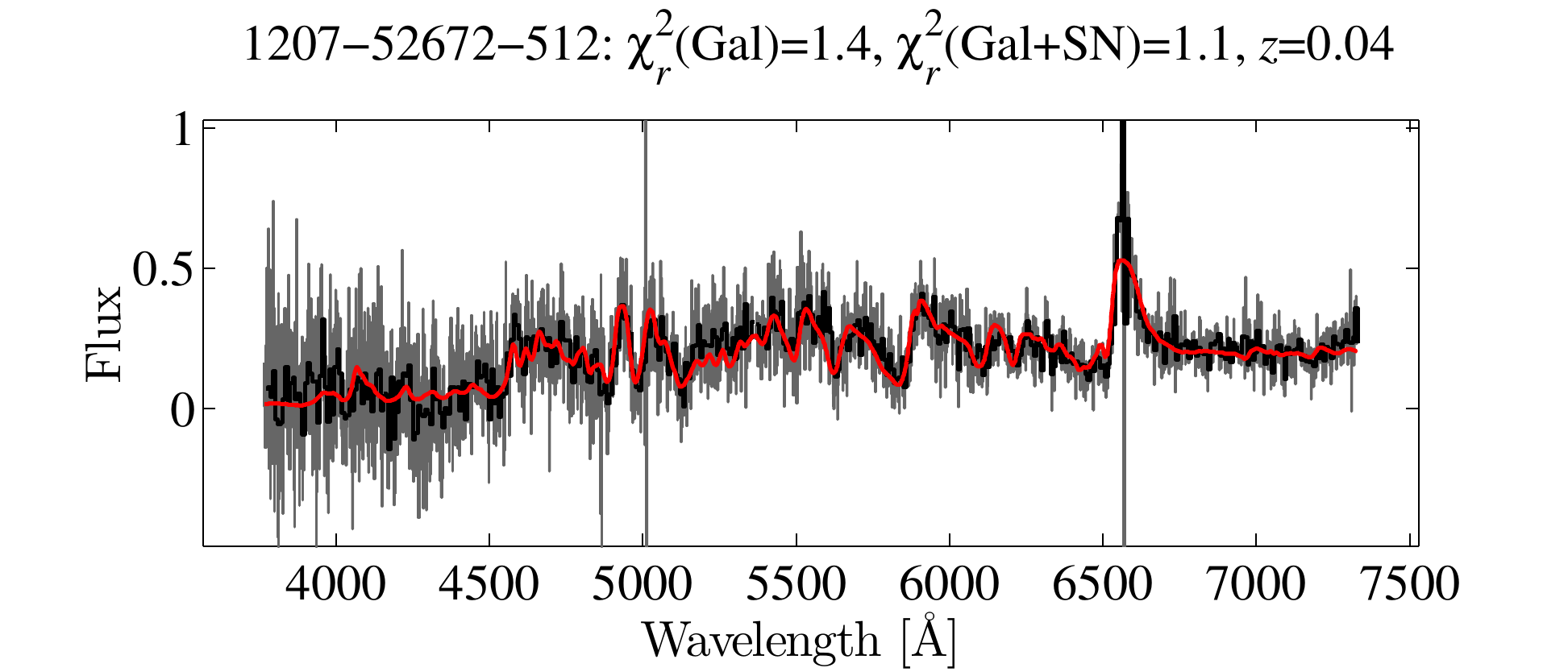} & 
   \includegraphics[width=0.5\textwidth]{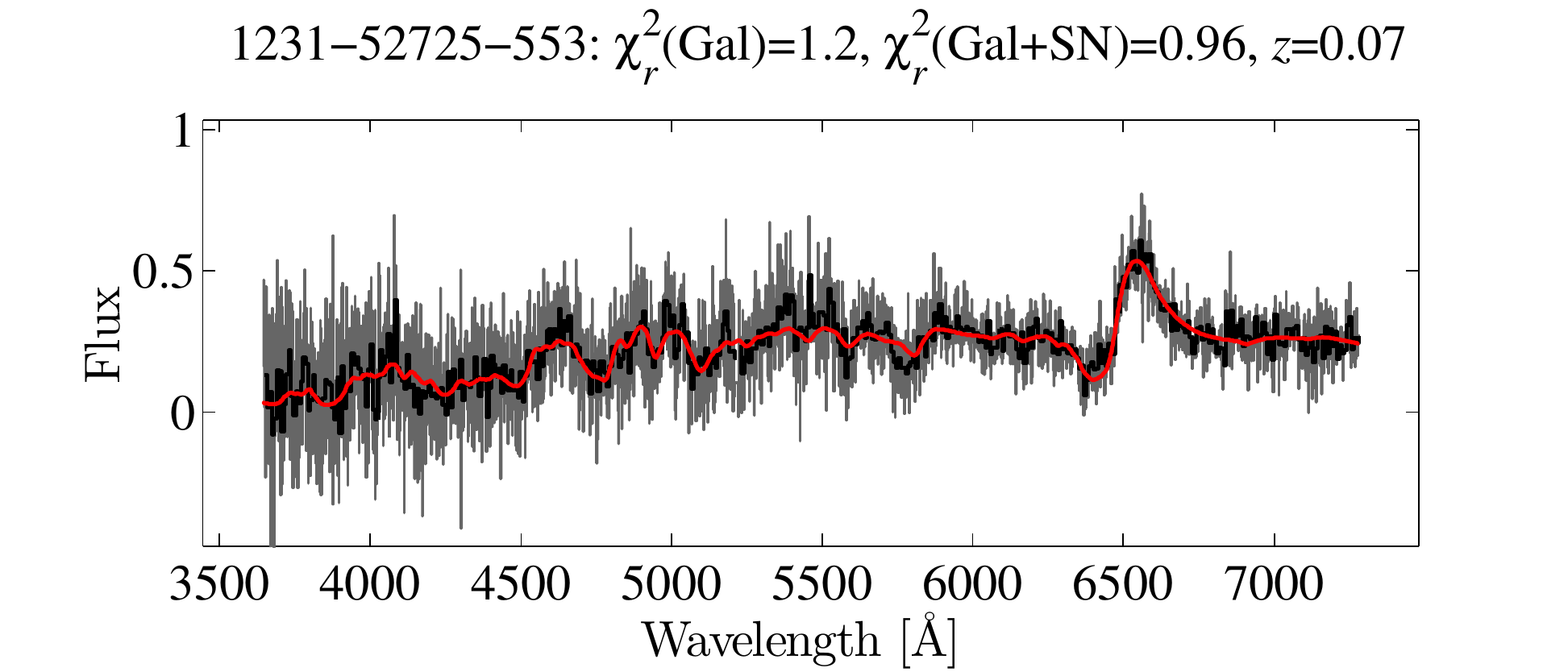} \\
   \includegraphics[width=0.5\textwidth]{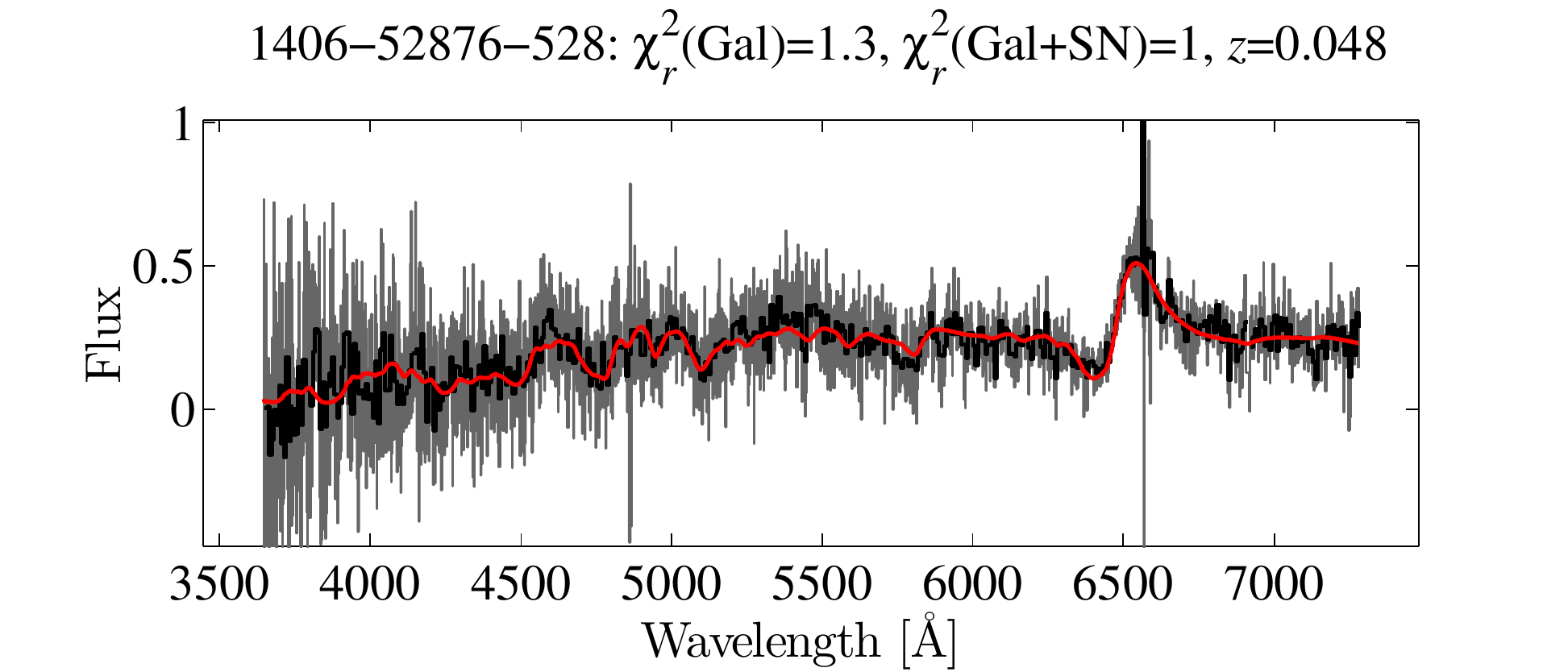} & 
   \includegraphics[width=0.5\textwidth]{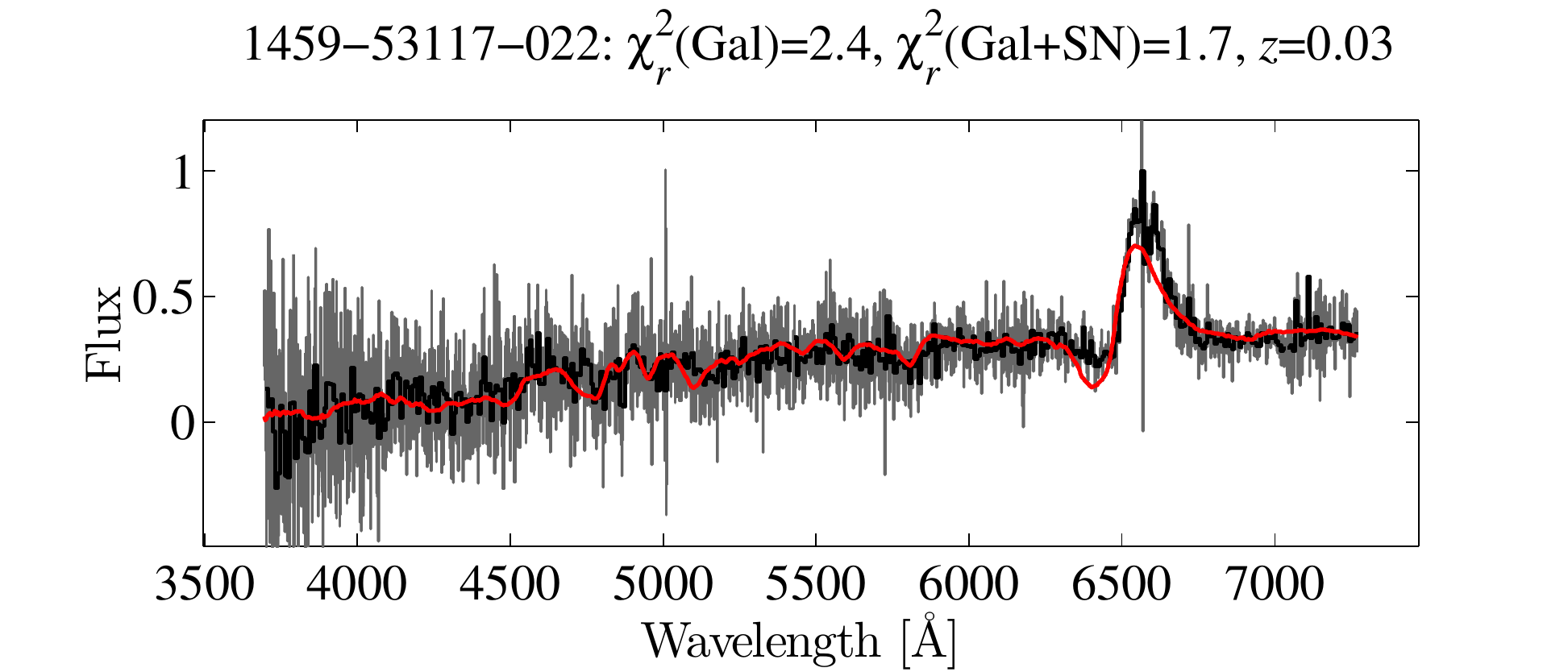} \\
   \includegraphics[width=0.5\textwidth]{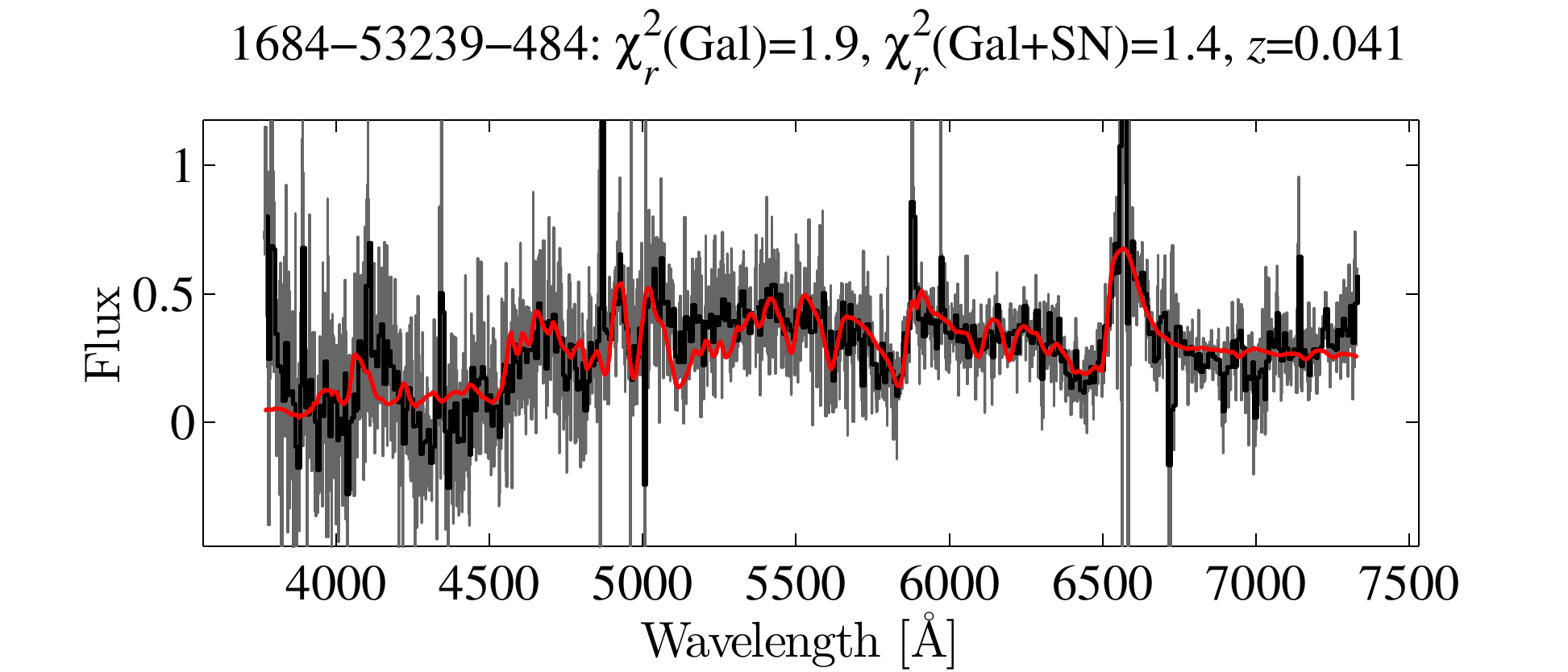} & 
   \includegraphics[width=0.5\textwidth]{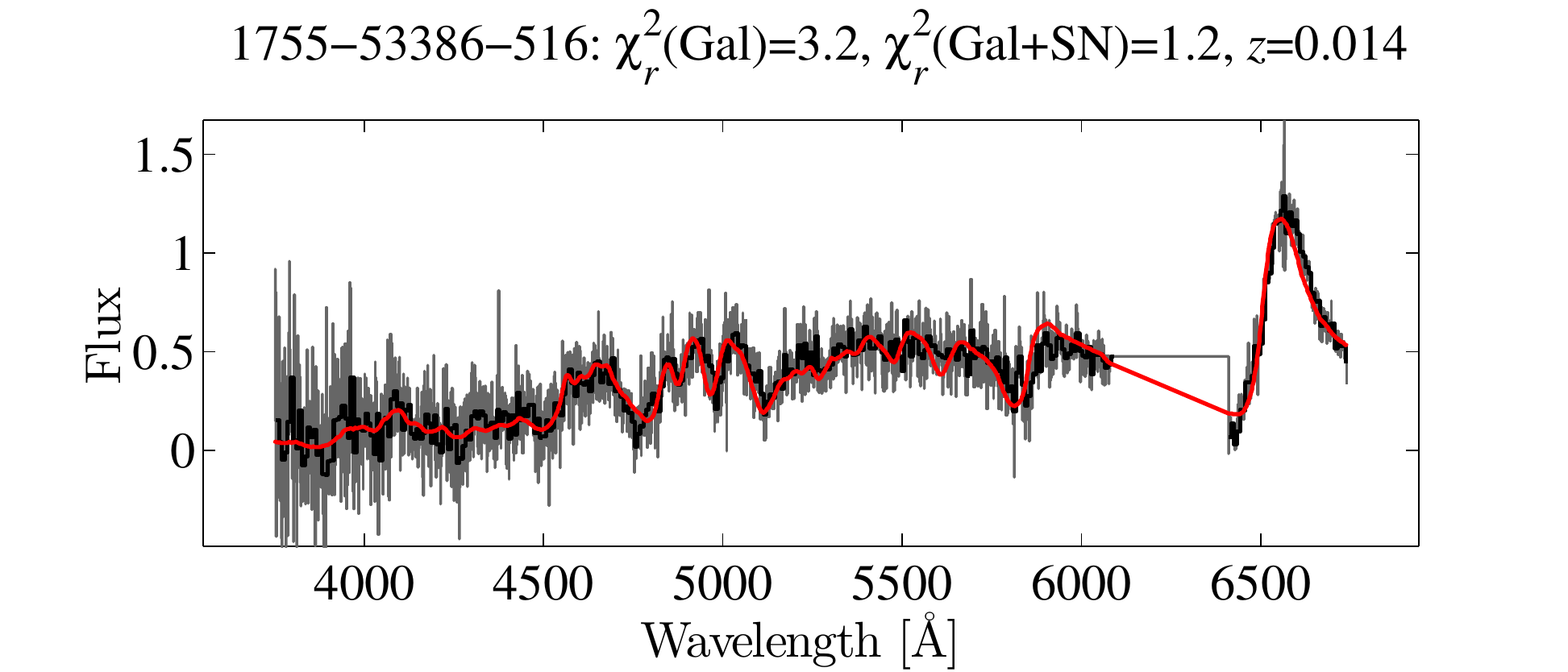} \\
   \includegraphics[width=0.5\textwidth]{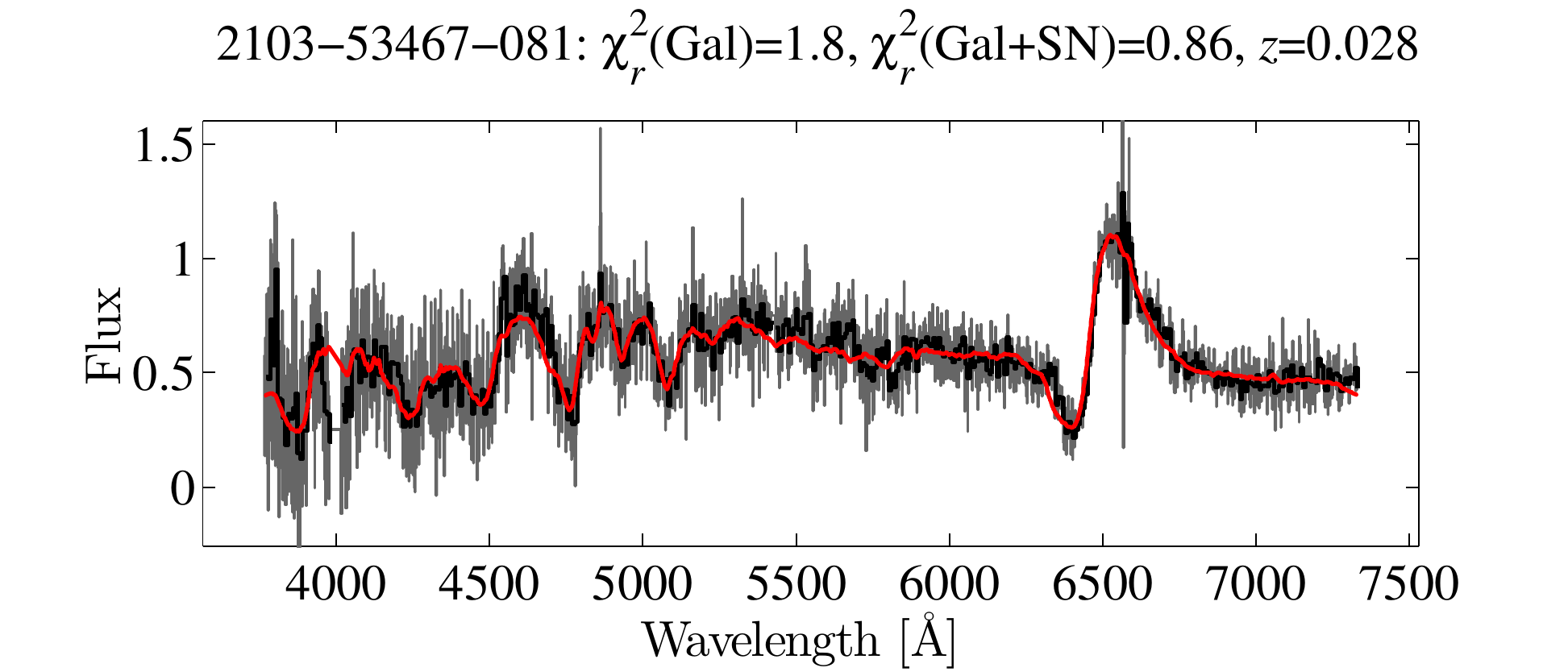} & 
   \includegraphics[width=0.5\textwidth]{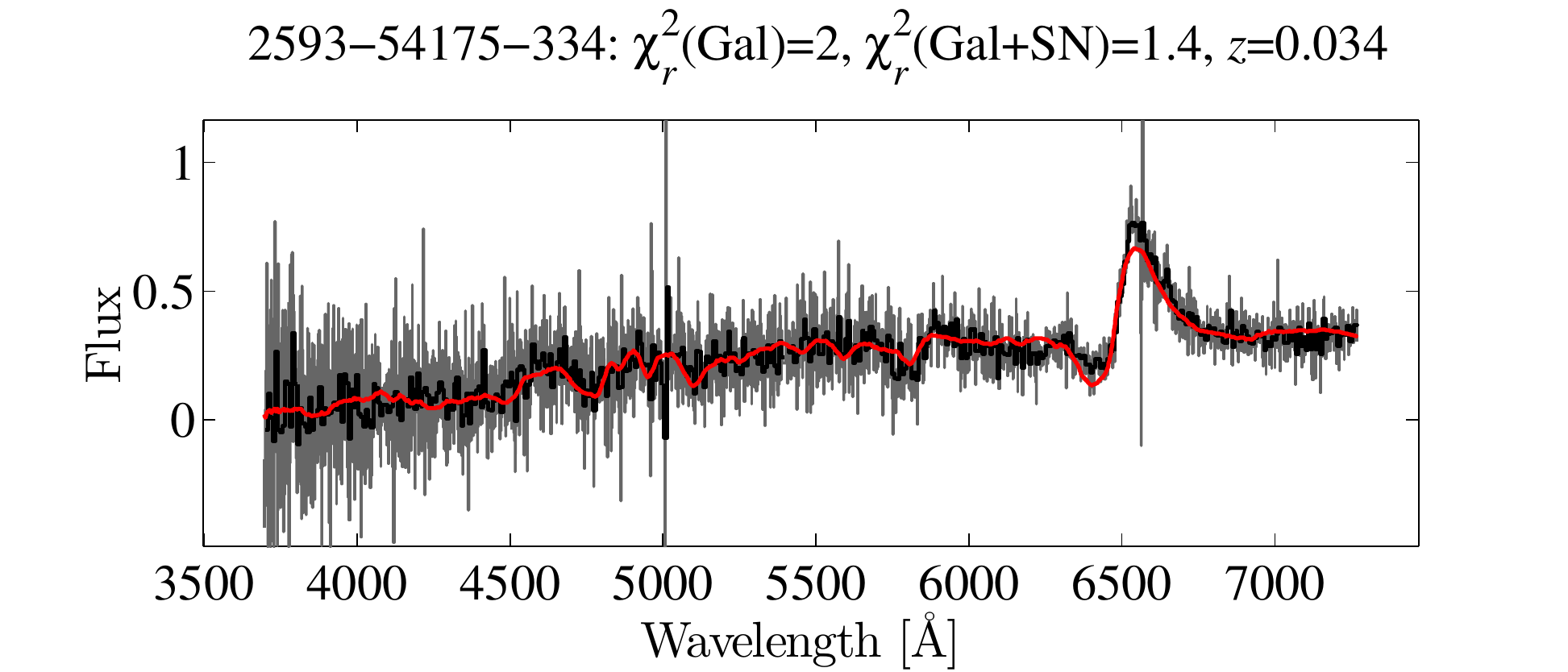} \\
  \end{tabular}
  \caption{Same as Fig.~\ref{fig:SNe_Ia}, for SNe~II discovered in SDSS DR7 galaxy spectra.}
  \label{fig:SNe_IIP}
 \end{minipage}
\end{figure*}

Figure~\ref{fig:sne_stretch} shows the stretch distributions of the 90 SNe~Ia in our sample.
Many previous studies have found that star-forming galaxies tend to host higher-stretch SNe~Ia than passive galaxies (e.g., \citealt{2000AJ....120.1479H,2009ApJ...707.1449N,2009ApJ...700.1097H}). 
For example, \citet{2006ApJ...648..868S} find a formal difference between the stretches of SNe~Ia hosted by passive and star-forming galaxies at $z>0.1$, with stretch
medians of $\sim 0.93$ and $\sim 0.98$, respectively. 
Our sample has stretch medians of 0.95 (1.0) for the 43 (47) SNe~Ia hosted by passive (star-forming) galaxies, which is a stretch difference of the same sense and similar magnitude to those seen by previous studies.
However, a number of statistical tests ($\chi^2$, Kolmogorov-Smirnov, Student) all indicate that, in the case of our sample, these differences are not significant, which is not surprising given that our stretch-recovery uncertainty is about twice as large as the expected difference between the populations (see Section~\ref{subsec:stretch}).

When using the figure of merit $\chi^2_\lambda$ introduced in Section~\ref{subsec:how}, we optimize our detection and classification method towards the discovery of SNe~Ia, and recover only two of the ten SNe~II in Table~\ref{table:SNe}, due to the relatively smaller wavelength ranges covered by the \snid\ SN~II templates. 
This, together with the overall small size of our SN~II sample, precludes deriving a SN~II rate in the context of this work.

\begin{table*}
\caption{SNe discovered in SDSS DR7}\label{table:SNe}
\begin{tabular}{c c c c c c c c c c c c c}
\hline
\hline
SDSS ID & Date & $\alpha$ & $\delta$ & $z$ & Age$_1$ & Age$_2$ & $s$ & $\R_{SN}$ & $\R_{H}$ & $\chi^2_{gal}$ & $\chi^2_{SN}$ & Type \\
(1) & (2) & (3) & (4) & (5) & (6) & (7) & (8) & (9) & (10) & (11) & (12) & (13) \\
\hline

0271-51883-171 & 05/12/00 & 154.5020 & $-0.0328$ & 0.065 & 27 & 42 & 0.91 & 19.54 & 18.11 & 2.4 & 1.2 & Ia \\

0291-51928-076 & 19/01/01 & 191.8892 & 0.0992 & 0.086 & 28 & 21 & 1.06 & 20.03 & 17.87 & 1.7 & 1.0 & Ia \\

0313-51673-154 & 09/05/00 & 231.8958 & $-0.0595$ & 0.044 & 33 & 38 & 1.11 & 19.85 & 17.43 & 1.6 & 1.1 & Ia \\

0328-52282-570 & 08/01/02 & 175.9218 & $-1.4770$ & 0.125 & 27 & 30 & 0.84 & 19.94 & 17.99 & 1.5 & 1.1 & Ia \\

0358-51818-181 & 01/10/00 & 263.1189 & 56.0737 & 0.123 & 21 & 30 & 1.05 & 20.19 & 18.21 & 1.7 & 1.1 & Ia \\

0394-51812-554 & 25/09/00 & 14.0069 & 0.4353 & 0.146 & $-5$ & $-5$ & 1.12 & 20.03 & 18.08 & 1.3 & 1.0 & Ia \\

0418-51884-144 & 06/12/00 & 9.8403 & 14.4699 & 0.017 & 12 & 16 & 1.04 & 19.46 & 17.36 & 2.1 & 1.1 & Ia \\

0424-51893-355 & 15/12/00 & 19.6493 & 14.6835 & 0.054 & 7 & 7 & 0.94 & 17.97 & 16.98 & 7.6 & 1.1 & Ia \\

0438-51884-166 & 06/12/00 & 121.0888 & 46.7870 & 0.187 & 16 & 12 & 0.88 & 21.30 & 18.16 & 1.1 & 1.0 & Ia \\

0438-51884-462 & 06/12/00 & 120.8026 & 47.6138 & 0.117 & 2 & $-5$ & 0.91 & 19.54 & 18.06 & 2.2 & 1.0 & Ia \\

0452-51911-319 & 02/01/01 & 140.6214 & 57.9081 & 0.063 & 9 & 10 & 0.83 & 19.43 & 17.93 & 1.8 & 1.1 & Ia \\

0480-51989-024 & 21/03/01 & 147.9711 & 1.1016 & 0.063 & 2 & $-8$ & 1.00 & 18.77 & 17.77 & 4.4 & 1.5 & Ia \\

0498-51984-102 & 16/03/01 & 212.7430 & 64.8475 & 0.141 & 20 & 18 & 0.99 & 20.43 & 18.09 & 1.2 & 0.9 & Ia \\

0500-51994-100 & 26/03/01 & 149.8156 & 0.9673 & 0.088 & 11 & 9 & 0.98 & 19.64 & 18.43 & 1.8 & 1.0 & Ia \\

0578-52339-314 & 06/03/02 & 159.7062 & 4.0156 & 0.129 & 6 & 3 & 1.04 & 19.54 & 17.87 & 1.8 & 1.2 & Ia \\

0604-52079-209 & 19/06/01 & 204.9366 & 62.3936 & 0.136 & 2 & $-5$ & 0.95 & 20.11 & 18.38 & 1.2 & 1.0 & Ia \\

0606-52365-412 & 01/04/02 & 213.3167 & 62.0323 & 0.142 & 6 & $-2$ & 0.88 & 20.15 & 17.45 & 1.4 & 1.2 & Ia \\

0622-52054-011 & 25/05/01 & 244.3058 & 48.4744 & 0.104 & 16 & 17 & 1.07 & 19.88 & 18.21 & 1.7 & 1.3 & Ia \\

0738-52521-360 & 04/09/02 & 336.7925 & 13.6661 & 0.150 & 27 & 22 & 1.04 & 20.19 & 18.14 & 1.1 & 1.0 & Ia \\

0745-52258-092 & 15/12/01 & 350.7757 & 13.6668 & 0.041 & 54 & 68 & 0.89 & 20.16 & 17.40 & 1.1 & 0.9 & Ia \\

0762-52232-067 & 19/11/01 & 129.7765 & 43.7123 & 0.125 & 31 & 34 & 1.06 & 20.39 & 18.03 & 1.0 & 0.9 & Ia \\

0814-52443-249 & 18/06/02 & 243.4005 & 43.7256 & 0.112 & 8 & 7 & 0.96 & 20.01 & 18.18 & 1.5 & 0.9 & Ia \\

0844-52378-462 & 14/04/02 & 184.4150 & 5.3234 & 0.104 & 31 & 36 & 1.05 & 20.32 & 17.85 & 1.2 & 1.0 & Ia \\

0905-52643-213 & 04/01/03 & 157.4940 & 53.5018 & 0.137 & 42 & 38 & 1.02 & 20.60 & 18.25 & 1.0 & 0.9 & Ia \\

0915-52443-543 & 18/06/02 & 211.3174 & $-1.7113$ & 0.054 & 29 & 24 & 1.03 & 19.33 & 17.44 & 1.5 & 0.9 & Ia \\

0966-52642-221 & 03/01/03 & 172.2523 & 48.7331 & 0.074 & 3 & 2 & 0.89 & 19.46 & 18.26 & 1.8 & 1.1 & Ia \\

1038-52673-135 & 03/02/03 & 191.8526 & 53.7308 & 0.153 & 12 & 11 & 0.88 & 20.59 & 19.08 & 1.2 & 1.0 & Ia \\

1059-52618-144 & 10/12/02 & 116.8937 & 27.4465 & 0.063 & 33 & 32 & 1.04 & 19.43 & 17.51 & 1.9 & 0.9 & Ia \\

1171-52753-185 & 24/04/03 & 244.8403 & 41.0899 & 0.038 & 10 & 8 & 0.89 & 18.00 & 16.16 & 3.9 & 0.9 & Ia \\

1189-52668-239 & 29/01/03 & 132.2645 & 5.8377 & 0.126 & 3 & 4 & 0.78 & 20.01 & 18.07 & 1.4 & 1.0 & Ia \\

1205-52670-632 & 31/01/03 & 122.8267 & 26.1661 & 0.144 & 12 & 13 & 0.97 & 20.25 & 17.58 & 1.3 & 1.1 & Ia \\

1278-52735-425 & 06/04/03 & 189.1061 & 50.6116 & 0.106 & 6 & 6 & 0.95 & 19.51 & 17.64 & 1.5 & 1.0 & Ia \\

1289-52734-413 & 05/04/03 & 220.2447 & 45.1308 & 0.074 & 8 & 8 & 0.93 & 19.53 & 16.57 & 1.0 & 0.8 & Ia \\

1298-52964-304 & 21/11/03 & 129.7903 & 7.4088 & 0.046 & 32 & 48 & 0.96 & 18.92 & 17.47 & 2.0 & 1.0 & Ia \\

1310-53033-459 & 29/01/04 & 173.5529 & 58.2622 & 0.122 & 16 & 14 & 1.06 & 19.56 & 17.67 & 1.9 & 0.9 & Ia \\

1324-53088-169 & 24/03/04 & 211.9955 & 54.3631 & 0.067 & 52 & 52 & 1.06 & 19.87 & 17.60 & 1.5 & 1.0 & Ia \\

1337-52767-480 & 08/05/03 & 245.0462 & 38.1115 & 0.130 & 2 & 4 & 0.84 & 19.80 & 18.12 & 1.5 & 1.0 & Ia \\

1392-52822-147 & 02/07/03 & 240.3856 & 26.9875 & 0.050 & 14 & 14 & 0.93 & 17.84 & 16.90 & 6.4 & 1.1 & Ia \\

1400-53470-234 & 10/04/05 & 230.1879 & 36.8118 & 0.103 & 6 & 3 & 1.03 & 18.89 & 16.58 & 1.8 & 0.9 & Ia \\

1400-53470-351 & 10/04/05 & 229.2268 & 37.1240 & 0.116 & $-9$ & $-14$ & 1.16 & 19.66 & 17.21 & 1.1 & 0.9 & Ia \\

1403-53227-456 & 10/08/04 & 237.2391 & 33.9571 & 0.128 & 27 & 33 & 0.84 & 20.26 & 18.39 & 1.2 & 1.0 & Ia \\

1445-53062-067 & 27/02/04 & 177.1144 & 42.1320 & 0.086 & 3 & $-3$ & 0.89 & 19.65 & 17.41 & 1.3 & 0.9 & Ia \\

1462-53112-638 & 17/04/04 & 202.1417 & 41.8523 & 0.028 & 8 & 7 & 0.99 & 16.52 & 16.53 & 13.3 & 1.1 & Ia \\

1581-53149-470 & 24/05/04 & 235.1032 & 32.8659 & 0.054 & 30 & 40 & 0.90 & 18.97 & 17.69 & 3.3 & 1.0 & Ia \\

1598-53033-380 & 29/01/04 & 155.7088 & 11.7030 & 0.102 & 2 & $-1$ & 0.92 & 19.67 & 17.99 & 1.6 & 1.0 & Ia \\

1645-53172-349 & 16/06/04 & 217.2314 & 35.0823 & 0.121 & 21 & 34 & 1.14 & 19.97 & 16.95 & 1.3 & 1.0 & Ia \\

1697-53142-506 & 17/05/04 & 199.1256 & 12.6769 & 0.151 & 3 & 4 & 1.06 & 19.92 & 17.66 & 1.3 & 0.9 & Ia \\

1700-53502-302 & 12/05/05 & 203.6364 & 11.1324 & 0.095 & 3 & 1 & 1.21 & 19.59 & 16.96 & 1.3 & 1.1 & Ia \\

1710-53504-488 & 14/05/05 & 219.0440 & 12.4451 & 0.085 & 27 & 24 & 0.90 & 19.70 & 18.03 & 1.2 & 1.0 & Ia \\

1744-53055-210 & 20/02/04 & 149.9506 & 11.4737 & 0.060 & 9 & 9 & 0.94 & 18.37 & 19.20 & 6.6 & 1.1 & Ia \\

\hline

\multicolumn{13}{l}{(1) -- SDSS DR7 plate, MJD, and fiber in which the SN was discovered.} \\                                                                                      
\multicolumn{13}{l}{(2) -- Date on which the SN was captured, in dd/mm/yy.} \\
\multicolumn{13}{l}{(3)--(4) -- Right ascensions and declinations (J2000).} \\
\multicolumn{13}{l}{(5) -- SN host-galaxy redshift.} \\
\multicolumn{13}{l}{(6)--(7) -- SN age, in days, as measured by SVD and \snid, respectively. SN~Ia (II) ages have an uncertainty of $\pm 6~(33)$ days.} \\
\multicolumn{13}{l}{(8) -- SN stretch, as measured with the \salt~ templates. All stretches have an uncertainty of $^{+0.10}_{-0.14}$.} \\
\multicolumn{13}{l}{(9)--(10) -- SN and host-galaxy \R-band magnitudes.} \\
\multicolumn{13}{l}{(11)--(12) -- Reduced $\chi^2$ value of galaxy and galaxy+transient fits.} \\
\multicolumn{13}{l}{(13) -- SN type.}
\end{tabular}
\end{table*}

\begin{table*}
\begin{tabular}{c c c c c c c c c c c c c}
\multicolumn{12}{l}{{\bf Table 4.} SNe discovered in SDSS DR7 -- {\it continued}} \\
\hline
\hline
SDSS ID & Date & $\alpha$ & $\delta$ & $z$ & Age$_1$ & Age$_2$ & $s$ & $\R_{SN}$ & $\R_{H}$ & $\chi^2_{gal}$ & $\chi^2_{SN}$ & Type \\
(1) & (2) & (3) & (4) & (5) & (6) & (7) & (8) & (9) & (10) & (11) & (12) & (13) \\
\hline

1755-53386-309 & 16/01/05 & 173.1052 & 14.6202 & 0.082 & 21 & 21 & 1.10 & 19.32 & 17.13 & 1.7 & 0.9 & Ia \\

1758-53084-523 & 20/03/04 & 127.3894 & 8.8682 & 0.112 & 2 & $-3$ & 0.85 & 19.62 & 17.75 & 1.6 & 1.0 & Ia \\

1782-53383-517 & 13/01/05 & 126.1666 & 54.6689 & 0.063 & 55 & 43 & 1.02 & 20.02 & 17.33 & 1.5 & 1.0 & Ia \\

1788-54468-126 & 03/01/08 & 145.0818 & 63.2817 & 0.120 & 2 & $-1$ & 1.00 & 20.46 & 18.02 & 1.2 & 1.0 & Ia \\

1791-54266-583 & 15/06/07 & 194.1119 & 10.2277 & 0.107 & 12 & 10 & 1.01 & 19.31 & 17.65 & 2.0 & 1.2 & Ia \\

1793-53883-040 & 28/05/06 & 196.4333 & 9.5797 & 0.055 & 52 & 50 & 1.01 & 19.68 & 17.03 & 1.4 & 0.9 & Ia \\

1801-54156-371 & 25/02/07 & 202.8033 & 7.9573 & 0.123 & 12 & 34 & 1.00 & 19.81 & 17.79 & 1.4 & 1.1 & Ia \\

1803-54152-260 & 21/02/07 & 205.4068 & 5.8710 & 0.060 & 33 & 40 & 0.83 & 19.80 & 17.11 & 1.3 & 1.0 & Ia \\

1843-53816-491 & 22/03/06 & 223.2384 & 31.0742 & 0.094 & 55 & 50 & 1.08 & 20.01 & 17.73 & 1.5 & 1.0 & Ia \\

1944-53385-434 & 15/01/05 & 144.4477 & 28.2876 & 0.153 & $-1$ & $-5$ & 0.86 & 20.10 & 18.49 & 1.6 & 1.0 & Ia \\

1949-53433-080 & 04/03/05 & 150.8604 & 32.1330 & 0.166 & 3 & $-5$ & 1.06 & 20.23 & 18.08 & 1.1 & 0.9 & Ia \\

1957-53415-232 & 14/02/05 & 155.6271 & 35.6764 & 0.128 & 21 & 17 & 0.95 & 20.23 & 18.00 & 1.2 & 0.9 & Ia \\

2019-53430-010 & 01/03/05 & 161.1689 & 30.6343 & 0.072 & 2 & 3 & 1.04 & 18.87 & 18.68 & 4.8 & 1.6 & Ia \\

2159-54328-161 & 16/08/07 & 230.6900 & 19.7057 & 0.109 & 12 & 12 & 0.89 & 19.70 & 17.65 & 1.2 & 0.9 & Ia \\

2165-53917-406 & 01/07/06 & 234.5504 & 25.0456 & 0.067 & 27 & 35 & 1.07 & 19.22 & 17.69 & 1.3 & 0.9 & Ia \\

2173-53874-154 & 19/05/06 & 240.4273 & 20.5769 & 0.123 & 2 & $-1$ & 0.87 & 20.30 & 17.80 & 1.4 & 1.1 & Ia \\

2199-53556-232 & 05/07/05 & 240.3188 & 17.7678 & 0.045 & 13 & 15 & 0.94 & 17.84 & 16.65 & 10.9 & 1.5 & Ia \\

2202-53566-403 & 15/07/05 & 246.0988 & 15.6767 & 0.084 & 4 & 4 & 0.99 & 18.54 & 18.45 & 8.1 & 1.2 & Ia \\

2218-53816-295 & 22/03/06 & 171.5008 & 26.0536 & 0.158 & 6 & 9 & 1.01 & 20.24 & 17.96 & 1.4 & 1.2 & Ia \\

2222-53799-480 & 05/03/06 & 176.1602 & 29.8899 & 0.076 & 12 & 15 & 0.90 & 19.29 & 18.23 & 3.3 & 1.1 & Ia \\

2376-53770-183 & 04/02/06 & 158.4883 & 20.3405 & 0.087 & 2 & $-3$ & 1.02 & 18.88 & 16.90 & 2.0 & 1.0 & Ia \\

2420-54086-142 & 17/12/06 & 123.7092 & 11.7155 & 0.088 & $-3$ & $-1$ & 0.94 & 20.77 & 18.05 & 1.5 & 1.2 & Ia \\

2429-53799-033 & 05/03/06 & 132.1170 & 14.4232 & 0.069 & 38 & 36 & 1.03 & 19.75 & 17.71 & 1.5 & 1.0 & Ia \\

2430-53815-267 & 21/03/06 & 132.4331 & 12.2987 & 0.050 & 33 & 36 & 1.03 & 19.47 & 18.77 & 2.4 & 1.1 & Ia \\

2499-54176-550 & 17/03/07 & 172.2386 & 17.2243 & 0.143 & $-3$ & $-9$ & 0.90 & 20.11 & 18.10 & 1.3 & 0.9 & Ia \\

2594-54177-348 & 18/03/07 & 158.4080 & 16.6030 & 0.052 & 21 & 23 & 1.13 & 20.69 & 17.48 & 1.5 & 1.1 & Ia \\

2603-54479-486 & 14/01/08 & 196.4791 & 17.8272 & 0.078 & 28 & 26 & 1.06 & 20.91 & 18.56 & 1.2 & 0.9 & Ia \\

2609-54476-295 & 11/01/08 & 181.4544 & 18.9829 & 0.168 & 3 & $-5$ & 1.12 & 19.96 & 18.08 & 1.6 & 1.1 & Ia \\

2614-54481-257 & 16/01/08 & 190.8844 & 18.9626 & 0.168 & 8 & 4 & 1.00 & 20.42 & 18.37 & 1.2 & 1.0 & Ia \\

2664-54524-468 & 28/02/08 & 200.7558 & 24.5066 & 0.073 & 6 & 6 & 0.93 & 19.11 & 17.63 & 2.4 & 1.2 & Ia \\

2744-54272-561 & 21/06/07 & 211.5930 & 16.4834 & 0.014 & 40 & 45 & 0.99 & 17.40 & 15.23 & 4.5 & 0.8 & Ia \\

2747-54233-613 & 13/05/07 & 218.8254 & 15.5351 & 0.107 & 12 & 10 & 0.87 & 19.90 & 17.88 & 1.6 & 1.1 & Ia \\

2754-54240-593 & 20/05/07 & 234.2158 & 12.1352 & 0.094 & 27 & 26 & 0.90 & 19.90 & 18.09 & 1.9 & 1.1 & Ia \\

2758-54523-082 & 27/02/08 & 213.1021 & 17.0065 & 0.174 & 6 & 6 & 0.93 & 20.25 & 18.28 & 1.4 & 1.1 & Ia \\

2768-54265-233 & 14/06/07 & 233.2249 & 13.1177 & 0.073 & 27 & 24 & 1.10 & 19.54 & 17.29 & 1.3 & 1.0 & Ia \\

2771-54527-005 & 02/03/08 & 212.4780 & 19.6616 & 0.077 & 16 & 17 & 1.09 & 20.21 & 16.91 & 1.2 & 1.0 & Ia \\

2792-54556-210 & 31/03/08 & 226.3821 & 17.9847 & 0.036 & 13 & 15 & 0.93 & 17.35 & 16.58 & 15.7 & 1.1 & Ia \\

2886-54498-598 & 02/02/08 & 164.2943 & 9.4009 & 0.088 & 6 & 10 & 0.96 & 19.86 & 18.25 & 2.0 & 1.3 & Ia \\

2949-54557-440 & 01/04/08 & 249.2445 & 27.1098 & 0.140 & 21 & 22 & 0.89 & 20.45 & 18.52 & 1.2 & 1.0 & Ia \\

2954-54561-572 & 05/04/08 & 233.5161 & 2.2105 & 0.034 & $-1$ & 5 & 0.83 & 18.82 & 16.76 & 1.3 & 0.8 & Ia \\

0437-51869-328 & 21/11/00 & 119.5556 & 44.0190 & 0.047 & 19 & 0 & $\cdots$ & 20.00 & 17.48 & 1.5 & 1.1 & II \\

0864-52320-082 & 15/02/02 & 129.9163 & 35.6542 & 0.160 & 4 & 5 & $\cdots$ & 20.00 & 18.31 & 1.2 & 1.0 & II \\

1207-52672-512 & 02/02/03 & 126.2081 & 29.6123 & 0.040 & 87 & 36 & $\cdots$ & 20.23 & 18.06 & 1.4 & 1.1 & II \\

1231-52725-553 & 27/03/03 & 186.9265 & 9.9579 & 0.070 & 34 & 49 & $\cdots$ & 20.15 & 18.41 & 1.2 & 1.0 & II \\

1406-52876-528 & 25/08/03 & 243.2197 & 30.8496 & 0.048 & 34 & 49 & $\cdots$ & 20.25 & 18.18 & 1.3 & 1.0 & II \\

1459-53117-022 & 22/04/04 & 198.2796 & 46.0984 & 0.030 & 35 & 50 & $\cdots$ & 19.88 & 19.33 & 2.4 & 1.7 & II \\

1684-53239-484 & 22/08/04 & 245.6866 & 32.6592 & 0.041 & 62 & 36 & $\cdots$ & 19.85 & 17.01 & 1.9 & 1.4 & II \\

1755-53386-516 & 16/01/05 & 174.8064 & 15.0377 & 0.014 & 63 & 82 & $\cdots$ & 19.51 & 18.71 & 3.2 & 1.2 & II \\

2103-53467-081 & 07/04/05 & 180.8496 & 35.3258 & 0.028 & 19 & 34 & $\cdots$ & 19.22 & 17.48 & 1.8 & 0.9 & II \\

2593-54175-334 & 16/03/07 & 157.8944 & 19.0687 & 0.034 & 35 & 113 & $\cdots$ & 19.98 & 19.68 & 2.0 & 1.4 & II \\

\hline

\multicolumn{13}{l}{(1) -- SDSS DR7 plate, MJD, and fiber in which the SN was discovered.} \\
\multicolumn{13}{l}{(2) -- Date on which the SN was captured, in dd/mm/yy.} \\
\multicolumn{13}{l}{(3)--(4) -- Right ascensions and declinations (J2000).} \\
\multicolumn{13}{l}{(5) -- SN host-galaxy redshift.} \\
\multicolumn{13}{l}{(6)--(7) -- SN age, in days, as measured by SVD and \snid, respectively. SN~Ia (II) ages have an uncertainty of $\pm 6~(33)$ days.} \\
\multicolumn{13}{l}{(8) -- SN stretch, as measured with the \salt~ templates. All stretches have an uncertainty of $^{+0.10}_{-0.14}$.} \\
\multicolumn{13}{l}{(9)--(10) -- SN and host-galaxy \R-band magnitudes.} \\
\multicolumn{13}{l}{(11)--(12) -- Reduced $\chi^2$ value of galaxy and galaxy+transient fits.} \\
\multicolumn{13}{l}{(13) -- SN type.}
\end{tabular}
\end{table*}


\section{Supernova Rate Analysis}
\label{sec:results}

\subsection{Visibility time}
\label{subsec:vistime}

In order to measure SN~Ia rates, we calculate a visibility time for each galaxy in our sample.
This is the period of time during which we could have detected a SN~Ia at that galaxy's redshift, with specific stretch and host-galaxy extinction values, and the SN~Ia detection efficiency curves we measured in Section~\ref{subsec:eff}. 
We construct SN~Ia light curves using the \citet{Hsiao2007} SN~Ia template spectra.
Each spectrum is reddened to simulate the effect of dust in the host galaxy using the \citet{cardelli1989} extinction law and $A_V$ values drawn from the \citet{neill2006} extinction model: a one-sided positive Gaussian centred on $A_V=0$ with a standard deviation of $\sigma=0.62$.
Next, the spectra are redshifted according to the redshift of the galaxy.
We apply synthetic photometry to the reddened and redshifted spectra to construct an \R-band light curve.
This light curve is then stretched according to Equation~\ref{eq:stretch}.
Each galaxy in our final sample is assigned a random stretch value from a Gaussian distribution with a standard deviation of $\sigma_s=0.1$, and the mean of the Gaussian is selected according to the sSFR of the galaxy: $\mu_s=0.98$ ($\mu_s=0.93$) for star-forming (passive) galaxies \citep{2006AJ....131..960S}.
Following \citet{2006AJ....131..960S} and \citet{2012ApJ...750....1M}, the stretch values for star-forming (passive) galaxies are restricted to the range $0.6 < s < 1.4$ ($0.6 < s < 1.1$).
We use the \citet{yasuda2010} LF to assign absolute $B$-band magnitudes at maximum light to each light curve.
Each point in the light curve is then multiplied by the appropriate detection efficiency, according to the \R-band magnitude of its host galaxy, as shown in Fig.~\ref{fig:fakes_eff}.
The visibility time is
\begin{equation}\label{eq:vistime}
 t_v = \int\limits_{-\infty}^\infty \epsilon[m(t)]dt,
\end{equation}
where $m(t)$ is the redshifted and reddened template SN~Ia light curve in the \R\ band, and $\epsilon(m)$ is the detection efficiency, as a function of magnitude.

The choice of LF used in the derivation of the visibility time is another source of systematic uncertainty, which we take into account by using three different LFs.
\citet{yasuda2010} assumed that the colour variation of SNe~Ia is due to host-galaxy extinction, with $R_V$ values of either $3.1$, as in our Galaxy, or $1.92$, which is closer to the values found for the host galaxies of SNe~Ia \citep{2008A&A...487...19N,2009ApJS..185...32K}.
\citet{li2011LF} produced an observed LF, which is not corrected for host-galaxy extinction, for SNe~Ia in E--Sa and Sb--Irr galaxies, which we use for the passive and star-forming galaxies in our sample, respectively. 
As it is as yet unclear whether the lower $R_V$ values found in SN~Ia host galaxies are real or due to systematic effects (such as intrinsic SN~Ia colour variance that is not taken into account in SN~Ia light-curve modelers; \citealt{Chotard2011}), we use the $R_V=3.1$ \citet{yasuda2010} LF as the fiducial LF and propagate the systematic uncertainty resultant from using the other LFs into the mass-normalised and volumetric SN Ia rates.
We find that the $R_V=1.92$ \citet{yasuda2010} and \citet{li2011LF} LFs add $+6.5$ and $-8.5$ per cent systematic uncertainties, respectively, to the mass-normalised and volumetric SN~Ia rates.

\subsection{The rate-mass relation}
\label{subsec:rate-mass}

The mass-normalised SN~Ia rate is the number of SNe~Ia, $N_{{\rm Ia},i}$, in galaxies within a specific mass range $i$, divided by the sum of the visibility times, $t_{v,j}$, of the $n$ galaxies within that mass range, weighted by their monitored stellar mass, $M_{*,j}$ (i.e., the mass of the galaxy within the spectral aperture, as observed at its present redshift, after mass loss during stellar evolution):
\begin{equation}\label{eq:mass_rate}
 R_{{\rm Ia},i}=\frac{N_{{\rm Ia},i}}{\sum\limits_{j=1}^n t_{v,j} M_{*,j}}.
\end{equation}
For the total stellar mass of the galaxies in our sample (used only for the purpose of binning the galaxies into various total-mass ranges), we use the \vespa~values from T09, with the 0.55 correction, and with the original T09 aperture correction.
As in other such studies, we present our rates in units of SNuM ($10^{-12}~{\rm M_{\odot}}^{-1}~{\rm yr}^{-1}$).
We measure the mass-normalised SN~Ia rates in four total mass bins, where the limits of the bins are chosen so that each bin contains approximately the same number of SNe~Ia.
We measure the rates of all galaxies, and specifically of star-forming and passive galaxies, distinguished according to their sSFR, using the previously defined borders.
Over the redshift range of our SN~Ia sample, $0.014<z<0.19$, we do not find a dependence of the rate on redshift, whether for all galaxies or for galaxies separated by sSFR.
We therefore analyse together all redshifts in our sample.
The cited uncertainties of the rates are the Poisson uncertainties on the number of SNe~Ia in each mass bin and the systematic uncertainties propagated from the different LFs used in the calculation of the visibility time, together with the number of possibly misclassified or extra-aperturial SNe Ia.
The systematic uncertainty stemming from the extra-aperturial SNe are added only to the bins that include the two possible SNe: 418-51884-144 is a $\sim 50 \times 10^{10}~{\rm M_{\odot}}$ passive galaxy, and 2594-54177-348 is a $\sim 2 \times 10^{10}~{\rm M_{\odot}}$ star-forming galaxy.
These rates are summarized in Table~\ref{table:rates}.

\begin{table}
\center
\hspace{1.5in}\parbox{5.8in}{\caption{Mass-normalised SN~Ia rates}\label{table:rates}}
 \begin{tabular}{c c c c}
  \hline
  \hline
  mass range & median$^a$ & SNe Ia & SN~Ia rate$^b$ \\
  {[$10^{10}~{\rm M_{\odot}}$]} & {[$10^{10}~{\rm M_{\odot}}$]} & & {[SNuM]} \\
  \hline
  \multicolumn{4}{c}{All galaxies} \\
  \hline
  $0 < M_* < 3$       & $1.3_{-1.0}^{+1.1}$   & 23 & $0.171_{-0.035,-0.011}^{+0.044,+0.009}$ \\
  $3 \leq M_* < 5.5$  & $4.1_{-0.8}^{+0.9}$   & 22 & $0.132_{-0.028,-0.006}^{+0.035,+0.008}$ \\
  $5.5 \leq M_* < 10$ & $7.4_{-1.3}^{+1.7}$   & 23 & $0.096_{-0.020,-0.007}^{+0.024,+0.005}$ \\
  $M_* \geq 10$       & $20_{-8}^{+24}$       & 22 & $0.057_{-0.012,-0.010}^{+0.015,+0.004}$ \\
  All masses          & $6_{-5}^{+16}$        & 90 & $0.10 \pm 0.01 \pm 0.01$ \\
  \hline
  \multicolumn{4}{c}{Star-forming galaxies} \\
  \hline
  $0 < M_* < 2$    & $0.8_{-0.6}^{+0.8}$  & 11 & $0.22_{-0.06,-0.02}^{+0.09,+0.01}$ \\
  $2 \leq M_* < 4$ & $2.9_{-0.6}^{+0.7}$  & 13 & $0.18_{-0.05,-0.00}^{+0.06,+0.01}$ \\
  $4 \leq M_* < 8$ & $5.6_{-1.1}^{+1.5}$  & 13 & $0.11_{-0.03,-0.001}^{+0.04,+0.006}$ \\
  $M_* \geq 8$     & $12_{-3}^{+9}$       & 10 & $0.061_{-0.019,-0.001}^{+0.026,+0.006}$ \\
  All masses       & $3_{-2}^{+6}$        & 47 & $0.118_{-0.017,-0.004}^{+0.020,+0.008}$ \\
  \hline
  \multicolumn{4}{c}{Passive galaxies} \\
  \hline
  $0 < M_* < 4$     & $2.3_{-1.3}^{+1.1}$    & 10 & $0.13_{-0.04,-0.008}^{+0.05,+0.008}$ \\
  $4 \leq M_* < 8$  & $5.9_{-1.3}^{+1.4}$    & 10 & $0.076_{-0.024,-0.007}^{+0.032,+0.004}$ \\
  $8 \leq M_* < 12$ & $9.8_{-1.3}^{+1.4}$    & 10 & $0.10_{-0.03,-0.011}^{+0.02,+0.004}$ \\
  $M_* \geq 12$     & $25_{-10}^{+25}$       & 13 & $0.062_{-0.017,-0.017}^{+0.022,+0.004}$ \\
  All masses        & $11_{-7}^{+22}$        & 43 & $0.082_{-0.012,-0.003}^{+0.015,+0.008}$ \\
  \hline
  \multicolumn{4}{l}{$^a$ Uncertainties represent the mass region occupied by 68 per} \\
  \multicolumn{4}{l}{cent of the galaxies in that bin. Stellar masses are based on an} \\
  \multicolumn{4}{l}{assumed \citet{Kroupa2007} IMF, via the T09 SFH reconstruction.} \\
  \multicolumn{4}{l}{$^b$ Rate uncertainties are Poisson uncertainties on the number} \\ 
  \multicolumn{4}{l}{of SNe~Ia in each mass bin. Systematic uncertainties, from} \\
  \multicolumn{4}{l}{using different LFs and from extra-aperturial or possibly} \\
  \multicolumn{4}{l}{misclassified SNe, are separated by commas.} \\
 \end{tabular}
\end{table}

The mass-normalised SN~Ia rate, averaged over all masses and redshifts in our sample, is $R_{{\rm Ia,M}} = \snumrate$.
For passive galaxies alone, the mass-normalised SN~Ia rate is $R_{{\rm Ia,M}}=0.082^{+0.015}_{-0.012}~{\rm(statistical)}~^{+0.008}_{-0.003}~{\rm(systematic)}$~SNuM.
This last value is consistent with the values obtained by the two other main galaxy-targeted surveys, $0.044^{+0.016}_{-0.014}$~SNuM for E/S0 galaxies \citep{2005A&A...433..807M}, and $0.125^{+0.026}_{-0.022}(0.028)~[0.104^{+0.016}_{-0.014}(0.024)]$~SNuM for elliptical (S0) galaxies (L11), where the uncertainties are statistical and systematic, respectively.

\begin{figure}
  \centering
  \includegraphics[width=0.5\textwidth]{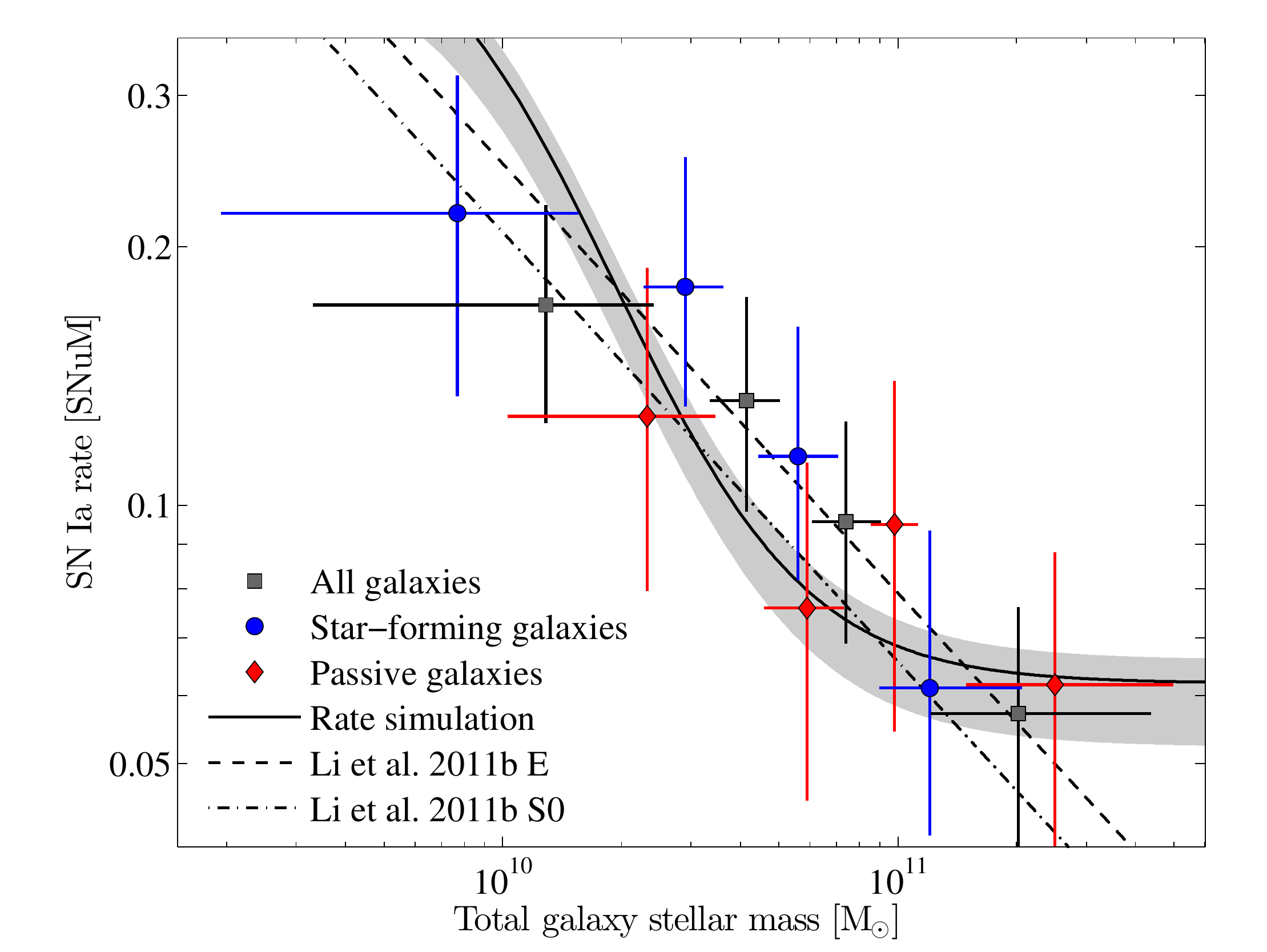}
  \caption{SN~Ia rates per unit stellar mass, as a function of total galaxy stellar mass. The mass-normalised SN~Ia rates for all galaxies in each mass bin are shown as black squares, while the rates for star-forming (passive) galaxies are shown as blue circles (red diamonds). Vertical error bars are based on the Poisson uncertainty on the number of SNe~Ia in the specific mass bin together with the systematic uncertainty stemming from misclassification and extra-aperturial SNe, along with using different LFs in the calculation of the visibility time; and the horizontal error bars denote the range within which 68 per cent of the galaxies fall within the mass bin. The solid curve shows the best-fitting SN~Ia rate as a function of stellar mass, as derived from a combination of a $t^{-1}$ DTD and the \citet{2005MNRAS.362...41G} relation between mass and galaxy age. The shaded area is the confidence region resulting from the 68 per cent statistical uncertainty of the DTD amplitude $\Psi_{1\rm Gyr}$. The dashed (dot-dashed) curves are the L11 power-law fits to their mass-normalised SN~Ia rates in local elliptical (S0) galaxies.}
  \label{fig:SNuM_dtd}
\end{figure}

Fig.~\ref{fig:SNuM_dtd} shows the SN~Ia rate per unit mass, as a function of galaxy mass, for all galaxies in our sample, and separately for passive and star-forming galaxies. 
The mass-normalised SN~Ia rates decrease with increasing galaxy mass.
We thus confirm, with the present sample at $z \sim 0.1$, the similar dependence found by L11 for SNe~Ia discovered in the local LOSS sample and by \citet{2012ApJ...755...61S} for SDSS-II SNe.

\citet{2011arXiv1106.3115K} pointed out that the rate-mass relation found by L11 can be explained by a $t^{-1}$ DTD, combined with the fact that massive galaxies formed earlier in the cosmic history.
We repeat here the \citet{2011arXiv1106.3115K} calculation, with some small differences.
The mass-normalised SN~Ia rate of galaxy $i$ at cosmic time $t$ can be expressed as the convolution of the SFH, $S(t)$, and a DTD, $\Psi(t)$, divided by the total stellar mass of the galaxy, $M_{*,i}$, after mass loss due to stellar evolution:
\begin{equation}\label{eq:rate_simulation}
 R_{{\rm Ia},i}(t) =\frac{1}{M_{*,i}} \int_{0}^t S(t')\Psi(t-t')dt'.
\end{equation}
Following \citet{2011arXiv1106.3115K}, we use the \citet{2005MNRAS.362...41G} relation between the stellar mass of a galaxy and its age to derive ages for a random selection of $10{,}000$ galaxies with redshifts drawn from the redshift distribution of the SDSS DR7 sample.
For each galaxy, we draw a galaxy age from a Gaussian distribution centred on the median values in table 2 from \citet{2005MNRAS.362...41G}, with the 16/84 per cent values acting as the distribution's lower and upper standard deviations.
Following \citet{2005MNRAS.362...41G}, we use an exponential SFH of the form $e^{-\alpha t}$, with indices $\alpha$ drawn from a flat distribution between 0 and 1.
The SFH is scaled to produce the galaxy's formed mass, $M_f$, over the period of time between a galaxy's formation time, $t_g$, and $t$.
We assume a power-law DTD with index $-1$, with the amplitude $\Psi_{1\rm Gyr}$ at $t=1$~Gyr left as a free parameter.
Equation~\ref{eq:rate_simulation} thus becomes:
\begin{equation}\label{eq:rate_simulation_all}
 R_{{\rm Ia},i}(t) = \frac{M_f}{M_*} \Psi_{1\rm Gyr} \int_{0}^t e^{-\alpha t'} (t-t')^{-1} dt'.
\end{equation}
As our measurements cover mainly old galaxies in the mass range $\sim 10^9$--$10^{12}~{\rm M_\odot}$, we set $M_f/M_*=2.3$ \citep{2003MNRAS.344.1000B}.
The best-fitting DTD amplitude, with $\chi^2_r = 1.5$, is $\Psi_{1\rm Gyr} = 0.070 \pm 0.016 \times 10^{-12}~{\rm M_\odot}^{-1}~{\rm yr}^{-1}$, where the uncertainty of $\Psi_{1\rm Gyr}$ is the 68 per cent confidence region, defined as the range of $\Psi_{1\rm Gyr}$ values within $\Delta \chi^2 = \pm 1$ of the minimal $\chi^2$ value.
The form of the mass-normalised SN~Ia rate as a function of mass is similar to the one obtained by \citet{2011arXiv1106.3115K}, though with a shallower decline at the high-mass end.
Fig.~\ref{fig:SNuM_dtd} shows the result of this simulation.

\subsection{The Type Ia supernova volumetric rate}
\label{subsec:rate_vol}

We can convert the mass-normalised SN~Ia rate, averaged over all masses and redshifts in our sample, to a volumetric rate, by multiplying it by the total cosmic mass density, as inferred from the galaxy stellar mass function (GSMF) measured at $z < 0.06$ by \citet{2012MNRAS.421..621B}.
However, the SDSS was a galaxy-targeted survey, and the distribution of galaxy mass is biased towards high-mass galaxies, as shown in Fig.~\ref{fig:DR7_gals}.
This, in turn, biases the size of our SN~Ia sample, as SNe~Ia are more common in low-mass galaxies (L11 and Fig.~\ref{fig:SNuM_dtd}).
Therefore, we must account for our galaxy sample's particular distribution of mass, $M$, weighted by the rate-mass relation we have found for these galaxies.
We do this by multiplying our mass-normalised rate by the ratio of the integrated cosmic GSMF, $B(M)$, to the SDSS DR7 galaxy-mass distribution, $D(M)$ (which is normalised so that $\int D(M) M dM = 1$), where both mass functions are weighted by the rate-mass relation $R(M)$ described in Section~\ref{subsec:rate-mass}:
\begin{equation}\label{eq:rate_vol}
 R_{{\rm Ia,V}} = R_{{\rm Ia,M}} \frac{\int B(M) R(M) M dM}{\int D(M) R(M) M dM}.
\end{equation}
The stellar estimates at the basis of the \citet{2012MNRAS.421..621B} GSMF assumed a \citet{2003PASP..115..763C} IMF, which is similar to the \citet{Kroupa2007} IMF used in \vespa.

\begin{figure}
 \begin{center}
 \includegraphics[width=0.5\textwidth]{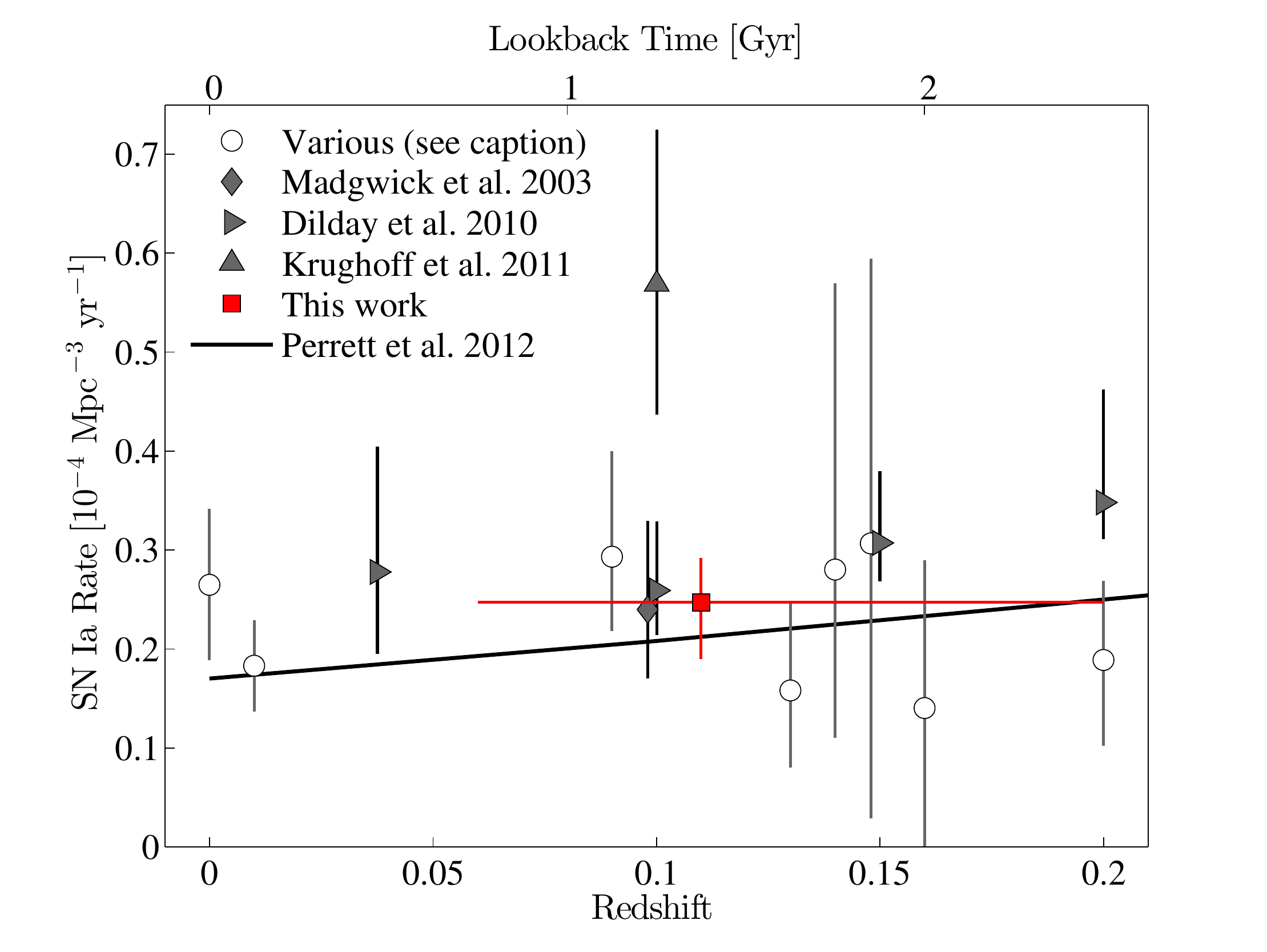} \\
 \includegraphics[width=0.5\textwidth]{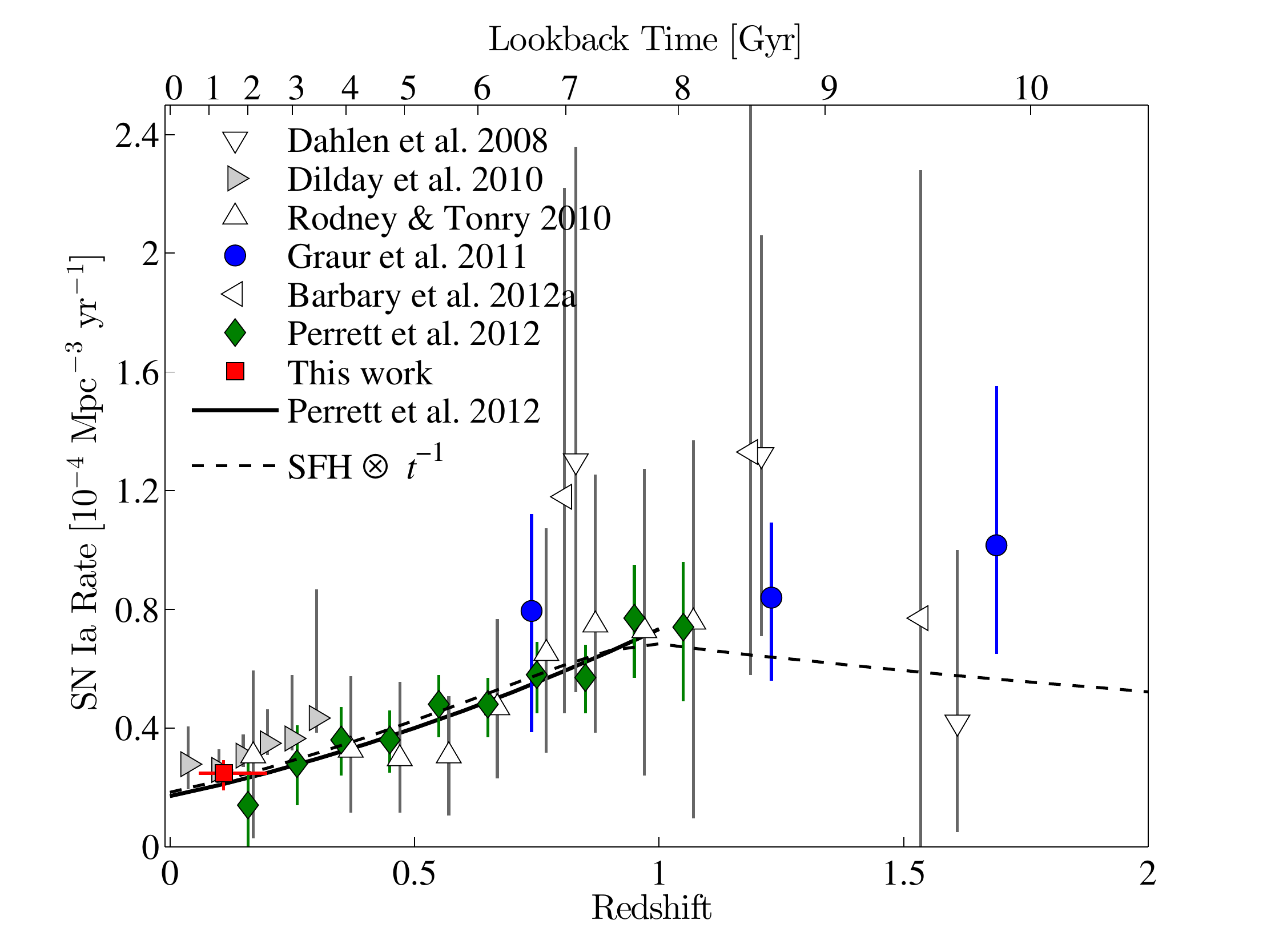}
 \caption{SN~Ia volumetric rate at $z=0.11$ (red square) compared to rates from the literature. Upper panel: circles denote results by \citet*{cappellaro1999}; \citet{hardin2000,blanc2004,horesh2008,dilday2010a,2010ApJ...723...47R}; L11; and P12. The M03 measurement, based on a spectroscopic SN search in SDSS DR1, is the black diamond, and the K11 measurement, based on SDSS DR5, is the black upright triangle. The \citet{dilday2010a} SDSS-II measurements are the right-facing triangles. The solid curve is the best-fitting power-law dependence on redshift found by P12 for their measured SN~Ia rates between $z=0.1$ and 1. All vertical error bars include statistical and systematic uncertainties. The horizontal error bar indicates the range that encompasses 68 per cent of the SDSS DR7 galaxies around the median redshift of $z=0.11$.
 Lower panel: our low-$z$ SN~Ia rate (red square) compared to recent high-$z$ rates -- \citet*[downturned triangles]{dahlen2008}, \citet[upturned triangles]{2010ApJ...723...47R}, \citet[right-facing triangles]{dilday2010a}, G11 (blue circles), \citet[left-facing triangles]{2012ApJ...745...31B}, and P12 (green diamonds). The solid curve is the same as above, and the dashed curve is the best-fitting SN~Ia rate evolution derived from convolving the \citet{2008ApJ...683L...5Y} cosmic SFH with a power-law DTD and fitting our volumetric rate measurement, together with those of P12 and G11. The most precise measurements in each redshift range are marked with filled symbols.}
 \label{fig:vol_rate}
\end{center}
\end{figure}

The volumetric SN~Ia rate, at a median redshift of $z=0.11$, is $R_{{\rm Ia,V}} = \volrate$.
The statistical uncertainty in the rate derives from the Poisson uncertainty from the size of the SN~Ia sample.
The systematic uncertainty is composed of two elements: the uncertainty propagated from using various LFs in the calcualtion of the visibility time; and an uncertainty of $-3.5$ SNe~Ia, as derived from the purity of our discovery and classification method and the possibility that some of our SNe exploded outside of the spectral aperture, as detailed in Sections~\ref{subsec:eff} and ~\ref{sec:sample}.
Varying the GSMF according to the uncertainties reported in table 1 of \citet{2012MNRAS.421..621B} adds a statistical uncertainty an order of magnitude smaller than the Poisson uncertainty, and is thus negligible.
With the systematic uncertainties, the total uncertainty is thus $\sim 20$ per cent.
The uncertainty, however, does not include the systematics in the mass estimates of the individual galaxies by T09, which are difficult to estimate.
Our volumetric SN~Ia rate measurement, along with measurements from other SN surveys in this redshift range, are shown in the upper panel of Fig.~\ref{fig:vol_rate}.
Our volumetric rate is in excellent agreement with most previous measurements (e.g., M03; \citealt{dilday2010a}).
Furthermore, our rate merges smoothly with the extrapolation to low redshift of the rates measured by P12 (whose own low-$z$ measurement has large errors).
This is shown in the lower panel of Fig.~\ref{fig:vol_rate}.
Our low-$z$ rate measurement, together with the P12 measurements, thus provides now a precise picture of the SN~Ia rate evolution from $z=0$ to 1.
Following G11 and P12, we convolve a power-law DTD with different cosmic SFHs (see G11, section 8) and fit our volumetric SN~Ia rate, together with those of P12 and G11 (see Fig.~\ref{fig:vol_rate}).
We find a best-fitting DTD power-law index of $-1 \pm 0.1 \pm 0.16$, where the uncertainties are statistical and systematic, respectively, and the systematic uncertainty derives from the different SFHs used in the fit.

Our rate calculation is improved compared to the SN~Ia rates derived by the two previous studies that have searched for SNe in SDSS spectra.
M03 estimated the $B$-band-luminosity-weighted SN~Ia rate at $z \approx 0.1$ using the average galaxy luminosity value of their sample and a fixed visibility time estimated at $t_v=20$~days.
They obtained $0.4 \pm 0.2~h^2$~SNuB ($10^{-12}~{\rm L_{B,\odot}^{-1}~yr^{-1}}$), where $h=100~{\rm km~s^{-1}~Mpc^{-1}}$.
\citet{horesh2008} converted this rate into a volumetric rate of $0.24 \pm 0.12 \times 10^{-4}~{\rm SNe}~{\rm yr}^{-1}~{\rm Mpc}^{-3}$ using the redshift-dependent \citet{botticella2008} luminosity density function, but without taking into account the then-unknown rate-mass relation, which weights the particular luminosity distribution of the galaxies in the SDSS DR1 sample (see Eq.~\ref{eq:rate_vol}).
Similarly, K11 obtained their volumetric rate, $0.569^{+0.098,+0.058}_{-0.085,-0.047} \times 10^{-4}~{\rm SNe}~{\rm yr}^{-1}~{\rm Mpc}^{-3}$ (the errors are statistical and systematic, respectively) by converting their $B$-band-luminosity-weighted rate of $0.472^{+0.081,+0.048}_{-0.071,-0.039}$~SNuB using the same redshift-dependent luminosity density function, again without accounting for the effect of the rate-mass relation.
Finally, neither M03 nor K11 made use of the detailed spectral information available for every galaxy in the sample, which can give the stellar masses of the galaxies (and which we have used for calculating rates, above) and the SFHs of the galaxies (which we use for recovering the DTD, below).

\subsection{The Type Ia supernova delay-time distribution}
\label{subsec:dtd}
The \vespa\ SFHs that are available for each of the individual galaxies that were effectively `monitored' by our SN survey permit applying to the data the M11 algorithm for recovering the DTD.
In this approach, briefly, Eq.~\ref{eq:rate_simulation} is approximated by a discretised form,
\begin{equation}\label{eq:discrete}
N_i=\sum_j m_{ij} \Psi_j \Delta t_i, 
\end{equation}
where $N_i$ is the expectation value of the SNe~Ia in the $i$th galaxy, $m_{ij}$ is the stellar mass formed in the galaxy over the $j$th time interval, ${\bf \Psi}$ is a binned version of the DTD, and $\Delta t_i$ is the visibility time of the $i$th galaxy.
The sum is over the three \vespa\ time bins: $<0.42$~Gyr (`prompt'), 0.42--2.4~Gyr (`intermediate'), and $>2.4$~Gyr (`delayed').
The values of $\Psi_j$ are treated as free parameters that are scanned so as to best match $N_i$ to the Poisson statistics of the observed sample, which has zero SNe in most of the monitored galaxies, and one SN in a small fraction of the galaxies.
M11, analysing the LOSS sample of L11, detected significant prompt and delayed components to the DTD.
Applying the method to a sample of SNe~Ia from the SDSS-II survey \citep{dilday2010a,2011ApJ...738..162S}, M12 obtained $4\sigma$ detections of all three DTD components.

For the current SN~Ia sample and the \vespa-derived SFHs for our galaxies, we obtain a strong detection of a delayed component in the age bin $>2.4$~Gyr, of \dtdrate, where the systematic uncertainty derives from the different LFs used in the calculation of the visibility time and the $-3.5$ SNe Ia due to possible misclassification and extra-aperturial SNe. 
Our recovered delayed component of the SN~Ia DTD, along with measurements from other SN surveys, are shown in Fig.~\ref{fig:direct_dtd}.
For the purpose of this compilation, the DTD values from M11, M12, and this work, have been scaled down by a factor of 0.7.
This factor converts the \citet{Kroupa2007} IMF assumed by \vespa, and therefore implicit in those DTDs, to the ``diet Salpeter'' IMF \citep{2003ApJS..149..289B} assumed in the other measurements shown.
Our delayed component measurement is 2--3 times higher than the corresponding value in M12, but similar to the value obtained by M11.
It is possible that this result is affected systematically by the `flux flow' problem we have identified in galaxies that hosted SNe.
Some of the blue light of a host galaxy, light from a young or intermediate-age stellar population, is removed from the galaxy spectrum by the SN~Ia template subtraction.
As a result, some of the DTD signal could `drift' from the prompt and intermediate bins to the delayed bin.
Alternatively, the high level of the delayed component could be real, a result of the dominance of massive, early-type galaxies in the sample (see Section~\ref{subsec:N/M*}, below).

We do not detect prompt and intermediate components to the DTD.
In retrospect, this is unsurprising, since our galaxy sample is dominate by old and massive galaxies, with little ongoing star formation.
90 SNe~Ia were hosted by the $707{,}792$ galaxies in our sample.
Of these, only $5{,}394$ galaxies have ${\rm log(sSFR/yr^{-1})} > -9.5$, none of which hosted a SN~Ia.
The SDSS-II sample used in M12, on the other hand, is composed of $66{,}400$ galaxies, which hosted a total of 132 SNe~Ia.
$3{,}867$ galaxies had ${\rm log(sSFR/yr^{-1})} > -9.5$, seven of which hosted a SN.
Comparing between the two samples, we would have expected $(66{,}400/707{,}792) \times (5{,}394/3{,}867) \times (90/132) \times 7 = 0.6$ SNe~Ia in the highly-star-forming galaxies in our sample, which is consistent with our having found none.
With no detected SNe~Ia in young galaxies, our method is unable to reconstruct the short- and intermediate-delay DTD components.

\subsection{The Type Ia supernova production efficiency}
\label{subsec:N/M*}

Integrating the SN~Ia DTD over a Hubble time yields the number of SNe~Ia per unit stellar mass formed in a short burst of star formation, $N/M_*$.
For intercomparing with other results, we again assume consistently a ``diet Salpeter'' IMF, which is similar to the \citet{1955ApJ...121..161S} IMF, but with 70 per cent of the mass \citep{2003ApJS..149..289B}.
The \citet{Kroupa2007} IMF used in \vespa\ has, yet again, 70 per cent of the mass in the diet Salpeter IMF \citep{2003ApJS..149..289B}, so we correct the $N/M_*$ values derived from our DTD reconstructions accordingly.

Each of our DTD reconstructions leads to different $N/M_*$ values.
In Section~\ref{subsec:rate-mass}, we set the index of the power-law DTD to $-1$ and fit for the normalisation.
This results in $N/M_* = \NMkis$.
This value is lower than those obtained by M11, M12, G11, and P12, but does not take into account the uncertainty in the slope of the DTD.

G11 found that $N/M_*$ lay in the range $N/M_* = (0.5$--$1.5) \times 10^{-3}~{\rm SNe}~{\rm M_\odot}^{-1}$.
Taking into account the statistical and systematic uncertainties derived from the cosmic SFHs used in G11, but fitting only the volumetric rate derived in Section~\ref{subsec:rate_vol}, together with those from G11 and P12, we find that $N/M_*$ lies in the range $N/M_* = (0.4$--$1.2) \times 10^{-3}~{\rm SNe}~{\rm M_\odot}^{-1}$.

\begin{figure}
 \begin{center}
 \includegraphics[width=0.5\textwidth]{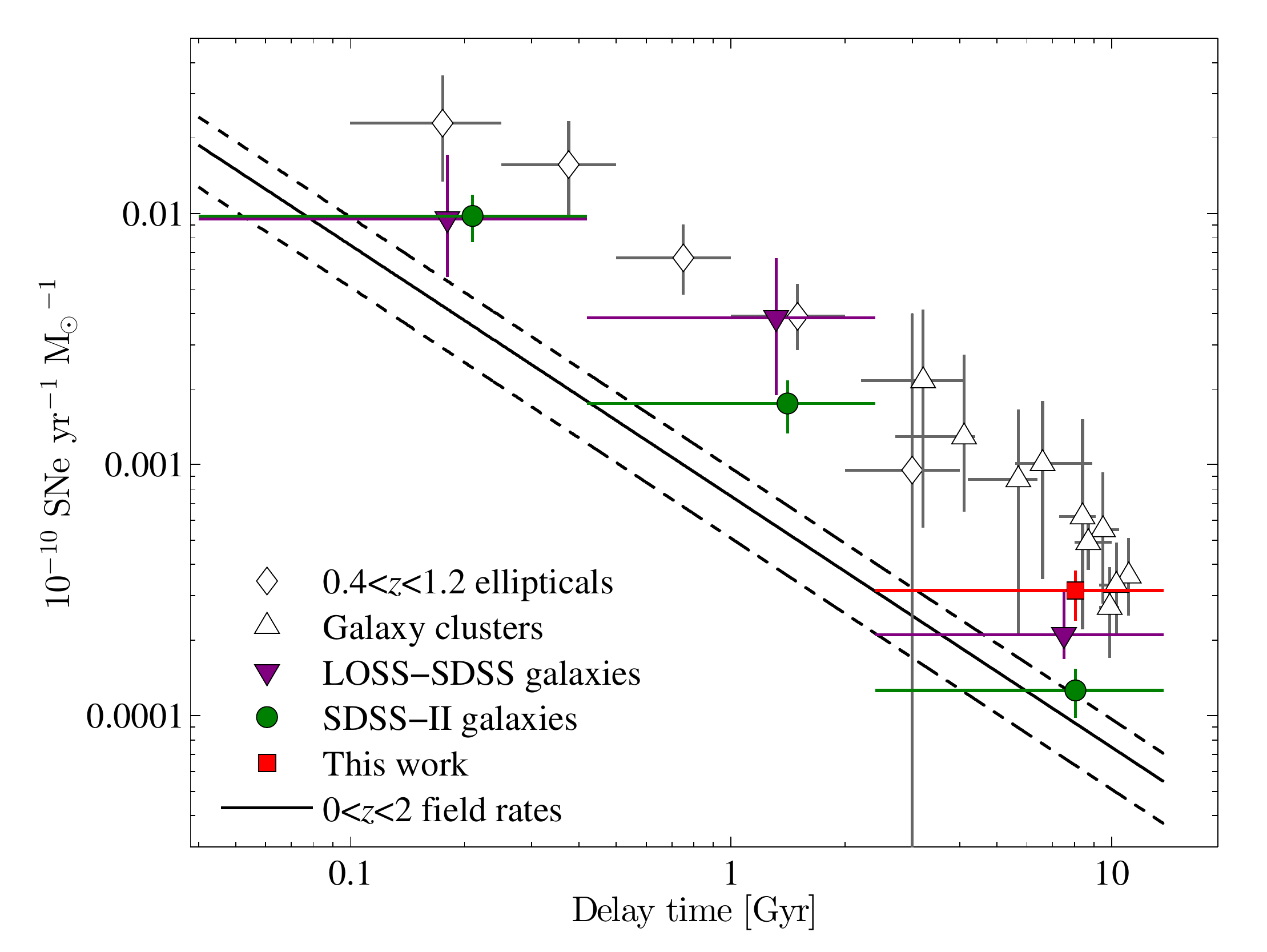}
 \caption{Recovered `delayed' SN~Ia DTD component (red square) compared to previous measurements from the literature: \citet{2008PASJ...60.1327T} DTD measurements from SNe~Ia discovered in $0.4<z<1.2$ elliptical galaxies (diamonds); galaxy cluster surveys (triangles; \citealt{maoz2010clusters} and references therein; \citealt{2012ApJ...746..163S}); and the `prompt', `intermediate', and `delayed' components recovered with the M11 method: purple downturned triangles for M11 and green circles for M12. The solid curve is a $t^{-1}$ DTD that, when convolved with the \citet{2008ApJ...683L...5Y} cosmic SFH, provides the best fit to our volumetric rate, together with those of P12 and G11 (see Fig.~\ref{fig:vol_rate}). The dashed curves are the systematic uncertainty on the amplitude of the DTD, derived from using different cosmic SFHs. Where necessary, measurements have been converted to account for the diet Salpeter IMF used here.}
 \label{fig:direct_dtd}
\end{center}
\end{figure}

Because our galaxy sample is composed mainly of old, massive galaxies, in Section~\ref{subsec:dtd} we recover only the delayed ($>2.4$~Gyr) component of the DTD.
When multiplied by the range of delay times it spans, $11.3$~Gyr, this component gives $N/M_* = \NMdtd$.
This is only one component of the overall $N/M_*$ (see M12, equation 10), which is why it is several times lower than the overall values found by M11 and M12.
By itself, however, this component of $N/M_*$ is larger than the respective components in M11 and M12.
This may be due to the flux flow problem described in Section~\ref{subsec:eff} and shown in Fig.~\ref{fig:fakes_mags}, which steals blue light from the host galaxies and biases the \vespa-derived stellar masses of galaxies hosting bright SNe.
Alternatively, a higher $N/M_*$ or a higher delayed DTD component in massive, old galaxies, as suggested by Fig.~\ref{fig:direct_dtd}, may be a real phenomenon, perhaps related to IMF effects (see M12).

\section{SUMMARY AND CONCLUSIONS}
\label{sec:discuss}

We have surveyed the $\sim 700{,}000$ galaxy spectra in SDSS DR7 that have \vespa-derived SFHs and discovered and classified 90 SNe~Ia and 10 SNe~II.
Using our SN~Ia sample, together with the \vespa-derived stellar masses of the galaxies we survey, we measure the mass-normalised SN~Ia rates at a median redshift of $z=0.11$ and find that they follow the so-called `rate-size', or rate-mass, relation, which was previously observed in the local Universe by L11.
We confirm that more massive galaxies have lower SN~Ia rates per unit stellar mass, and that, for galaxies of a given total mass, the mass-normalised SN~Ia rate is weakly dependent on the galaxy's sSFR.
Following \citet{2011arXiv1106.3115K}, we also confirm that this rate-mass relation is essentially a coincidental `conspiracy.'
On the one hand, less massive galaxies tend to form later in the history of the Universe (`downsizing').
On the other hand, numerous recent SN~Ia rate studies point to a DTD that peaks at short delays and decreases monotonically with time (\citealt{maoz2010clusters,maoz2010magellan}; G11; M11; M12; \citealt{2012ApJ...745...32B,2012ApJ...749L..11B}; P12).
The rate-mass relation results from the fact that, for more massive galaxies, we are probing further down along the delayed tail of the DTD.
Our mass-normalised SN~Ia rates cover a mass range of $\sim 10^9$--$10^{12}~{\rm M_\odot}$.
In the future, it would be interesting to extend the mass-normalised SN~Ia rates to both lower and higher galaxy masses, where our (albeit simplified) calculation predicts that the rate per unit mass will plateau.

We have also derived a mass-normalised SN~Ia rate, averaged over all masses and redshifts in our galaxy sample, $\snumrate$, and a volumetric rate, $R_{{\rm Ia,V}}(z=0.11) = \volrate$.
The latter is consistent both with previous measurements from other surveys and with the low-redshift extrapolation of the power-law dependence on redshift fit by P12 to their SN~Ia rate measurements.

As the SDSS DR7 galaxy sample is composed mainly of massive, old galaxies, applying the direct DTD recovery method of M11 yields only a `delayed' component of \dtdrate\ in the delay-time range $>2.4$~Gyr.
This value is 2--3 times higher than the corresponding value in M12, but similar to the value obtained by M11.
The differences may be due to the systematic flux flow problem that exists in our SN discovery process.
We find that the time-integrated number of SNe~Ia per unit formed stellar mass derived from a power-law DTD with index $-1$ and scaled to yield the above value, is $N/M_* = \NMkis$, on the lower end of the range of $(0.4$--$1.2) \times 10^{-3}~{\rm SNe}~{\rm M_\odot}^{-1}$ that we recover from fitting our volumetric SN~Ia rate, together with those from G11 and P12.
These $N/M_*$ values are consistent with the values obtained by G11 and P12, but lower than those obtained by M11.
It is as of yet unclear whether the difference in SN~Ia production efficiency seen in volumetric field surveys and surveys of old stellar populations reveals a real difference (see M12 for a discussion), or may be a systematic effect caused by, for example, overstimation of the cosmic SFH because of over-correction of dust extinction.

In terms of the progenitor question, our results are consistent with previous indications for a power-law DTD with an index $\sim -1$.
This comes mainly through the rate-mass relation that we measure (which is naturally explained with such a DTD), but also through the delayed DTD component we recover directly (this component is part of the above DTD), and through the precise volumetric rate we find (which merges smoothly with rates at higher redshifts and, compared to the cosmic SFH, implies such a DTD).
It remains to be seen whether theoretical SD models may produce such a DTD, or whether this result indicates that many, most, or all SNe~Ia arise through the DD channel.
Further studies, using both SN~Ia rates and other relevant techniques, will shed light on this issue.

While large photometric SN surveys, such as the SDSS SN Survey \citep{dilday2010a,2011ApJ...738..162S}, Supernova Legacy Survey (\citealt{neill2006}; P12), and the ongoing PTF have generated large samples of SNe of all types, they require costly spectroscopic follow-up in order to robustly classify their candidates.
Massive spectroscopic surveys, such as previous data releases of the SDSS or the ongoing SDSS-III \citep{2011AJ....142...72E}, provide an ideal platform for our code, with which one can assemble large transient samples at no extra cost.
Spectroscopic SN surveys, however, also have several disadvantages when compared to traditional imaging-based surveys:
\begin{enumerate}
 \item\noindent The SN yield per galaxy or per exposure time is lower, for several reasons: (a) spectroscopy requires longer exposure times than imaging, hence fewer galaxies can be targeted in a given amount of time; (b) the spectral apertures used in surveys such as SDSS cover only certain areas of specific galaxies in the telescope's field of view (though this can be obviated with the use of integral-field or grism spectroscopy); and (c) the two-dimensional brightness contrast of a SN over its host galaxy in an image is greater than the one-dimensional contrast in a spectrum, which leads to shorter visibility times (the mean SN~Ia visibility time of our survey is 39 days).
 \item\noindent Imaging-based SN surveys can be done in rolling mode, allowing for the measurement of light curves (though these can also be measured using follow-up imaging).
 A rolling spectroscopic survey is impractical, though that may change in the future. 
\end{enumerate}

The detection and classification method introduced here can be improved upon by, e.g., optimizing it for other types of SNe and by replacing the template library with bases of SN eigenspectra, similar to the \salt\ templates. 
Furthermore, we will search for ways to discover new types of transients that cannot be fitted with current spectral templates.
Finally, although the current work has been an archival search, there is no obstacle to discovering SNe in spectra in real time, while the SNe can still be followed up. 
We are currently applying such an approach to  spectra from the SDSS-III Baryon Oscillation Spectroscopic Survey \citep{2012arXiv1208.0022D}.


\section*{Acknowledgments}
We thank Iair Arcavi, Carles Badenes, St\'{e}phane Blondin, Avishay Gal-Yam, Yuqian Liu, Filippo Mannucci, Tom Matheson, Maryam Modjaz, Hans-Walter Rix, Rita Tojeiro, Benny Trakhtenbrot, and the anonymous referee for helpful discussions and comments.
OG thanks Robert Quimby and the Kavli IPMU for their hospitality in the course of this work.
This work was supported by a grant from the Israel Science Foundation.

Funding for the SDSS and SDSS-II has been provided by the Alfred P. Sloan Foundation, the Participating Institutions, the National Science Foundation, the U.S. Department of Energy, the National Aeronautics and Space Administration, the Japanese Monbukagakusho, the Max Planck Society, and the Higher Education Funding Council for England. The SDSS Web Site is http://www.sdss.org/.

The SDSS is managed by the Astrophysical Research Consortium for the Participating Institutions. The Participating Institutions are the American Museum of Natural History, Astrophysical Institute Potsdam, University of Basel, University of Cambridge, Case Western Reserve University, University of Chicago, Drexel University, Fermilab, the Institute for Advanced Study, the Japan Participation Group, Johns Hopkins University, the Joint Institute for Nuclear Astrophysics, the Kavli Institute for Particle Astrophysics and Cosmology, the Korean Scientist Group, the Chinese Academy of Sciences (LAMOST), Los Alamos National Laboratory, the Max-Planck-Institute for Astronomy (MPIA), the Max-Planck-Institute for Astrophysics (MPA), New Mexico State University, Ohio State University, University of Pittsburgh, University of Portsmouth, Princeton University, the United States Naval Observatory, and the University of Washington.


\end{document}